\documentclass[11pt,twoside,table,vi]{mitthesis}
\usepackage{lgrind}
\usepackage{cmap}
\usepackage[T1]{fontenc}
\usepackage{graphicx}
\usepackage{amsmath}
\usepackage{multirow}
\usepackage[breaklinks=true]{hyperref}

\usepackage{color}
\usepackage{soul}    
\usepackage{gensymb} 
\usepackage[table]{xcolor}
\usepackage{hhline} 

\usepackage{enumitem}

\usepackage[a-1b]{pdfx}   

\usepackage{pifont}

\graphicspath{ {./images/} }

\def\all{all}
\ifx\files\all  \else 

\begin{document}

\title{Ubiquitous Metadata: Design and Fabrication of \\ Embedded Markers for Real-World Object \\ Identification and Interaction}

\author{Mustafa Doğa Doğan}
\prevdegrees{B.S., Boğaziçi University (2018) \\
S.M., Massachusetts Institute of Technology (2020)}
\department{Department of Electrical Engineering and Computer Science}

\degreenew{Doctor of Philosophy}

\degreemonth{February}
\degreeyear{2024}
\thesisdate{January 26, 2024}


\supervisor{Stefanie Mueller}{Associate Professor of Electrical Engineering and Computer Science}

\chairman{Leslie A. Kolodziejski}{Professor of Electrical Engineering and Computer Science \\ Chair, Department Committee on Graduate Students}

\maketitle
 


\cleardoublepage
\setcounter{savepage}{\thepage}
\begin{abstractpage}
%
%
%

The convergence of the physical and digital realms has ushered in a new era of immersive experiences and seamless interactions. As the boundaries between the real world and virtual environments blur
and result in a "mixed reality,"
there arises a need for robust and efficient methods to connect physical objects with their virtual counterparts. In this thesis, we present a novel approach to bridging this gap through the design, fabrication, and detection of embedded machine-readable markers.

The vision of mixed reality relies on mobile and wearable devices being aware of the surroundings to enhance real-world experiences with contextual information. For individual object identification, machine-readable tags such as barcodes and RFID labels are typically used. Barcodes, though cost-effective, tend to be obtrusive, less durable, and less secure than RFID labels. Regardless of their type, such tags are usually added to objects after fabrication, rather than being integrated into the original design.


This thesis attempts to overcome 
the shortcomings of traditional post-hoc augmentation processes by proposing novel tagging approaches that extract hidden, integrated features of objects and employ them as machine-detectable markers.
Our research focuses on the design, implementation, and evaluation of comprehensive systems for embedding and interacting with embedded markers.
We categorize the proposed tagging approaches into three distinct categories: \textit{natural markers}, \textit{structural markers}, and \textit{internal markers}. 
Natural markers, such as those used in \textit{SensiCut}, are inherent fingerprints of objects repurposed as machine-readable identifiers,
while structural markers, such as \textit{StructCode} and \textit{G-ID}, leverage the structural artifacts in objects that emerge during the fabrication process itself.
Internal markers, such as \textit{InfraredTag} and \textit{BrightMarker}, are embedded \textit{inside} fabricated objects using specialized materials. 
Leveraging a combination of methods from computer vision, machine learning, computational imaging, and material science, the presented approaches offer robust and versatile solutions for object identification, tracking, and interaction.

These markers, seamlessly integrated into real-world objects, effectively communicate an object’s identity, origin, function, and interaction, functioning as gateways to "ubiquitous metadata" – a concept where metadata is embedded into physical objects, similar to metadata in digital files.
Across the different chapters, we demonstrate the applications of the presented methods in diverse domains, including product design, manufacturing, retail, logistics, education, entertainment, security, and sustainability.
Finally, we discuss the challenges and opportunities associated with deploying embedded machine-readable markers at scale, such as integration with mass manufacturing, privacy considerations, consumer product applications, and novel interactions for AR personalization.



In conclusion, this thesis presents a comprehensive exploration of embedded machine-readable markers as a means to connect real-world objects to virtual worlds to achieve ubiquitous metadata. 
The thesis forges ahead with
a future where objects come alive, environments become interactive, and virtual worlds seamlessly merge with our everyday lives.

\end{abstractpage}


\cleardoublepage

\section*{Acknowledgments}


Reflecting upon the incredible odyssey of doing a PhD at MIT, I am compelled to acknowledge the exceptional individuals whose support, guidance, and expertise have shaped my growth and enriched my scholarly pursuits.

First and foremost, I am forever indebted to my thesis advisor, Stefanie Mueller. Her mentorship, scholarly brilliance, and firm belief in my potential have been pivotal in my intellectual and professional development.
Under her guidance, I have had the privilege to be part of the distinguished Human-Computer Interaction (HCI) Engineering group, likely the most international research team at CSAIL.
I am incredibly honored that I got to hang out with the coolest and most interesting individuals with so many different interests, both in research and life. Thank you, Cedric, Dishita, Faraz, Junyi, Martin, Marwa, Ticha, and Yunyi, for being constant sources of inspiration and encouragement, and not hesitating to offer me last-minute feedback whenever I needed it.
A special note of gratitude goes to Ticha, who has gone above and beyond to support us, whether it is through offering us fancy teas and snacks, or hosting cozy meetups with live music.
After almost 6 years, 
it is astounding to think that time flew so quickly; looking back, I feel privileged to have worked with brilliant postdocs who have gone on to be at top research institutions around the globe~--~Junichi, Michael, Moji, Mackenzie, Yoonji~--~thank you for your mentorship at MIT and beyond. And Linda, as the one-and-only administrative assistant of our lab, the biggest thanks go to you, for coming to our rescue whenever we struggled.

I would like to express my earnest gratitude to my academic advisor Daniela Rus, and my thesis advisors and mentors Arvind Satyanarayan and Michael Nebeling. Their guidance and mentorship have been crucial in shaping my research journey, and their insights and feedback have deeply enriched my understanding of the broader scientific landscape.

One of the most fun and rewarding experiences during my time at MIT was having the opportunity to work with numerous undergraduate researchers, \textit{UROP}s, as well as Master's students, which made me a better mentor.
Boundless gratitude to and admiration for Ahmad Taka, Steven Acevedo, and Veerapatr "Vic" Yotamornsunthorn, who won Best UROP and Master's Thesis awards, and made me and Stefanie so proud.

Steven, you are greatly missed, you will be remembered forever.
May you rest in peace.
To the reader of this thesis: Please consider supporting research for treatments and a cure for DSRCT, an aggressive and rare type of cancer that took Steven from us,  by donating to the MD Anderson Cancer Center through the \href{https://www.corymonzingofoundation.org/}{Cory Monzingo Foundation}.



I extend my heartfelt appreciation to my other esteemed colleagues and friends at MIT and in Boston, whose camaraderie, collaborative spirit, and intellectual exchange
have been a constant source of strength during challenging times.
I am indebted to Cathy, Dylan, Jack, Kru, Lucio, Miranda, Raul, Reza, Safa, Tuna, and Ulya for making MIT and Boston feel like a second home, even during the freezing winters.

I remain genuinely thankful for my friends Ecem and Ege, who have been beacons of motivation and support since \textit{İstanbul Erkek Lisesi}, our high school.
I am still moved to tears when I think about all we accomplished together, from Technology Student Association trophies, to film contests and \textit{İcat Çıkar}.
Also enormous thanks to my dear friends from the HCI community and beyond, including Eric, Frederik, Jasper, Jingyi, Mert, and Parastoo, whose presence has been a pillar of support throughout my PhD journey.
And heartfelt thanks to Marcel, who made Delft feel like a true home during times of uncertainty.



I consider myself extremely fortunate to have had the opportunity to collaborate with exceptional mentors and role models from institutions around the world. 
I am grateful to Sriram Subramanian and Kaan Akşit from University College London, Elvin Karana, and Zjenja Doubrovski from TU Delft,
Thijs Roumen from Cornell Tech, 
Alexa Siu, Jennifer Healey, Tong Sun, Chang Xiao, and Eunyee Koh from Adobe Research, David Kim, Mar-Gonzalez Franco, and Ruofei Du from Google, 
and 
Aakar Gupta from Meta Reality Labs.
Their invaluable contributions have elevated the impact and rigor of my research.

I am also immensely grateful to the professors and mentors from my college years, with whom I have not only forged a strong academic bond but also developed lasting friendships over the years. I owe Metin Sitti, İlke Ercan, Ali Emre Pusane, Donghoon Son, and Ankur Mehta a profound debt of gratitude for instilling in me the love for learning and 
the belief in my own potential. Their mentorship laid the foundation upon which I stand today.

I am profoundly thankful to my parents, Neşe and Rifat, and my sister, Bengü, I owe an immeasurable debt of gratitude for their unconditional love, understanding, and belief in my abilities. Their support has been a constant source of motivation throughout the years.

In closing, I would like to express my deep appreciation to all individuals and institutions that have played a role in shaping my doctoral journey. Your collective support and encouragement have been invaluable in realizing my ambitions and aspirations.


\pagestyle{plain}
\tableofcontents
\newpage
\listoffigures
\newpage
\listoftables

\chapter{Introduction}


Mobile and wearable devices promise to enrich our daily interactions by providing information to us whenever we need it. These devices, especially those intended for augmented reality (AR), or more broadly, mixed reality (MR) applications, need to constantly identify what is around us and display relevant digital content.
Imagine yourself in a store, browsing various items, and desiring further information about the ones that caught your attention: the unit price, calorie information, recipes, or perhaps more affordable options.
To facilitate such interactions, the objects around us need to carry labels that describe what they are and communicate this information to digital devices.
While paper-based labels like QR codes are cost-effective to manufacture, they often prove visually distracting and lack the durability and security of electromagnetic alternatives like radio-frequency identification tags (RFID), which can be discreetly embedded within the object.
In this thesis, we study the design, fabrication, and detection of \textbf{novel, low-cost, and durable physical tagging mechanisms} that allow for \textbf{unobtrusive identification} of everyday objects, products, and materials to overcome these challenges.

\begin{figure}[t!]
  \centering
  \includegraphics[width=0.5\linewidth]{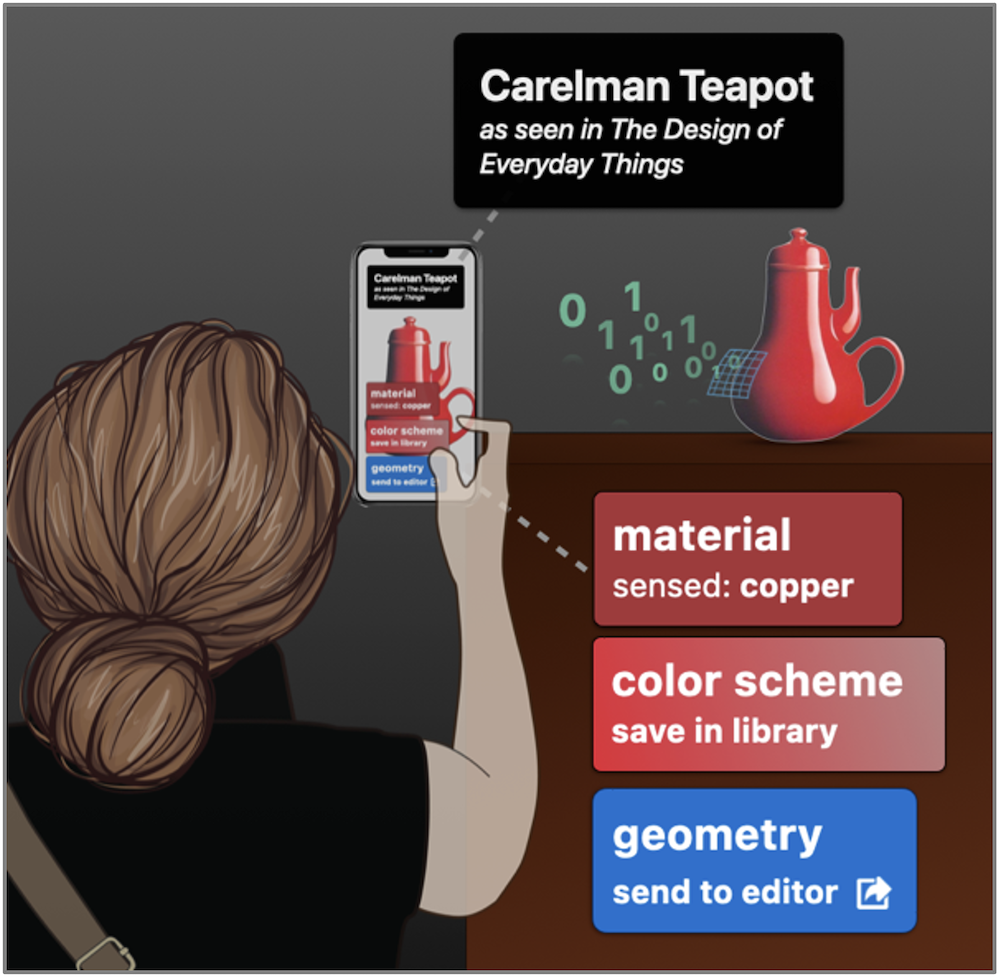}
  \caption{My research vision. Reading metadata embedded unobtrusively into objects and materials in the physical world and reflecting it in the digital world.
 }
  \label{fig:vision}
\end{figure}

We envision that one day, every real-world object in our lives will contain some sort of metadata that describes its identity, look, origin, function, or instructions on how to use it — and we will be able to access these just by pointing our phones or augmented reality glasses at the objects (Figure~\ref{fig:vision}). This is familiar to us in the digital ecosystem, where audio or image files include embedded metadata that helps the user or software seamlessly identify not just what they are but also when and by whom they were captured or recorded. We aspire to transfer this paradigm to \textbf{physical objects -- using embedded physical markers}. We call this "\textbf{ubiquitous metadata}."

Ubiquitous metadata refers to the concept of embedding metadata into physical objects, allowing them to be identified and augmented with relevant information when viewed through digital devices.
This approach could allow information to weave itself into the "fabric of everyday life" until it is indistinguishable from it~\cite{weiser_computer_1999}.
The implementation of ubiquitous metadata through physical tagging mechanisms requires robust fabrication and detection pipelines. If implemented correctly, this can open up new possibilities for immersive spatial computing, where the paradigm will shift from \textit{content inside the screen}, to \textit{users inside the content}, where one will be  able to directly and seamlessly interact with and contribute to the digital content associated with physical objects.

\section{Design Criteria}
\label{DesignCriteria}

As we progress in our exploration of machine-readable markers, we have identified pivotal design criteria that
serve as the guiding principles for 
to our proposed tagging methods.
This list is derived from the limitations of existing solutions and literature, which we will detail in Section~\ref{thesis-Related-Work}. 
The criteria consider the practicality, robustness, and widespread adoption of the markers in real-world applications, and are listed as follows.

\begin{enumerate}[topsep=0pt,itemsep=-1ex]

    \item \textit{Embeddedness}:
    We advocate for these markers to be integral components of objects right from the fabrication stage, rather than being considered as afterthoughts or additions.
    This seamless integration into the physical object's design ensures both unobtrusiveness and resistance to tampering.
    The embedding strategy may be more \textit{opportunistic}, implying a reliance on existing features for identification, or more \textit{engineered}, involving more sophisticated marker embedding processes to achieve specific design objectives.
    Some of our methods demonstrate the feasibility of integrating the marker embedding process into existing manufacturing pipelines, which enables the creation of the object and its marker in a single, harmonious 
    process, thus preserving the object's form and appearance.
    
    \item \textit{Uniqueness}: Our proposed tagging mechanisms feature a unique identifier (ID) or metadata for precise object or product identification.
    A unique ID can come in the form of a distinctive pattern in each object instance for facilitating straightforward identification and accurate data retrieval from digital databases. On the other hand, a unique metadata approach involves embedding more comprehensive information directly into the object, providing a richer context beyond a mere identifier. The choice between these two options depends on the specific requirements of the application and the desired depth of self-contained information associated with each physical entity.
    

    
    \item \textit{Cost-effectiveness}: The methods prioritize cost-effectiveness, leveraging readily available materials and standard manufacturing processes. By operating without an embedded power source, i.e., adopting a \textit{passive} design, these markers also become more durable and less susceptible to failure, accommodating diverse usage scenarios and environmental conditions. Further, this approach reduces the complexity of tag design and streamlines the object's fabrication process.
    

    \item \textit{Detectability}: Regardless of their embedded nature, these markers should be easily detectable by commonly available sensors, such as RGB and near-infrared cameras on devices. This not only enhances user experience but also promotes widespread adoption, as it eliminates the need for specialized equipment. The choice of sensors should align with the intended application to find a balance between accuracy and accessibility.
    
    
    \item \textit{Spatiality}: 
    Spatial awareness plays a pivotal role in the successful integration of ubiquitous metadata, especially in the context of AR.
    Some of the markers we will introduce in this thesis provide spatial information about the objects that they are attached to. Knowing the exact position and orientation of the objects is critical for accurately overlaying and anchoring virtual elements in AR applications. Additionally, the markers should retain their machine-readability from various angles and distances to accommodate the dynamic interactions characteristic of AR applications.

\end{enumerate}

\vspace{0.2cm}


In summary, the design criteria of the proposed embedded machine-readable markers encompass important considerations, including the integrated embedding of markers as part of the object design, the choice between unique ID and metadata, the cost-effectiveness of passive solutions, detectability through accessible sensors, and spatial anchoring in AR. 
Establishing a balance to meet these criteria is essential for creating effective and versatile systems.

\section{Categories of Machine-Readable Markers}
Aligned with the established design criteria, this thesis centers around the exploration and development of three distinct categories of embedded machine-detectable markers. 
These categories are strategically designed to comprehensively address the identified criteria through different technical methods.

The categories, presented below based on the design criterion of embeddedness, range from the most 
\textit{opportunistic}, implying a reliance on existing features for identification, 
to the most \textit{engineered}, meaning more sophisticated marker embedding processes.

\begin{enumerate}[topsep=0pt,itemsep=-1ex]

    \item \textit{Natural markers}: Leveraging the inherent fingerprints of objects and materials, this category explores the use of natural features for identification.
    \item \textit{Structural markers}: Exploiting the structural artifacts that emerge during the fabrication process, this category aims to repurpose inherent elements in the object's construction as markers.
    \item \textit{Internal markers}: Using specialized materials, this category involves embedding custom markers seamlessly \textit{inside} the fabricated items.
    
    
\end{enumerate}

\vspace{0.2cm}

In the pursuit of these categories, novel methods were developed and evaluated, optimizing the design of the markers to have minimal impact on the object's look and form (i.e., unobtrusiveness).
The fabrication processes were designed to be compatible with low-cost tools, such as commonly available fused deposition modeling (FDM) 3D printers. Subsequently, the markers can be detected using readily accessible sensors such as RGB and near-infrared cameras.



Throughout the following sections, we will study these tagging approaches in detail and elaborate on the methodologies used to ensure their effectiveness. By combining techniques from computer vision, machine learning, computational imaging, and material science, our presented approaches offer robust and versatile solutions for object identification, tracking, and interaction.

\section{Natural Markers}

Natural markers allow us to leverage objects’ natural properties, such as their micron-scale surface texture, as identifiers.

To make use of naturally available markers, we can employ optical methods such as laser speckle imaging, where an image sensor detects how the laser reflects off the object’s raw surface. In this method, differences in surface structure result in unique speckle patterns for each material type, which allows us to classify objects. 
While we can integrate this technology in any digital device, it comes in handy for tools that already have a laser source, e.g., laser cutters.

\begin{figure*}[t]
  \centering
  \includegraphics[width=1\linewidth]{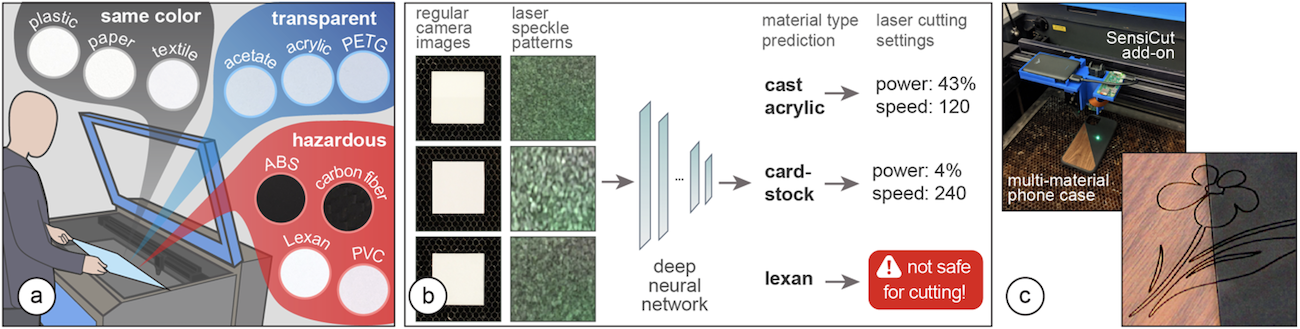}
  \caption{SensiCut. (a) Many laser cutting materials look alike and are hard to distinguish visually by users or regular cameras. Furthermore, there are many hazardous materials that are often confused for safe ones. (b) SensiCut instead senses the sheet’s unique surface structure using speckle imaging and deep learning. This enables us to select the correct machine settings, which prevents material waste and ensures user safety, as well as (c) precisely engrave multi-material objects.}
  \label{fig:intro-sensicut}
\end{figure*}

For the project \textbf{\textit{SensiCut}}~\cite{dogan_sensicut_2021}, we leverage laser speckle imaging to sense material sheets in laser cutters without any pre-labeled stickers (Figure~\ref{fig:intro-sensicut}). To identify materials in traditional laser cutting and set the laser power or speed settings accordingly, users either manually select the type from a database, or an on-board camera takes pictures of material sheets that come with a QR code. Our imaging method, on the other hand, can automate this mundane but risky task, and is not subject to the limitations of conventional cameras, which may confuse visually similar materials, or QR codes, which may be cut off from sheets.

\vspace{0.2cm}

Our technical evaluation shows that \textit{SensiCut} can accurately sense laser cutting materials most commonly used by designers under different environmental conditions, such as illumination and sheet orientation. \textit{SensiCut}’s sensing hardware can be attached to existing laser cutters with our compact sensing hardware add-on, enabling a new workflow for using laser cutters and applications, e.g., providing user safety alerts when hazardous materials are detected, facilitating rapid prototyping, and engraving multi-material objects.

While \textit{SensiCut} focuses on addressing these important laser cutting challenges as identified in recent human-computer interaction literature (Yildirim et al.~\cite{yildirim_digital_2020}), our technique may be used within everyday tools and consumer electronics such as smartphones, many of which now come with infrared lasers for facial recognition~\cite{hallereau_apple_2017}.
More recently, research prototypes have used speckle imaging hardware directly on mixed reality  headsets to amplify detection capabilities, showing the potential for sensing-enhanced AR interactions in the future~\cite{streli_structured_2023}.



\section{Structural Markers}

A distinct category within embedded markers closely aligns  with the fabrication process itself. These markers, which we refer to as \textit{structural markers}, leverage the inherent characteristics and structures that emerge during the fabrication process to create markers.
In this section, we will introduce two methods that utilize different digital fabrication processes and enable different use cases.


\subsection{Laser Cutting Artifacts as Machine-Readable Data}

\begin{figure}[t]
  \centering
  \includegraphics[width=0.89\linewidth]{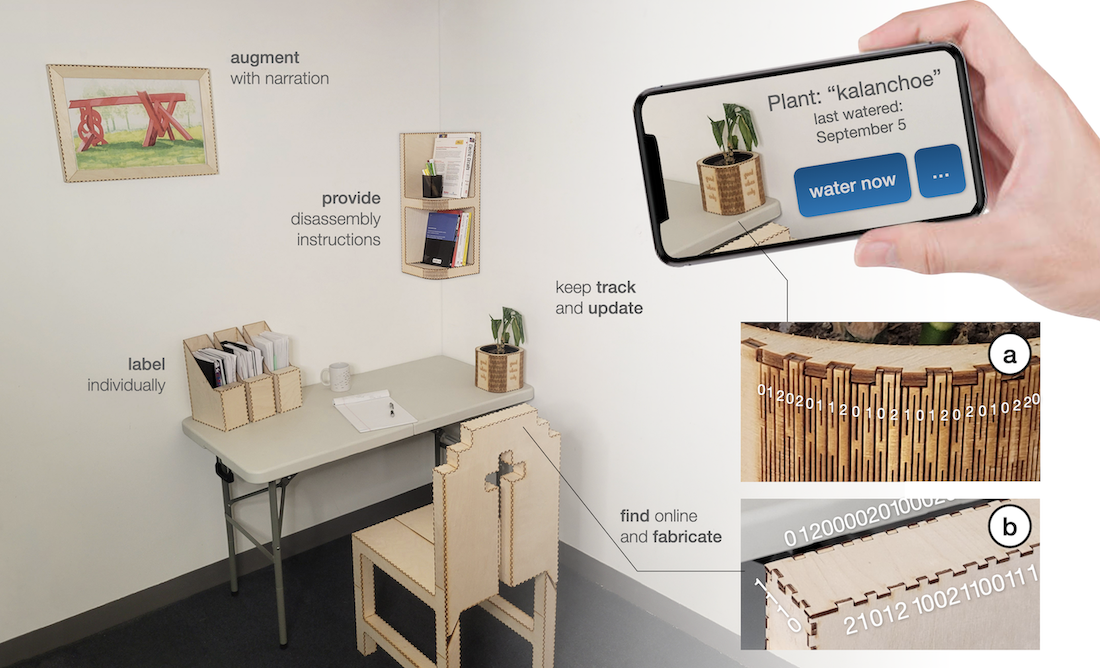}
  \caption{StructCode embeds data in the fabrication artifacts of laser-cut objects, such as the patterns of (a) living hinges and (b) finger joints, to augment objects with data.
  Here, the embedded StructCodes allow narration for a painting and status updates for a potted plant, among others.}
  \label{fig:StructCodeTeaser}
  \vspace{-0.2cm}
\end{figure}

\textbf{\textit{StructCode}}~\cite{dogan_structcode_2023}
allows users to store machine-readable metadata within laser-cut objects' structure by harnessing the fabrication artifacts themselves. By selectively modifying the lengths of laser-burnt finger joints and living hinges, \textit{StructCode} represents different bits of information without requiring additional parts or materials. These modifications, inherent to the fabrication process, enable the integration of meta-information such as labels, instructions, and narration directly into the physical objects (Figure~\ref{fig:StructCodeTeaser}). \textit{StructCode} provides a reliable and unobtrusive means of capturing information, and can be read with standard device cameras, such as those available on smartphones and AR/VR headsets.

We present and evaluate a marker decoding pipeline that is robust to various backgrounds, viewing angles, and wood types. In our mechanical evaluation, we show that StructCodes preserve the structural integrity of laser-cut objects.


\subsection{3D Printing Artifacts as Identifiers}

Compared to fabrications methods such as laser cutting,  3D printing provides greater flexibility in achieving complex 3D geometries for objects.
To integrate identifiers into 3D printed items, we may exploit the artifacts resulting from the extrusion-based FDM process.
By manipulating the 3D printer’s path through custom CNC machine instructions ("G-code"), we can obtain unique and subtle surface textures for each instance of the same 3D model.

\begin{figure}[t]
  \centering
  \includegraphics[width=0.72\linewidth]{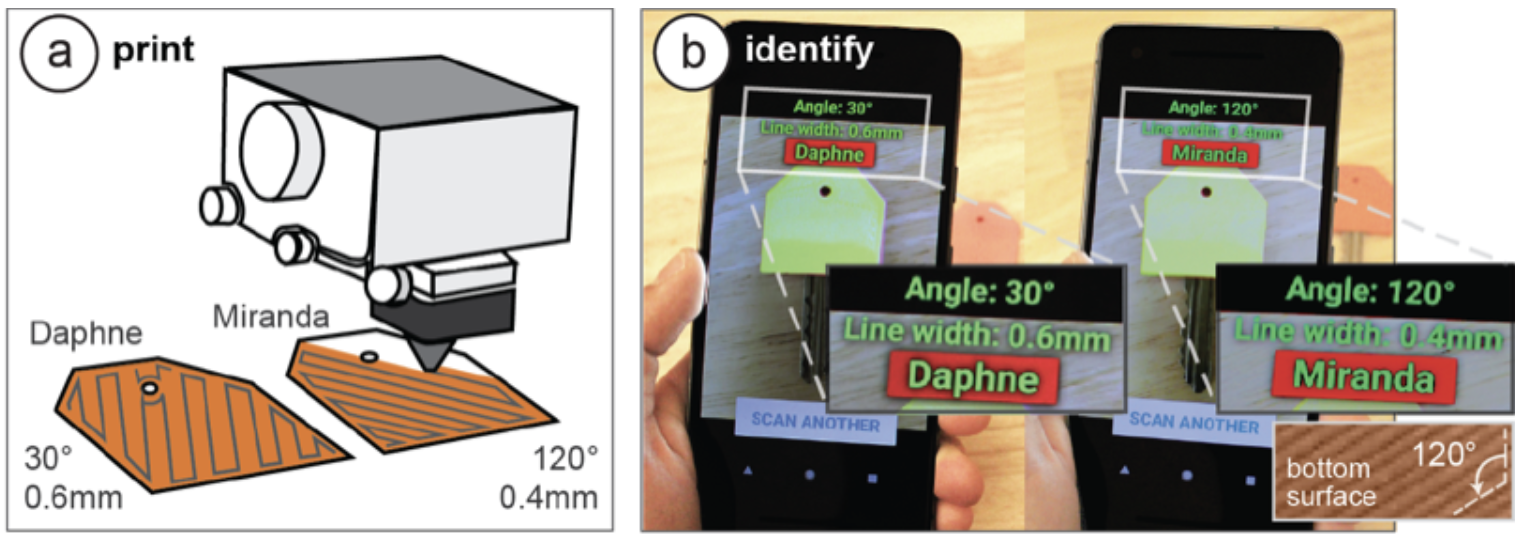}
  \caption{G-ID. (a) 3D printed objects inherently possess surface patterns  due to the print path. G-ID exploits such features that would normally go unnoticed to identify unique instances of an object. Our mobile app (b) uses image processing to detect them.
 }
  \label{fig:gid}
  \vspace{-0.2cm}
\end{figure}

By strategically adjusting slicing parameters that do not alter the geometry of the object, our method \textbf{\textit{G-ID}}~\cite{dogan_g-id_2020, dogan_demonstration_2020} creates unobtrusive machine-readable differences. These variations serve as distinct markers for identifying and distinguishing similar-looking objects from a single photograph captured using a conventional smartphone (Figure~\ref{fig:gid}).

We evaluated how finely these texture-related parameter differences can be differentiated between, 
and built a mobile application that uses image processing techniques to retrieve these parameters from photos and their associated labels. This enables low-cost, interactive applications for user identification.




\section{Internal Markers}

While \textit{G-ID} is powerful for distinguishing between different copies of 3D printed objects, an important limitation is that the objects do not carry a large amount of data, but rather a single, uniquely identifiable texture. This restricts users from embedding larger amounts of custom information.

On the other hand, internal markers enable users to embed data in the form of 2D markers directly into 3D~printed objects.
Previous works in internal markers require the use of expensive or specialized equipment.
For instance, \textit{AirCode}~\cite{li_aircode_2017} uses a camera and projector setup to detect gaps beneath the object surface. \textit{InfraStructs}~\cite{willis_infrastructs_2013} uses a terahertz scanner to detect even deeper holes. 
Our goal is to develop internal markers that are easy to fabricate and that can be detected using low-cost scanning tools.

\subsection{Invisible Markers in 3D Printed Objects}

\begin{figure}[t]
  \centering
  \includegraphics[width=1\linewidth]{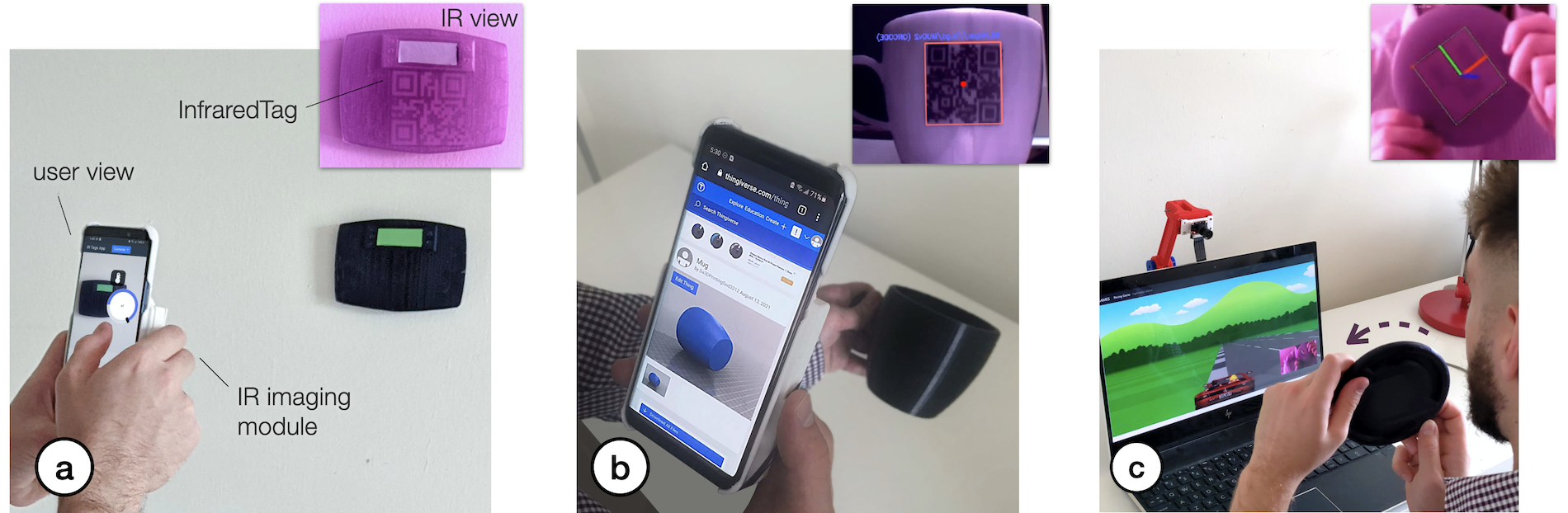}
  \caption{InfraredTags are 2D markers and barcodes embedded unobtrusively into 3D printed objects and can be detected using infrared cameras (top-right images). This allows new applications for (a) identifying and controlling devices in AR interfaces, (b) embedding metadata such as 3D model URLs into objects, and (c) tracking passive objects for tangible interactions.}
  \label{fig:infraredtags}
\end{figure}

\textbf{\textit{InfraredTags}}~\cite{dogan_infraredtags_2022, dogan_infraredtags_2022-2} are 2D markers and barcodes imperceptible to the naked eye that can be 3D printed as part of objects, and detected rapidly by low-cost near-infrared (NIR) cameras (Figure~\ref{fig:infraredtags}). 
These 2D markers and barcodes consists of multiple bits, as opposed to a single texture demonstrated in \textit{G-ID}.
We achieve this by printing objects from an infrared-translucent filament, which infrared cameras can see through, and by having air gaps inside for the tag’s bits, which appear at a different intensity in the infrared image.

By integrating \textit{InfraredTags} into 3D printed objects during the fabrication process, we enable interactive applications for identifying and controlling devices in AR interfaces or embedding metadata such as links to download the object’s 3D model. 
We also show how convolutional neural networks can be used in conjunction with data augmentation methods to improve the detection of these tags from near-infrared camera streams~\cite{dogan_demonstrating_2022}.

\subsection{Increasing the Contrast of Markers}

\begin{figure*}[t]
  \centering
  \includegraphics[width=1\linewidth]{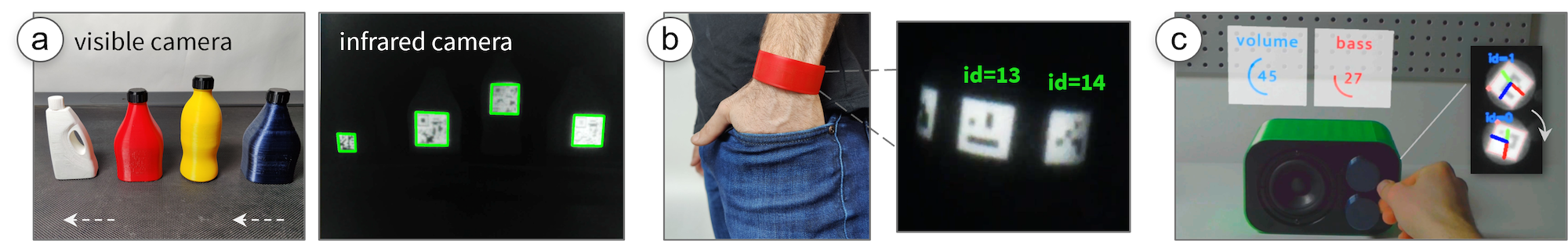}
  \caption{BrightMarkers are embedded into objects using a NIR-fluorescent filament. (a) When viewed with a NIR camera with the matching filter, the markers appear with high contrast, which allows them to be tracked even when the objects are in motion, e.g., on a conveyor belt.
  (b) BrightMarker can be used to fabricate custom wearables for tracking, or (c) for transforming physical controls into precise input methods in MR environments.}
  \label{fig:intro-brightmarker}
\end{figure*}

We built on \textit{InfraredTag} to create markers that have higher contrast. Called \textbf{\textit{BrightMarker}}~\cite{dogan_brightmarker_2023, dogan_demonstrating_2023}, these markers are are fabricated using NIR-fluorescent filaments to allow real-time tracking of 3D~printed color objects. 
By using an infrared-fluorescent filament that "shifts" the wavelength of the incident light, our optical detection setup filters out all the noise to only have the markers present in the infrared camera image.
The high contrast of the markers allows us to track them robustly regardless of the moving objects' surface color (Figure~\ref{fig:intro-brightmarker}). 


\textit{BrightMarker} can be used in a variety of applications, such as custom fabricated wearables for motion capture, tangible interfaces for AR/VR, rapid product tracking, and privacy-preserving night vision.
\textit{BrightMarker} exceeds the detection rate of state-of-the-art invisible marking, and even small markers (1"x1") can be tracked at distances exceeding 2m.

\vspace{0.5cm}

The integration of these embedded machine-readable markers into real-world objects serves as a gateway to virtual worlds, enabling users to access a wealth of multimedia content, interactive experiences, and valuable contextual information. We will showcase the applications of our methods across diverse domains, including product design, manufacturing, entertainment, marketing, logistics, security, and sustainability.

\section{Structure of the Thesis}

In the remainder of the thesis, we first introduce the landscape of related work in the identification, tagging, and digital fabrication domains (Chapter~\ref{thesis-Related-Work}). We then explain each of the previously introduced methods in individual chapters, including natural markers (\textit{SensiCut} in Chapter~\ref{thesis-SensiCut}, and structural markers (\textit{StructCodes} in Chapter~\ref{thesis-StructCodes}, \textit{G-ID} in Chapter~\ref{thesis-G-ID}), internal markers (\textit{InfraredTags} in Chapter~\ref{thesis-InfraredTags}, \textit{BrightMarkers} in Chapter~\ref{thesis-BrightMarkers}).


Lastly, we address the challenges and opportunities associated with deploying embedded machine-readable tags at scale (Chapter~\ref{thesis-Discussion}). These considerations encompass aspects such as generalization to faster fabrication methods, integration into consumer products, AR interactions, and privacy concerns. By addressing these challenges, we aim to pave the way for a future where objects come alive, environments become interactive, and virtual worlds seamlessly merge with our everyday lives.

In conclusion, this thesis represents a comprehensive exploration of embedded machine-readable tags as a means to bridge the gap between real-world objects and virtual worlds. Through our research, we strive to advance a future where ubiquitous metadata enhances our interactions, enabling us to effortlessly access information and engage with the physical and digital realms.

\chapter{Related Work}
\label{thesis-Related-Work}

In the realm of identifying and tagging physical objects, a wide range of approaches has been explored to bridge the gap between the physical and digital domains.
This section provides an overview of existing methodologies, and highlights their strengths, limitations, as well as contributions to HCI.

\section{Computer Vision-Based Approaches}

Computer vision, as a powerful and versatile tool, has significantly advanced object recognition capabilities~\cite{szeliski_computer_2010}
Machine-learning (ML) classifiers, particularly convolutional neural networks (CNNs)~\cite{lecun_gradient-based_1998}, have proven highly capable at assigning general labels to objects based on visual features~\cite{goodfellow_deep_2016}.
This made various computational solutions possible in the field of semantic understanding.
Techniques like \textit{YOLO}~\cite{redmon_you_2016} and \textit{SSD}~\cite{liu_ssd_2016} provide real-time \textit{object detection} by identifying the bounding boxes of objects in the scene.
\textit{Semantic segmentation} involves classifying each pixel in an image into predefined categories, while \textit{instance segmentation} takes this a step further by distinguishing between individual instances of the same object category.
These techniques, as employed in popular models such as \textit{Mask R-CNN}~\cite{he_mask_2017} and \textit{U-Net}~\cite{ronneberger_u-net_2015}, as well as frameworks such as \textit{MediaPipe}~\cite{lugaresi_mediapipe_2019}, 
allow for more precise identification and localization of objects.

While these classifiers excel at recognizing and categorizing objects into broader classes, they often fall short in capturing fine-grained details necessary for a comprehensive understanding of the scene~\cite{lin_microsoft_2014, hoiem_putting_2008}. The incapacity to extract specific \textit{metadata} -- such as the subtype of an object, detailed product information, origin, or owner -- poses a challenge in achieving a deeper and more contextually rich comprehension of the environment.

Real-world scenarios often demand more granular information about objects that go beyond generic labels. For instance, in a retail context, being able to access detailed product information, including specifications or pricing, enhances the overall user experience. In industrial settings, knowing the specific model or version of a machine component holds critical importance. 
As we explore methods for ubiquitous metadata, we aim to address this limitation by capturing and associating specific metadata with each object instance, ensuring a more comprehensive and contextually relevant representation of the physical environment.

\section{Use Cases for Markers in HCI}
Markers have been used to allow instance-level object identification to extend beyond the capabilities of conventional computer vision~\cite{holmquist_tagging_2006}.

In HCI, visual markers, such as ArUco markers, barcodes, and QR codes have been used to mark objects and enable different interactive applications with them.
For instance, \textit{Printed Paper Markers~}\cite{zheng_tangible_2020} use different paper structures that conceal and reveal fiducial markers (i.e., \textit{ArUco}~\cite{romero-ramirez_speeded_2018}) to create physical inputs, such as buttons and sliders.
\textit{DodecaPen}~\cite{wu_dodecapen_2017} can transfer users' handwriting to the digital environment by tracking ArUco markers attached on a passive stylus.
\textit{Cooking with Robots}~\cite{sugiura_cooking_2010} uses detachable markers to label the real-world environment for human-robot collaboration. 
\textit{Position-Correcting Tools}~\cite{rivers_position-correcting_2012} scan QR code-like markers to precisely position CNC tools while users cut sheets.



\begin{figure}[h]
  \centering
  \includegraphics[width=0.89\linewidth]{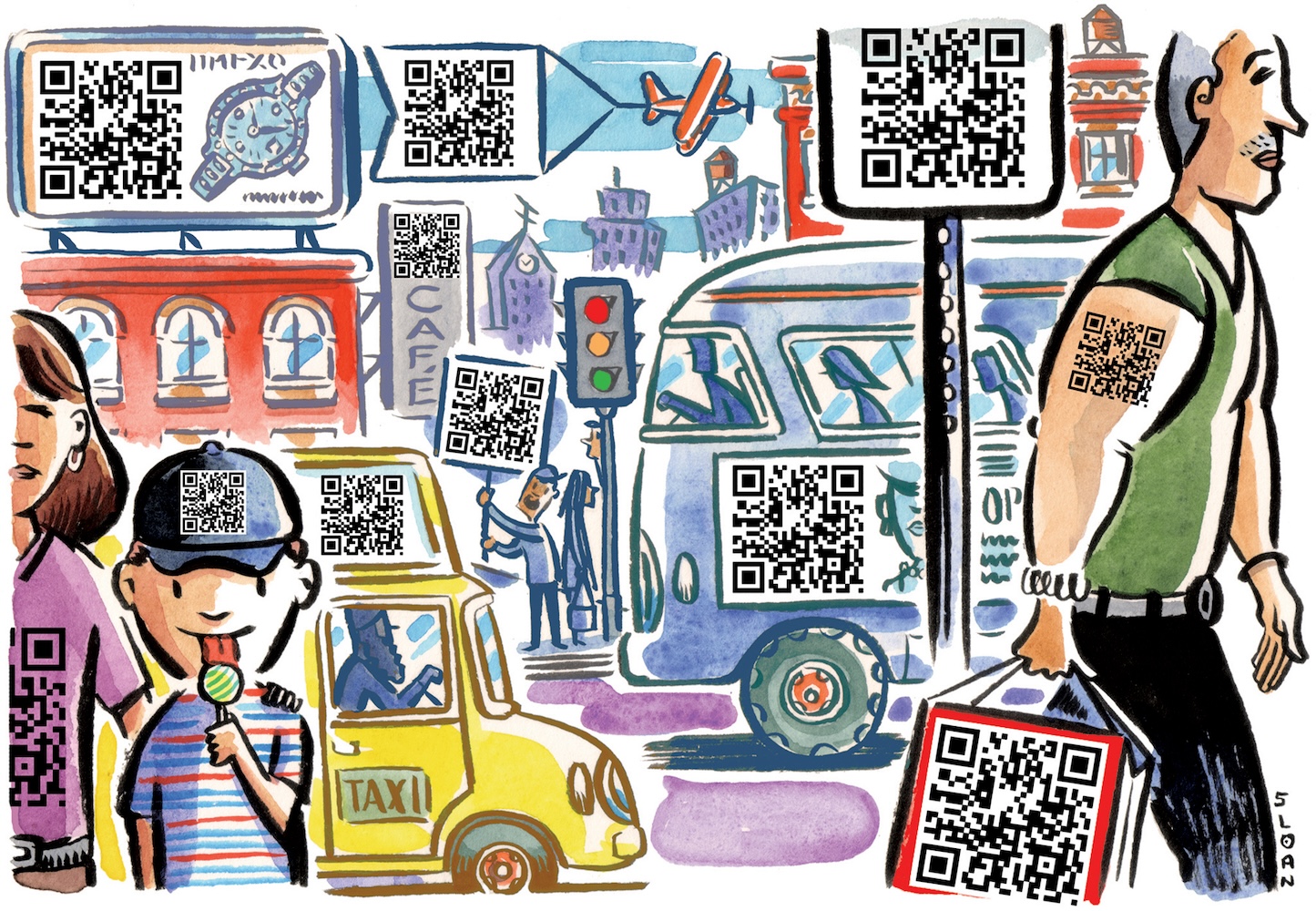}
  \caption{An exaggerated portrayal of an urban environment saturated with QR codes, showing the aesthetic and practical challenges of current object tagging methods.  Illustration used with permission from \href{https://www.michaelsloan.net/}{Michael Sloan}.}
  \label{fig:IllustrationQRCodes}
  \vspace{-0.2cm}
\end{figure}

While traditional barcodes and QR codes offer a cost-effective identification solution, such markers are visible to the human eye, which impacts object aesthetics and may reduce the usable area on the object, as depicted in Figure~\ref{fig:IllustrationQRCodes}. 
Furthermore, practical usage often sees visually unappealing labels removed by users, and visible codes may be tampered with by malicious third parties~\cite{mcgrath_fbi_2022}.
Conversely, RFID tags, equipped with unique identifiers, can be attached to physical objects for efficient tracking~\cite{spielberg_rapid_2016, hsieh_rfibricks_2018}. 
While RFID provides non-line-of-sight identification and finds applications in various environments, it has limitations in terms of range and scalability~\cite{michael_pros_2005}. The cost of RFID tags, coupled with the size of RFID readers, may pose challenges in large-scale deployments and integration with the object's geometry while objects during fabrication~\cite{white_comparison_2007}.

\section{Making Markers Less Obtrusive}
Researchers have investigated two primary approaches to make markers less obtrusive: making existing visible markers more \textit{aesthetic}~\cite{qiao_structure-aware_2015, bencina_improved_2005, preston_enabling_2017}, or developing unobtrusive markers that can be part of the objects ~\cite{maia_layercode_2019, li_aircode_2017}. 

To make existing codes more aesthetic, researchers have modified traditional QR codes (\textit{halftoned}) to look more like an aesthetic image (e.g., a photo) while still preserving its detectability~\cite{qiao_structure-aware_2015}.
\textit{ReacTIVision}~\cite{bencina_improved_2005} creates fiducial markers that look like amoeba to create an organic look.

Specifically for 3D objects, there are more specialized ways to abstract away the marker component. To achieve this, markers can be embedded \textit{inside} the objects, or integrated unobtrusively onto the \textit{surface} of 3D objects so that users do not perceive them as "markers."

Integrating markers with objects in a seamless and unobtrusive manner can provide meaningful contextualization and allows for sharing of information \cite{ettehadi_documented_2021}.
In HCI, the general idea has been explored for photographs~\cite{tancik_stegastamp_2020}, font characters~\cite{xiao_fontcode_2018}, charts~\cite{fu_chartem_2021}, documents~\cite{want_bridging_1999, dogan_standarone_2023}, and textiles~\cite{zhu_exploring_2023}.
In the next sections, we review different types of tags for 3D physical objects specifically and categorize the approaches based on where the tag is located on the object, how much data it can store, and how complex the associated fabrication and detection processes are.

\subsection{Marker Location: Inside, Surface, Structure}
Researchers have investigated different locations on objects to embed markers~\cite{baudisch_personal_2017}. One way to embed markers is on the inside of objects.  
For instance, \textit{AirCode}~\cite{li_aircode_2017}, \textit{InfraStructs}~\cite{willis_infrastructs_2013}, and \textit{InfoPrint}~\cite{jiang_infoprint_2021} embed markers by adding air gaps inside 3D printed objects. Another way is to embed tags on the surface of objects instead. 
For instance, \textit{Seedmarkers} \cite{getschmann_seedmarkers_2021} are visual markers optimized for aesthetics that can be placed on the surface of either laser-cut or 3D printed objects.
\textit{ObjGen}~\cite{matusik_objgen_2023} engraves \textit{Data Matrix} cells on the object to store the vector file it originated from. However, since multiple codes are needed, they typically occupy the whole object surface, resulting in an unnatural look.
Tenmoku et al. place small visual components (e.g., T-shaped or triangular) on objects to use them as visually elegant tracking markers for mixed reality, however, these markers do not carry any data~\cite{tenmoku_visually_2007},
\textit{Acoustic barcodes}~\cite{harrison_acoustic_2012} are physical notches that can be etched on the object surface using a laser cutter and that create unique, identifiable bursts of sound when swiped.
\textit{LayerCodes} \cite{maia_layercode_2019} are barcodes on the surface of 3D models printed from an infrared resin and then detected from smartphone camera images. 


\subsection{Data Complexity: Identifiers, Information}

A marker is a label that can be used either to identify items or to store information in the form of data.
Kubo et al.~\cite{kubo_3d-printed_2020} distinguish multiple copies printed with different infill structures by measuring their vibration characteristics. However, while such methods can store identifiers, they do not allow users to store information specifically defined by the user, such as custom texts or metadata, in the tag. 
\textit{AnisoTag}~\cite{ma_anisotag_2023} increases the tags' data capacity by combining different print textures.

To store information, \textit{AirCode}~\cite{li_aircode_2017}, \textit{LayerCode}~\cite{maia_layercode_2019}, and \textit{InfraStructs}~\cite{willis_infrastructs_2013} embed barcodes and QR codes in the form of air gaps or lines ("physical bits") within the objects. Since such codes store arrays of characters or digits, they can be used to embed texts or metadata within the object.  

\subsection{Complexity of Tagging Approach: Fabrication, Detection}

To embed unobtrusive markers that store information into objects, existing methods require specialized fabrication equipment or additional post-processing steps. For instance, \textit{LayerCodes} \cite{maia_layercode_2019} require a custom modification to a 3D printer and a special NIR resin. 
\textit{AirCodes}~\cite{li_aircode_2017} require that objects are printed in two parts and combined manually after washing away the support material. 


Another important consideration is the availability and complexity of the detection equipment. For example, \textit{InfraStructs}~\cite{willis_infrastructs_2013} employs a large and costly terahertz scanner, which is not readily available to consumers. \textit{AirCodes}~\cite{li_aircode_2017} use a camera and projector setup that are calibrated to each other, which takes time to set up and image the tag.  
\textit{InfoPrint}~\cite{jiang_infoprint_2021} uses a thermal camera to capture patterns embedded into objects.
\textit{AnisoTag}~\cite{ma_anisotag_2023} needs the object to be manually swiped using a collimated laser beam and photoresistors, making it difficult to integrate it with users' existing devices.

\vspace{0.3cm}

To facilitate the implementation of ubiquitous metadata,
the markers should ideally be detectable using commodity hardware, capturing the tags must follow a simple workflow, and the detection method needs to work across a variety of different environments.

\chapter{SensiCut: Material-Aware Laser Cutting Using Speckle Sensing and Deep Learning}
\label{thesis-SensiCut}

\section{Introduction}

While there have been many support tools for laser cutting that help users with tasks such as automatically packing parts onto sheets~\cite{saakes_paccam_2013,sethapakdi_fabricaide_2021}, 
 systems that support users with the different \textit{material types} available for laser cutting are largely unexplored~\cite{baudisch_personal_2017}.

For users, working with the various materials available in a workshop comes with several challenges: First, identifying unlabeled sheets from scrap buckets or material stockpiles in a shared workshop is challenging since many materials are visually similar~\cite{katz_identification_1998}. As a result, users may take the wrong material from the stack and use it with another material's power and speed setting. This can lead to wasted material when the power setting is too low, causing the outline to not be cut through -- or worse the material may catch fire when the power is too high leading to safety risks. Further, there are many materials that are not safe to laser cut because they release toxic fumes~\cite{environmental_health_and_safety_laser_2019}. These hazardous materials may easily be mistaken for safe materials due to similarity in appearance (e.g., PVC vs. acrylic)~\cite{materials_engineering_pentonipc_how_2020}.

Because of the challenges outlined above, laser cutter users desire smarter machines that can "identify the materials they [are] working with, so that the system could [...] suggest settings based on material" as shown in a recent HCI study by Yildirim et al.~\cite{yildirim_digital_2020}.
One naive solution for this is to add a camera to laser cutters to automatically identify the sheets. 
However, a conventional camera can be easily fooled by visually similar materials or materials with printed decorative textures that imitate another material.

\begin{figure}[h]
  \centering
  \includegraphics[width=0.7\columnwidth]{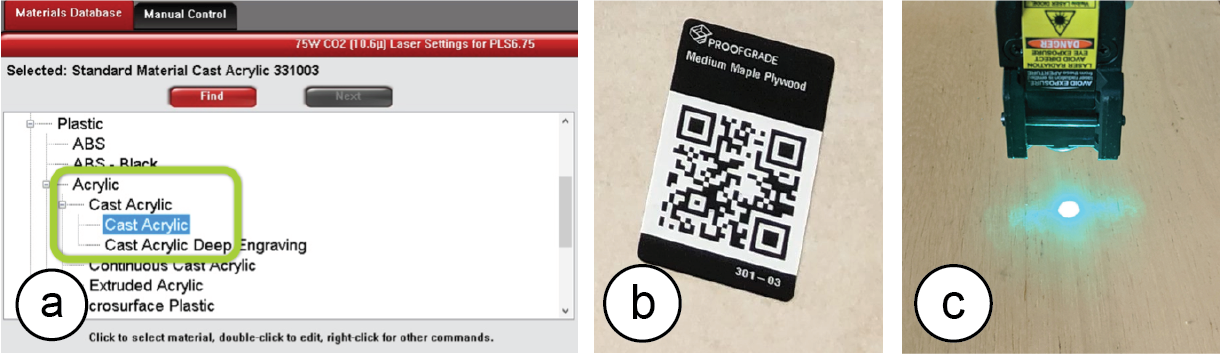}
  \caption{Existing material identification approaches: (a)~Manually selecting from a database (e.g., \textit{ULS}~\cite{universal_laser_systems_universal_2020}) or (b)~scanning QR code stickers on sheets~(\textit{Glowforge}). (c)~SensiCut uses \textit{speckle sensing} to identify the material based on its surface structure without the need for additional tags.}
  \label{fig:related-work}
\end{figure}

To ensure reliable identification, 
recent laser cutters use sticker tags attached to the sheets (e.g.,~QR codes on \textit{Glowforge  Proofgrade} sheets~\cite{glowforge_inc_glowforge_2019}).
As can be seen in Figure~\ref{fig:related-work}b, these tags can be detected by a camera 
even when materials look similar or are transparent.
However, scanning the tags to detect the material type has its own issues. First, a new tag has to be attached onto each new material sheet.
Second, laser cutter users need to be careful to not cut off the tag to ensure that the remaining part of the sheet can later still be identified. 
These issues exist because using tags for identification is not inherently material-aware as the laser cutter does not measure the physical properties of the  material.

In this project, we investigate how we can identify laser cutting materials by leveraging one of their inherent properties, i.e., \textit{surface structure}. A material's surface structure is unique even when it is visually similar to another type.
To achieve this, we use \textit{speckle sensing}. This imaging technique works by pointing a laser onto the material's surface and imaging the resulting speckle patterns. We built a hardware add-on consisting of a laser pointer and a lensless image sensor, which can be attached to the laser cutter head using a mount. We then use the captured speckle patterns to identify the material type with our trained neural network. Our user interface uses the material type information to support users in different ways, i.e. it automatically sets the power and speed settings for the detected material, it warns the user against hazardous materials, it automatically adjusts the shape of a design based on the kerf for the detected material, and finally, it automatically splits designs when engraving onto multi-material objects. We also discuss how speckle sensing can be used to estimate the thickness of sheets as another material-aware component for future laser cutters.

\begin{figure*}[t]
  \centering
  \includegraphics[width=1\textwidth]{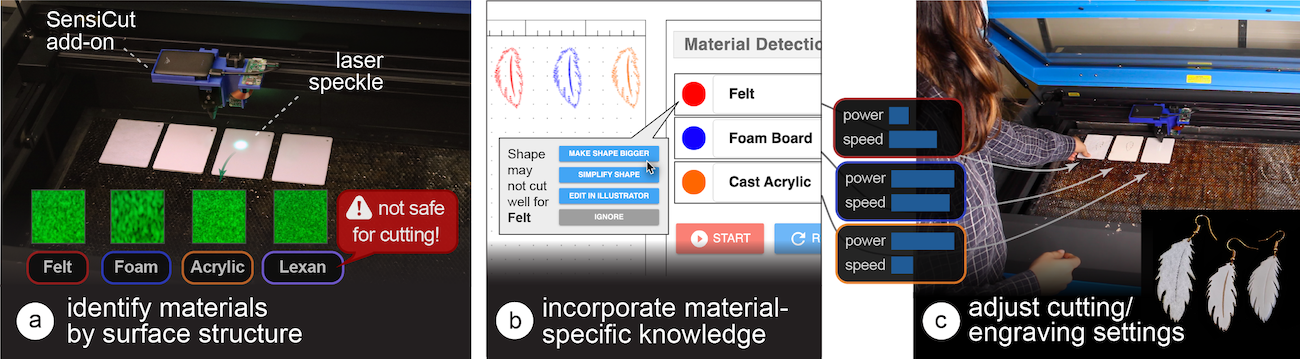}
  \caption{SensiCut augments standard laser cutters with a speckle sensing add-on that can (a)~identify materials often found in workshops, including visually similar ones. (b)~SensiCut's user interface integrates material identification into the laser cutting workflow and also offers suggestions on how to adjust a design's geometry based on the identified material (e.g., adjusting the size of an earring cut from felt since the kerf for felt is larger than for other materials). 
  (c)~Each identified sheet is cut with the correct power and speed settings.}
  \label{fig:teaser-sensicut}
\end{figure*}

In summary, by leveraging speckle sensing as an identification technique, we can improve the material awareness of existing laser cutters.
Our work enables safer and smarter
material usage, addresses common material identification-related challenges users face when laser cutting, and encourages makers to reuse laser-cut scraps to reduce waste~\cite{dew_designing_2019, vyas_making_2019}.
We note that this chapter was originally published at \textit{ACM UIST 2021}~\cite{dogan_sensicut_2021}. 
Our contributions are the following:

\begin{itemize}[leftmargin=0.5cm, noitemsep, topsep=2pt]
    \item An end-to-end laser cutting pipeline that helps users identify materials by sensing the material's surface structure using laser speckles to, e.g., automatically set the corresponding power/speed, warn against hazardous materials, adjust designs based on material-specific kerf, or split designs when engraving onto multi-material objects.

    \item A compact (114g) and low-cost material sensing add-on for laser cutters that simplifies hardware complexity over prior work by using deep learning.
    
    \item A speckle pattern dataset of 30~material types (38,232 images), which we used to train a convolutional neural network for robust laser cutter material classification (98.01\%~accuracy).
    
    \item A technical evaluation showing which visually similar materials speckle sensing can distinguish under various sheet orientation and illumination conditions.

\end{itemize}

\section{Motivation}

SensiCut addresses an important open challenge in the personal fabrication literature. 
A recent field study~\cite{yildirim_digital_2020} in HCI revealed an unaddressed user need for fabrication tools concerning the “awareness of material types”. In particular, users wished that the tools could “identify the materials they were working with [and] suggest settings.” The authors conclude that 
“HCI researchers could advance [these tools] by leveraging new sensing [...] capabilities”. 

To further understand what specific challenges exist, we surveyed five additional HCI publications ~\cite{annett_exploring_2019, hudson_understanding_2016, boll_designing_2020, palanque_designing_1998, knibbe_smart_2015}. 
We also conducted formative interviews with six expert users that we recruited by reaching out to makerspaces. Each expert user had several years of laser cutting experience working with different material types.
During the 1-hour semi-structured interviews, we interviewed them about their experiences using different material types, difficulties they had identifying materials, and how different material types affected their designs for laser cutting.
Additionally, we performed a study in which we gave 13 novice users, who had used laser cutters at least once but no more than four times, a list of 30~materials commonly found in workshops (list of materials in Section~\ref{Choosing-Material-Samples}), and asked them to match them to 30~unlabeled sheets.
For the interview responses, we took a bottom-up approach in our thematic analysis to identify four main challenges, which we report below.

\vspace{0.2cm}
\noindent\textit{Characterizing unlabeled sheets:} 
We found that users have a hard time identifying materials. In our study, novices were able to label on average only 29.23\%
(SD=6.41) of the sheets correctly. 
The ones that were correctly identified by most users were cardboard and cork. The top~10~mislabeled sheets were all different types of either plastic or wood.  However, this is not only an issue for novices, but also for experts. One senior maker we interviewed reported that certain materials are too similar to distinguish by only looking and touching. He added that he checks if a sheet is acrylic or Delrin by "breaking the sheet and seeing how brittle it is." Identifying materials by their surface structure eliminates these issues since even similar types of plastic have different surfaces structures.

\vspace{0.2cm}
\noindent\textit{Democratizing material knowledge:} One way to help novices identify materials is to ensure sheets are labeled at all times. However, in practice, this is infeasible to do for all sheets. One expert we interviewed, a manager of a large workshop, said that "there is no way to keep track of all the sheets [as] so many people contribute to the scrap piles."
Another option if sheets are unlabeled is that novice makers ask an experienced maker which type of material it is. However, Annett et al.~\cite{annett_exploring_2019} report that makers with “knowledge [of] material [were] often difficult to access” and that users need “intelligent sensing [of] materials.”
Hudson et al.~\cite{hudson_understanding_2016} show that “early in the casual makers’ learning process motivation appeared to be very fragile” and “early failures [can] result in them completely giving up.” A smart  system that provides access to reliable material identification would eliminate this issue, thereby lowering the entry barrier to laser cutting and democratizing its use.

\vspace{0.2cm}
\noindent\textit{Automating mundane work:} Laser cutting requires several steps that are mundane and would benefit from being automated. For instance, in today's workflow, users have to identify the sheet, select the correct material type from the material database, and then verify the power/speed settings. Yildirim et al.~\cite{yildirim_digital_2020} found that professional users want "automated [fabrication tools] that could pick up menial work, [e.g.] registering materials." They "find it frustrating when they have to monitor an autonomous [tool]." 
A material-aware sensing platform can remove the tedious overhead and allow users to focus on the essential work.

\vspace{0.2cm}
\noindent\textit{Enhancing safety of all users:} Laser cutting poses both safety and health hazards~\cite{herrick_emerging_2016, love_perceptions_2018}. In our interviews, all experts reported that they experienced multiple fires in the laser cutter at their workspaces.
One of them said that "all of the places [he has] worked at had a fire" and that it is a "huge safety risk." Concerning health, one of our interviewees, a class instructor, said someone almost cut a hazardous material that includes chlorine, which would release toxic fumes and corrode the machine. In addition, not adhering to the rules would have "revoked all class participants' access to the workshop." For safety-critical tasks, HCI researchers have looked into designing interfaces where the role of users is “mediated by computer technology”~\cite{boll_designing_2020, palanque_designing_1998}. For fabrication tools specifically, Knibbe et al.~\cite{knibbe_smart_2015} found the “implementation [of] safety alerts could provide significant benefits within group makerspaces.” A smart sensing platform can provide such safety alerts and prevent human error by determining if hazardous materials are used or when users accidentally select wrong laser settings, which can cause a fire.

\section{Speckle Sensing Hardware Add-On for Laser Cutters}

In this section, we first discuss the working principle behind laser speckle imaging. We will then show how we built a sensing add-on that can be mounted onto existing laser cutters and highlight the technical contributions of our add-on over prior work.

\subsection{Speckle Sensing Working Principle}

\begin{figure*}[t]
  \centering
  \includegraphics[width=1\textwidth]{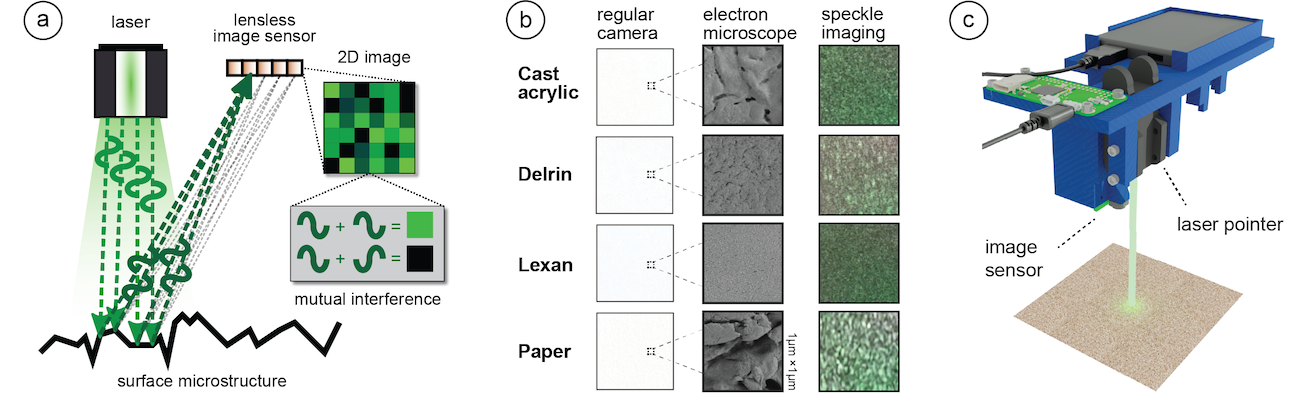}
  \caption{Speckle sensing. (a)~Laser rays reflect off the material surface and arrive at the image sensor. Phase differences between the rays result in mutual interference and thus dark or bright pixels in the captured image. 
 (b)~Different  materials viewed by a regular camera, a scanning electron microscope, 
   and our speckle sensing imaging setup.
   (c)~Our speckle sensing add-on consists of  a laser pointer, lensless image sensor, microprocessor, and battery.}
  \label{fig:speckle-concept}
\end{figure*}

Figure~\ref{fig:speckle-concept}a illustrates how laser speckle sensing works. It uses a coherent light source, i.e., a laser, to create the speckles and an image sensor for capturing them. To create the speckle pattern, the laser light reflects off the material surface, resulting in a reflectance pattern (speckle) of bright and dark spots that looks different depending on the material's surface structure. This occurs because the tiny features of the material surface lead to small deviations in the optical path of the reflected laser beam. To show this, we provide additional electron microscope images of different materials in Figure~\ref{fig:speckle-concept}b. Although the materials look visually similar to the human eye, the electron microscope images clearly show different surface structures, resulting in different speckle images that can be used for material identification. 

\subsection{Hardware Add-On}

To integrate speckle sensing into an existing laser cutter, we consider (1)~which light source and (2)~image sensor to use, as well as (3)~how to mount all required components on the platform. We provide additional  specifications for each component in the appendix.

\vspace{0.2cm}
\noindent\textit{Laser Pointer:} Our initial idea was to utilize the laser pointer present in our laser cutter (model: \textit{Universal Laser Systems} (ULS) PLS6.150D), which conventionally serves as a guide to align the material sheet. However, we found that laser pointers need to be sufficiently powerful to create speckles detectable by the image sensor. For this reason, green laser pointers work best given equal power since most commercial cameras have a Bayer mask with twice as many green elements as red or blue. Unfortunately, our laser cutter has a red laser pointer with a power of <1 mW, which according to our experiments was not sufficient to create detectable speckle patterns. 
We therefore decided to use an additional green laser pointer (515nm, <5mW). In the future, this additional laser pointer may not be necessary if manufacturers increase the power of their existing laser pointers.

\vspace{0.2cm}
\noindent\textit{Image Sensor:} Our goal 
was to choose a sensing setup that is compact, i.e., uses as few components as possible, yet provides sufficiently high resolution to detect the speckles. When surveying the related work, we found that existing setups consist of multiple sensors and/or LEDs~\cite{harrison_lightweight_2008, sato_spectrans_2015}.
Each component in these setups helps acquire a unique datapoint related to high-level statistics, such as the average brightness or overall spectral reflectivity, which are then input into a classifier as 1D data.
We found that we can reduce the hardware complexity over prior work to only a single image sensor if feed the raw  image, in which the 2D \textit{spatial} data correlates with the materials' surface \textit{structure}, directly into a convolutional deep neural network (see Section~\ref{section-CNN-training}).
Although the 2D image input requires additional time to compute the prediction result compared to 1D data, it does cause not a disruption to the laser cutting workflow (0.21s on a 2GHz Intel Core i5 processor).

For the image sensor, we chose an 8MP module (model: \textit{Raspberry Pi} \cite{pagnutti_laying_2017}.
As explained previously, this image sensor, like most commercial ones, has a higher sensitivity to the green region of the spectrum \cite{sony_imx219pqh5_2020}, which is beneficial for capturing speckle images created by our green laser pointer. We placed the laser pointer and image sensor as close as possible so that the speckles' intensity caused by the laser illumination is high enough when captured by the image sensor. 

Before mounting the image sensor, we removed the lens of the camera module using a lens focus adjustment tool. We did this because the laser pointer only illuminates a tiny area on the material sheet (i.e., size of the laser spot). When imaging the sheet with the off-the-shelf camera module that has the lens attached, the speckle is present in only a small portion of the entire image 
because the lens directs not just the laser light, but all the available light rays in the scene onto the image sensor.
This gives us less speckle data to work with.
When removing the lens from the camera module, however, it is mainly just the reflected laser light that hits the sensor, causing the speckle pattern to appear across the entire image (last column in Figure~\ref{fig:speckle-concept}b).
Thus, the camera module with no lens utilizes all the pixels of the bare image sensor and can capture a higher-resolution pattern.

\vspace{0.2cm}
\noindent\textit{Microprocessor and Battery:}
Since commercial laser cutters are closed source, we had to add a small and lightweight microprocessor and an external battery pack to allow our add-on to capture images.
Since the microprocessor has limited computational capacity, we send the captured images wirelessly to a computer for further processing. 
To make speckle sensing available in future laser cutters, manufacturers do not need to add these components since laser cutters already include processing hardware and a power supply.

\vspace{0.2cm}
\noindent\textit{Mounting on Laser Cutter:} To make our hardware add-on compact and easy to use, we designed and 3D printed the lightweight mechanical housing (60g) shown in Figure~\ref{fig:speckle-concept}c. The housing snaps onto the laser cutter’s head and can be mounted with a small rod. Attaching the sensing hardware to the laser head allows us to avoid additional calibration since because of the mount, the add-on is always located at a fixed offset from the laser. All together, the add-on weighs 114g.

\vspace{0.2cm}
In summary, our speckle-based hardware add-on consists of a laser pointer, a lensless image sensor, a microprocessor, and a battery pack. However, to integrate speckle sensing into future laser cutters, manufacturers only need to add the lensless image sensor -- all other components (power and computing infrastructure, laser pointer) already exist in laser cutters.

\section{User Interface and Applications}
\label{userInterface}

In this section, we describe our custom user interface (UI) that integrates laser speckle sensing into the laser cutting pipeline. In particular, we show how it helps users identify the material of a single sheet or multiple sheets at once, and supports users with cutting or engraving multi-material objects. Additionally, SensiCut can offer safety warnings, provide extra information on materials, and help with kerf-related geometry adjustments. We also describe how we use the interface and SensiCut's material sensing capabilities for different applications.

\begin{figure}[h]
  \centering
\includegraphics[width=0.72\linewidth]{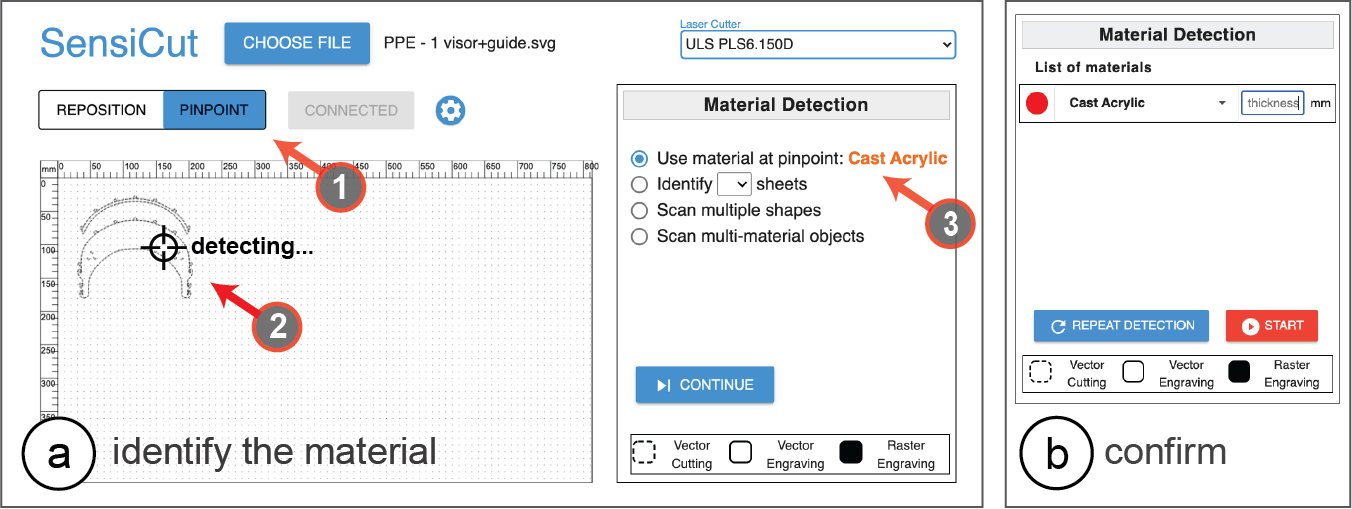}
  \caption{SensiCut UI. (a) The \textit{Pinpoint} tool allows users to identify the material at a desired location on the cutting bed. 
  (b) The user enters the thickness value corresponding to the detected sheet and starts cutting.}
  \label{fig:SensiCut-UI}
\end{figure}

Similar to the traditional laser cutting pipeline, users first start by loading their design (i.e., an SVG file) into the SensiCut UI, which subsequently shows it on the canvas (Figure~\ref{fig:SensiCut-UI}a). Next, users place the material sheet they intend to use inside the laser cutter.

\subsection{Identifying Single Material Sheets}

To identify a single material sheet, users point SensiCut's laser pointer to a desired location on the sheet as shown in Figure~\ref{fig:SensiCut-UI}a. They can do this by first choosing the \textit{Pinpoint} tool (arrow 1) and clicking the point on the canvas that corresponds to the physical location in the laser cutter (arrow 2). SensiCut then moves the laser over this location to capture the speckle pattern and classify the material. The resulting material name is then shown to the user (arrow 3).

After the user clicks \textit{Continue}, the shapes in the canvas are color-coded to reflect the material type
(e.g., \textit{red} corresponds to \textit{Cast Acrylic}).
The user then enters the material thickness in the text field next to the classification result (Figure~\ref{fig:SensiCut-UI}b).
Once ready for laser cutting, they can hit the \textit{Start} button and SensiCut automatically retrieves the appropriate laser power, speed, and pulse per inch (PPI) settings from the material settings database.

Using the identified material type, SensiCut can further support users via different functionalities:

\vspace{0.2cm}
\noindent\textit{Toxic and Flammable Material Warnings:} As mentioned in the introduction, there are many materials that should not be laser cut because they are toxic, flammable, and/or harmful to the machine. Based on the identified material type, SensiCut  displays a warning whenever the user requests to cut a material that is hazardous and should not be used in the laser cutter.

\vspace{0.2cm}
\noindent\textit{Showing Material-Relevant Information:} Even though some materials appear similar, they may exhibit different characteristics, which novice users may not be aware of. To address this, SensiCut displays additional information on each detected material to inform users about general characteristics of the material, ideal uses with sample pictures, and handling/care instructions. We referred to the laser cutting service \textit{Ponoko}~\cite{ponoko_inc_ponoko_2020} to retrieve this information. Workshop managers can edit and extend this information depending on the workshop type and its users (architecture vs. engineering).

\vspace{0.2cm}
\noindent\textit{Kerf Adjustments:} Design files can have details that are too intricate for certain material types, especially when the sheets are thin. Cutting these fine geometries can fuse details together because of \textit{kerf}. The kerf, i.e., the amount of the material removed due to the laser, depends on the type of the material~\cite{roumen_kerf-canceling_2020, roumen_springfit_2019}. When details are affected by kerf, SensiCut shows a warning to the user and then offers three options to address it: SensiCut can either slightly enlarge the design based on the material type, smooth out too intricate details, or ask the user to adjust the file manually in the drawing editor (e.g., \textit{Adobe Illustrator}).

\vspace{0.2cm}
\textit{\textbf{Application}: Fabricating a Face Shield From Different Unlabeled Scraps:}
In this application example, we would like to fabricate a face shield and use transparent plastic materials to ensure clear sight while wearing it. We start by surveying different designs online and after deciding on one~\cite{studio_eq_free_2020}, we download the parts, which consist of a visor and a shield. The design instructions highlight the importance of using the correct material for each part. In particular, it is recommended to use a transparent \textit{rigid} material for the visor (e.g., acrylic) and a transparent \textit{flexible} material for the face shield (e.g., acetate or PETG).

\begin{figure}[t]
  \centering
  \includegraphics[width=1\textwidth]{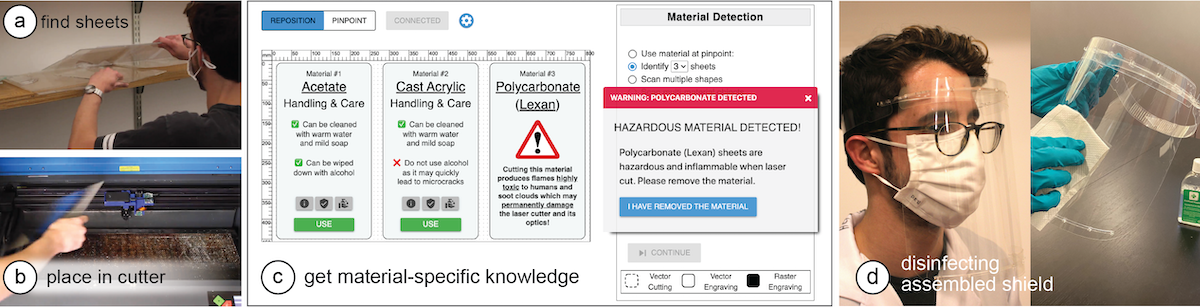}
  \caption{Making a face shield. (a)~Find flexible sheets from the stockpiles and (b)~place in the cutter. (c)~Upon identification, the UI labels all 3 materials and gives relevant information on, e.g., their handling and safety. (d)~After the cut parts are assembled, the shield can be safely sanitized with alcohol.}
  \label{fig:PPE-shield}
\end{figure}

We start by browsing through the leftover scrap materials from our workshop.
However, while going through the transparent scrap materials, we notice that almost all of them are unlabeled. To make the visor, we take the first rigid transparent material we find in the pile that has a sufficient size and place it inside the laser cutter. We then open SensiCut's UI (Figure~\ref{fig:SensiCut-UI}a) and select \textit{Pinpoint} to identify the material at a desired location.
The right-hand bar shows the material has been detected as \textit{cast acrylic}, which is a suitable rigid material for the visor. Once we confirm, SensiCut automatically retrieves the appropriate power/speed settings for this job and the laser cutter starts cutting the acrylic.

We repeat the procedure for the shield, which needs to be made from a flexible transparent material. We go back to the scrap materials and find three different flexible sheets (Figure~\ref{fig:PPE-shield}a). We read online that acetate and PETG may be more suitable than other plastics, but are not sure which one is which. We take all three and place them in the laser cutter. In the UI, we then choose \textit{Identify sheets} and set the number of sheets to \textit{3}. 
Next, we click on one point on each sheet to instruct SensiCut to identify the material there. Once the results are displayed, we realize that one of them is a \textit{polycarbonate (Lexan)}, which SensiCut labels with a "hazardous" warning (Figure~\ref{fig:PPE-shield}c). The other two sheets are identified as acetate and (thin) cast acrylic.

To learn more about the difference between the two materials, SensiCut shows information from its knowledge database and displays it on the corresponding material.
For example, it shows that acetate has high impact strength and is reasonably flexible, and that, in contrast to acrylic, it can be wiped down with alcohol, which is important to disinfect the shield. We remove the Lexan and acrylic sheets from the laser cutter, 
choose acetate in the UI, 
and start cutting. Now that all parts are cut, we can assemble the final face shield (Figure~\ref{fig:PPE-shield}d).

\subsection{Identifying Multiple Sheets of Different Materials at Once}
SensiCut also allows users to cut multiple sheets of different material types in rapid succession. The user first loads the design files that contain the shapes they want to cut from the different sheets. Next, they place the corresponding material sheets inside the laser cutter and initiate the \textit{Scan multiple shapes} mode. This causes SensiCut to go to the location of each shape and capture an image there for material identification. The resulting material names are then displayed in the \textit{Material Detection} sidebar and the shapes are similarly color-coded based on the material types.

\begin{figure}
  \centering
  \includegraphics[width=1\textwidth]{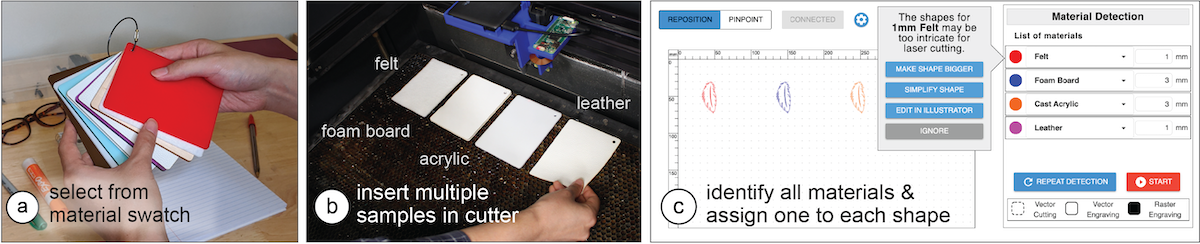}
  \caption{Rapid material testing for product design. (a)~Choosing multiple samples from a material swatch and (b)~inserting in the laser cutter. (c)~After sensing, the UI matches each shape with the corresponding material. It also warns that the kerf for felt will compromise the design.}
  \label{fig:Material-Sample-Swatch-Prototyping}
\end{figure}

\vspace{0.2cm}
\textit{\textbf{Application}: Rapid Testing of Multiple Material Types 
for Product Design: }
\label{app-multi-sample-product-design}
In our second example, we want to rapidly prototype a new earring design in a white color. We want to fabricate the earrings to test the look and feel of different materials to determine which one looks best when worn. To evaluate different material types, we pick a handful of white samples from a material swatch (Figure~\ref{fig:Material-Sample-Swatch-Prototyping}a). 

To speed up our prototyping process, we want to cut all the material samples at once. To do this, we place all our selected material samples on the laser cutter bed as shown in Figure~\ref{fig:Material-Sample-Swatch-Prototyping}b. We then load the earring design. Next, we position a copy of the earring design in the UI in the location where each material sample is placed. Next, we choose the \textit{Scan multiple shapes} option. After the scan, each earring's shape is color-coded to reflect the detected material type: felt, foam board, cast acrylic, and leather (Figure~\ref{fig:Material-Sample-Swatch-Prototyping}c).


\begin{figure}
  \centering
  \vspace{0.3cm}
  \includegraphics[width=1\textwidth]{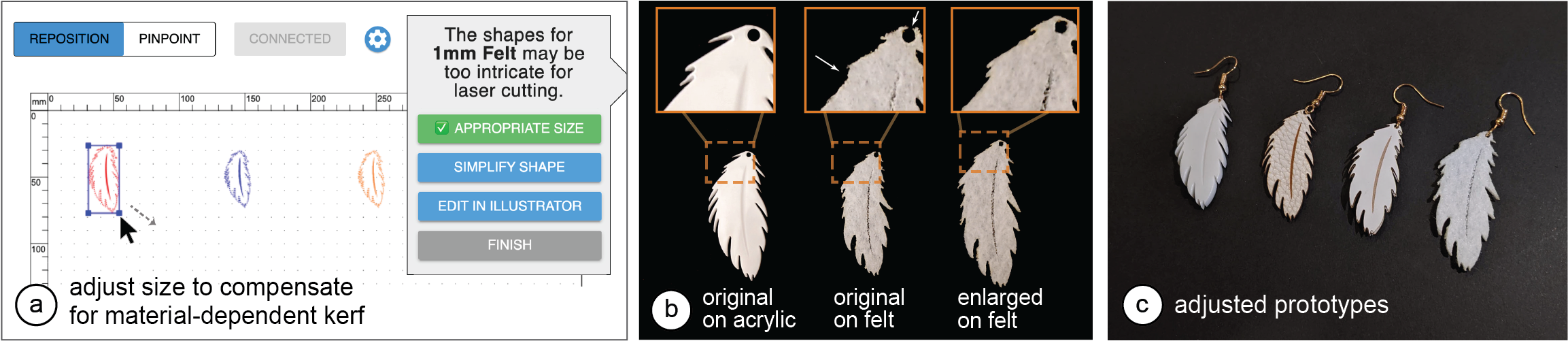}
  \caption{Adjusting the design for kerf. (a)~The user can enlarge the shape to compensate for the thicker kerf for felt. (b)~Without the adjustment, details like the hole and fine blades of the leaf disappear in the cut felt. (c)~The fabricated prototypes.}
  \label{fig:Material-Cut-Earrings}
\end{figure}

Next, we enter the thickness for each material sheet: 1mm for felt and leather, and 3mm for the others. For felt, SensiCut shows a notification that our design has details that are too intricate for a thickness of 1mm and may thus fuse together because of kerf (Figure~\ref{fig:Material-Cut-Earrings}b). We choose the \textit{Make shape bigger} option and enlarge the felt earring so that the minimum feature size no longer goes below the kerf limit (Figure~\ref{fig:Material-Cut-Earrings}a). 
Once adjusted, we cut and engrave all sheets in a single job. 
The finished earring prototypes are shown in Figure~\ref{fig:Material-Cut-Earrings}c.


\subsection{Engraving onto Multi-Material Objects}

\begin{figure*}[t]
  \centering
  \includegraphics[width=0.84\textwidth]{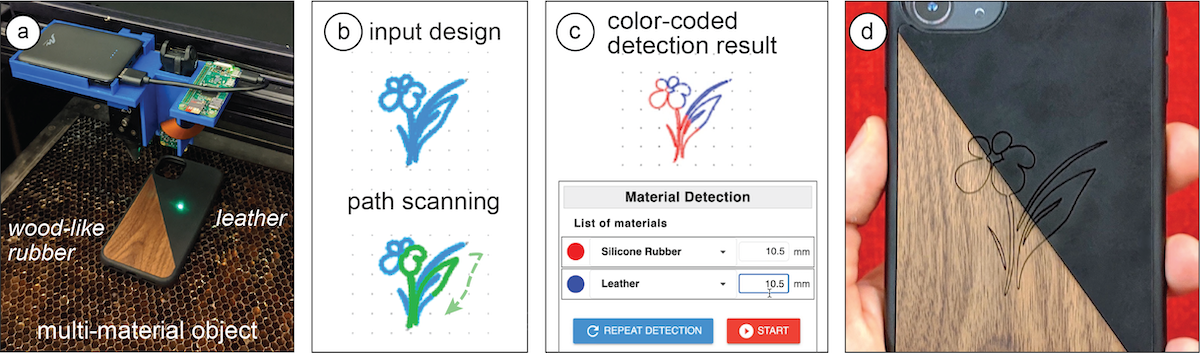}
  \caption{Engraving a multi-material phone case, which consists of (a)~wood-like rubber and leather. SensiCut scans (b)~the  input design's outline and (c)~splits it into two parts based on the material type detection. (d)~Engraved design.}
  \label{fig:multimaterial-case-engraving-scanning}
\end{figure*}

Compared to individual material sheets, cutting or engraving designs onto multi-material objects (e.g., the smartphone case in Figure~\ref{fig:multimaterial-case-engraving-scanning}a) is a particularly challenging task. It requires a cumbersome workflow where users first have to split the design into multiple files, one for each material. More specifically, proper alignment of the shapes in the digital design with the different parts of the physical object is challenging without knowing where the material borders are located.

SensiCut facilitates cutting and engraving on multi-material objects by automatically splitting the design precisely along the border of different materials by sensing the material type at each point in the design. For this, users start by loading a single file containing the entire design, insert the multi-material object into the laser cutter, position the design onto the multi-material object, and select the \textit{Scan multi-material objects} mode. SensiCut then samples points along the laser-cut path to identify the material at each point. The scanning progress is shown in the SensiCut UI by highlighting the scanned trajectory. After scanning is completed, SensiCut splits the design according to the detected material type at each point to ensure the correct laser settings will be used.

\vspace{0.2cm}
\noindent
\textit{\textbf{Application}: Personalizing Existing Multi-Material Products:}
\label{app-multimaterial-case-engraving}
In this example, we want to engrave a custom design at the center of a smartphone case that consists of two different materials across its surface, i.e., leather and wooden parts (Figure~\ref{fig:multimaterial-case-engraving-scanning}a).


First, we load the design file, position it on top of the phone case, and select \textit{Scan multi-material objects}, which then moves the laser head along the design's engraving path to detect the material at each point (Figure~\ref{fig:multimaterial-case-engraving-scanning}b). Once the scan is complete, SensiCut splits the design into two parts, one for each of the two materials (Figure~\ref{fig:multimaterial-case-engraving-scanning}c). 

Once SensiCut identified the materials, we realize that the part we had thought was wood is actually made of silicone rubber with a decorative wood pattern. SensiCut is not deceived by the disguise pattern because it measures surface structure and sets the correct laser engraving settings. Once we confirm, our design is engraved onto our multi-material phone case (Figure~\ref{fig:multimaterial-case-engraving-scanning}c).

\begin{figure}[h]
  \centering
  \includegraphics[width=0.75\linewidth]{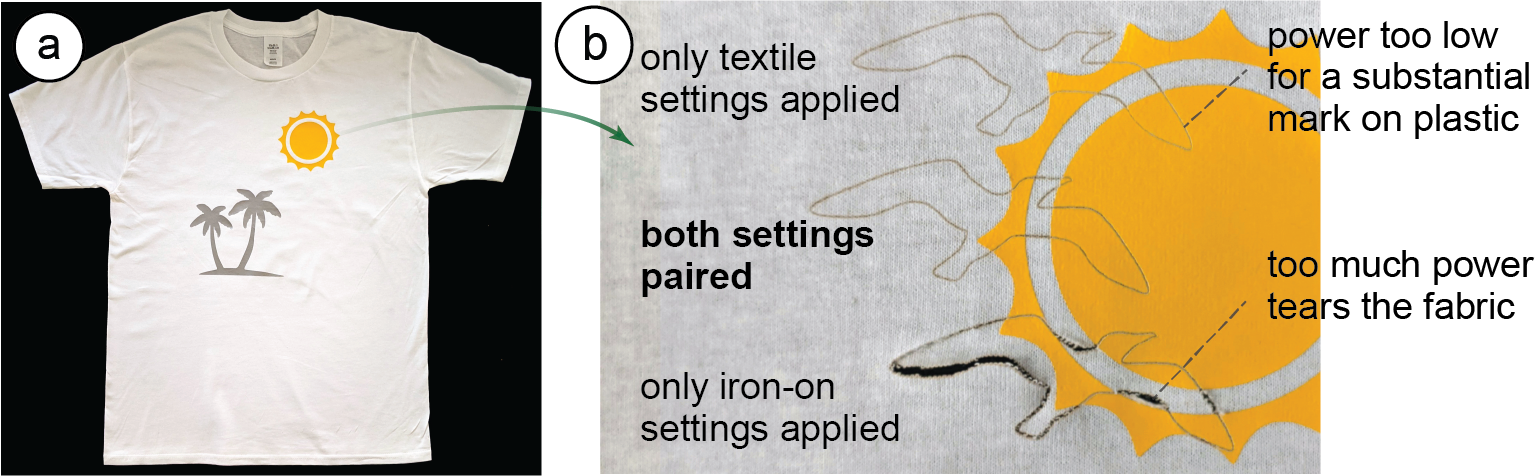}
  \caption{Engraving a pattern on a T-shirt that has (a)~plastic details on it. (b)~SensiCut  uses the right combination of laser settings after partitioning the design (middle). Top/bottom shows the outcome for singular settings.}
  \label{fig:EngraveTShirt}
\end{figure}

\vspace{0.2cm}
\noindent
\textit{\textbf{Application}: Customizing Multi-Material Garments:}
Figure~\ref{fig:EngraveTShirt}a shows another multi-material item, i.e. a T-shirt, that we want to engrave with a custom seagull design. The T-shirt has a plastic iron-on material applied on it. To engrave our design, SensiCut detects which parts are made of textile and which are made of plastic. It then splits the seagull design into multiple paths accordingly and assigns the correct laser power/speed settings for each one (Figure~\ref{fig:EngraveTShirt}b). If we had instead used only one set of laser power/speed settings for the entire seagull design, i.e., the settings for either textile or plastic, the lines would either not be visible on the yellow plastic or the textile would have been burned.
Further, it would be particularly difficult to achieve this without SensiCut: One would have to remove the iron-on plastic from the fabric itself, engrave the plastic and fabric separately, and put them back together precisely. This shows how SensiCut could help users further customize garments that have non-textile parts (e.g., ~\cite{forman_defextiles_2020}) quickly and on demand.

\section{Classification of Materials}

SensiCut can differentiate between 30 different materials 
 relevant to the challenges laser cutter users face.
In the next section, we discuss how we built a dataset of speckle patterns of these materials using an automated script, and how we trained a convolutional neural network (CNN) to be able to distinguish between them.

\subsection{Choosing Material Samples}
\label{Choosing-Material-Samples}
For our dataset, our goal was to choose materials that are most representative of the materials commonly found in makerspaces and workshops, with a particular focus on the ones that cause confusion because of their appearances.
Figure~\ref{fig:MaterialTree}a summarizes the list of materials we compiled by surveying a range of online communities (e.g., \textit{Thingiverse}~\cite{makerbot_thingiverse_2020}, \textit{Instructables}~\cite{autodesk_instructables_2020}), educational materials on laser cutting~\cite{bates-green_materials_2018}, supply vendors~\cite{ponoko_inc_ponoko_2020, trotec_laser_laser_2020}, as well as the laser cutter material databases that come with the default laser cutter control software (e.g., ULS \textit{Universal Control Panel}~\cite{universal_laser_systems_universal_2020}).

The resulting material list includes 30 different materials ranging from different types of paper, plastic, wood, fabric to (engraved) metal. In the next section, we discuss how these selected materials are representative of the challenges that laser cutter users face.

\vspace{0.2cm}
\noindent\textit{Different Laser Cutting Materials with Similar Appearance:} As our formative material labeling study showed, plastics are particularly challenging to distinguish for users due to their visual similarity ~(Figure~\ref{fig:MaterialTree}b).
To represent such cases, we purchased samples of cast acrylic, extruded acrylic, and Delrin (also known as \textit{acetal} or \textit{POM}) of the same color, which require different settings to properly cut/engrave a design~\cite{bates-green_materials_2018}. We also included samples of different transparent sheets, i.e., acrylic, PETG, and acetate. In addition, we included other materials that have slightly varying appearances, but still are difficult to distinguish for non-expert users who are not familiar with the specific nuances, such as different types of wood (e.g., maple, oak, bamboo, or birch)~\cite{wiedenhoeft_species_2014}.



\vspace{0.2cm}
\noindent\textit{Hazardous Materials that Look Similar to Safe Ones:} To represent cases where some of the commonly found materials in workshops are hazardous (flammable, toxic, or harmful to the machine) and cannot be safely laser cut~\cite{bates-green_materials_2018, kokosa_hazardous_1992}, we included polyvinyl chloride (PVC), Lexan (polycarbonate), acrylonitrile butadiene styrene (ABS), and carbon fiber sheets. PVC is often mistaken for the common laser-cut material acrylic. However, it is highly toxic as it releases hydrochloric acid fumes when heated, which also rapidly corrode the laser system~\cite{herrick_emerging_2016}. Lexan and ABS are also hazardous and easily flammable\footnote{\scriptsize \url{https://wiki.aalto.fi/display/AF/Laser+Cutter+Materials}} but look similar to safe plastics. However, whether a material is considered safe for laser cutting or not depends on the specific hardware setup (air filter type and volume, power of laser) as well as local regulations~\cite{haferkamp_air_1998}. For our setup,  materials in the ULS material database that comes with our laser cutter and its \textit{UAC 2000} filter (MERV 14, HEPA, 2 Carbon filters) are marked as safe. For instance, polystyrene is listed as safe for our setup but may not be safe for others. Thus, we recommend that when deployed in a new workshop, SensiCut's database be updated by the workshop manager locally after checking material safety data sheets (MSDS) for potential laser generated air contaminants (LGAC). Workshop managers should also talk to their local occupational health institution (e.g., \textit{NIOSH}\footnote{\scriptsize \url{https://www.cdc.gov/niosh/}} in the US).



\vspace{0.2cm}
To make the material composition of our dataset representative of a real-world workshop, where certain materials like acrylic and cardboard are much more available in terms of
quantity/color options, we included more than 1 sheet for these as seen in Figure~\ref{fig:MaterialTree}a.
This also allows us to evaluate our system for different colors and transparencies. In total, we used 59 material samples, the majority of which were purchased from \textit{Ponoko}~\cite{ponoko_inc_ponoko_2020}, except for the 5 hazardous material sheets (PVC, Lexan, etc.), which we purchased from other suppliers on \textit{Amazon.com}.
A list of these material samples and the associated vendors can be found in the appendix.

\subsection{Data Capture and Material Speckle Dataset}

After purchasing the different materials, we captured images of each sample to build a dataset for training our convolutional neural network. 

\vspace{0.2cm}
\noindent\textit{Preliminary Experiment:}
\label{DataCaptureExperiment}
Before capturing data for all materials, we ran a preliminary experiment to determine two values: (1)~the distance between the image sensor and the material surface at which the speckle pattern is most visible, and (2)~the number of images necessary for training the classifier with high accuracy. For the distance, we empirically found that 11cm between our image sensor and the material surface led to the best results. For the number of images, we placed material samples below the image sensor at the recommended distance and took images, 
moving the sample in the xy plane manually to simulate how the laser cutter would take images at various points of the sheet.
We found empirically that around 80-100 images are sufficient for each material to train a CNN for classification.

\vspace{0.2cm}
\noindent\textit{Data Collection:} After this manual exploration, we started the data collection of all materials. For this, we wrote a script to automate the laser cutter's movement and image capture.  
For our material samples (6.3cm x 6.3cm), we chose to capture a 9x9 grid of points leading to 81 images, which satisfies our criteria from the preliminary experiment.
For consistency, we kept the image sensor settings, i.e., exposure time, digital/analog gains, and white balance constant.

Additionally, we captured images at different heights (z-locations) to ensure that the network can classify materials of different thicknesses.
This is necessary since the speckle pattern changes with the distance between the material surface and the image sensor.
We chose 8 different heights ranging from 0mm (to support paper) to 7mm (thickest material sheet we were able to buy) spaced at 1mm increments. However, not every sheet has a thickness of a multiple of 1mm (e.g., some sheets are 2.5mm). We can generate this additional data using data augmentation methods as explained in Section~\ref{section-CNN-training}.
Since our model was trained for materials with a thickness of 0-7mm and the material surface was 11cm away from the image sensor, this leads to an effective detection range of 110-117mm. To integrate material identification into other cutting-based methods like \textit{LaserOrigami}~\cite{mueller_laserorigami_2013} or \textit{FoldTronics}~\cite{yamaoka_foldtronics_2019}, the model can be trained for larger distances in the future.

\vspace{0.2cm}
\noindent\textit{Dataset:} Our final data set contains 38,232 images from 59 material samples of 30 unique materials (14.93 GB, 800x800 pixels each). Each material sample includes 648 images (81 images/height x 8 heights), which took about 40 minutes to capture with our automated setup. The majority of this time is spent waiting for the laser head to stabilize after moving to a new location to ensure that the captured image is not blurry. The dataset is used for training the CNN and does not need to be stored on the user's computer. The trained CNN model that is used at detection time is 120MB. The dataset is publicly available\footnote{\url{https://hcie.csail.mit.edu/research/sensicut/sensicut.html}}.

\subsection{Training the Neural Network}
\label{section-CNN-training}

To train the CNN and build a detection model using the captured images, we used transfer learning with a ResNet-50
model~\cite{he_deep_2015} that was pre-trained on the \textit{ImageNet} dataset~\cite{russakovsky_imagenet_2015}.
We used the Adam optimizer with a learning rate of 0.003 and a batch size of~64. 
We used 80\% of images for the training set and reserved 20\% for the validation set.

\vspace{0.2cm}
\noindent\textit{Image Size Used for Training:} For the input image size, we chose 256x256 pixels. Although we captured the images in 800x800 pixels, we found that the  higher resolution caused lower accuracy as the model overfit to irrelevant details in the image. The lower resolution input also saves training time because the model has fewer nodes to compute. Moreover, it speeds up the detection during use (i.e., average prediction time: 0.21s for 256x256px vs. 0.51s~for 400x400px on a 2GHz \textit{Intel Core i5}). We still keep the full-size images in our dataset to enable future research.

\vspace{0.2cm}
\noindent\textit{Data Augmentation:} To make the model robust to different lighting conditions and intermediate sheet thicknesses (e.g., 2.5mm), we generated additional images during training using data augmentation. Every time the network starts training on a new batch of images, a portion of the images is transformed
by changing the brightness and the contrast of all pixels (by up to \( \pm \)30\%), as well as zooming into the image to enlarge the speckles as would be the case when the thickness of the sheet decreases (by up to \( \pm \)20\%). This allows our model to generalize better and also saves time by avoiding the capture of more images with the physical setup.


\vspace{0.2cm}
\noindent
In the future, new materials can be added to SensiCut by capturing more speckle images and adding them to the dataset. For this, the neural network needs to be retrained but the weights from this previous training can be used (transfer learning), which significantly speeds up the process , i.e., takes only 10-12 minutes vs. 6 hours  training from scratch.

\begin{figure*}[t]
  \centering
  \includegraphics[width=1\textwidth]{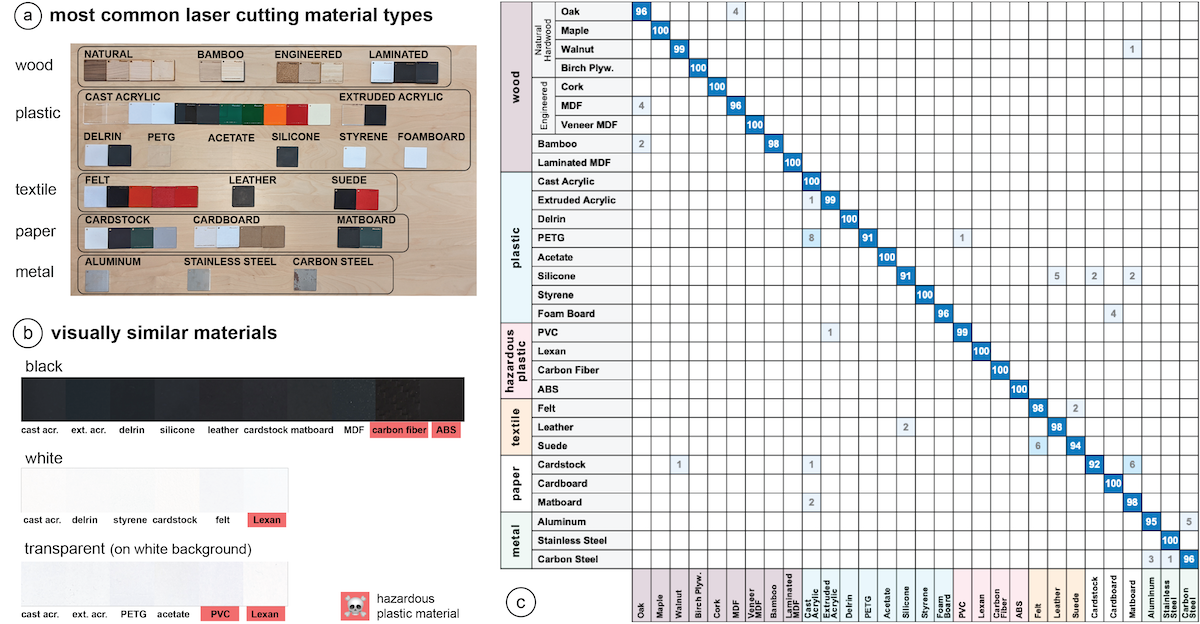}
  \caption{Material considerations and evaluation. (a) Most common material types for laser cutting. (b) Visually similar materials. 
  (c) Confusion matrix from our trained classifier for 30 material subtypes.
  }
  \label{fig:MaterialTree}
\end{figure*}

\section{Evaluation of Material Classification}

We conducted a technical evaluation to determine our trained classifier's accuracy. We also carried out additional tests
to understand how the model generalizes to different physical conditions (rotation of sheets, illumination variations) and material sheets purchased from different vendors.

\subsection{Detection Accuracy Results}
\label{analysis-one}



The results of the classification accuracy for the 30 different materials in our dataset are shown as a confusion matrix in Figure~\ref{fig:MaterialTree}c. Our average identification accuracy is 98.01\% (SD=0.20) across the different materials. This is based on a 5-fold cross-validation, which we ran to ensure consistency of the classification accuracy across different training and validation splits. The small standard deviation shows that training our model leads to similar results independent of how the dataset is split. For this reason and the fact that cross-validation is a time-consuming procedure (30 hours for 5-fold),
the remainder of the technical evaluation is based on a single run.

We further analyzed the results to understand which of the materials outlined in Section~\ref{Choosing-Material-Samples} are confused for each other most.
For instance, given a specific color (either white, black, or red), we evaluated the identification accuracy across cast acrylic, felt, paper, and laminated MDF. The accuracy was 100\% for white and red, and 92\% (SD=12.72) for black, on average. The latter is likely due to the fact that black reflects less light. Since the image sensor's exposure is the same for all photos, this causes the reduced accuracy for black materials. Enabling adaptive exposure when capturing images could eliminate this difference in the future~\cite{yeo_specam_2017}.

Figure~\ref{fig:MaterialTree}c also shows materials that were mistaken for each other. For instance, leather and silicone were confused with each other at a relatively high rate compared to other pairs. We believe this is because our full-grain leather piece is \textit{hot stuffed}, i.e., conditioned with unrefined oils and greases, which likely makes its surface structure closer to that of silicone.
One can also observe some confusion between walnut and paper-based materials, such as cardstock. The
similarity in their surfaces may be due to the fact that paper is produced using cellulose fibres derived from wood.

We also evaluated the accuracy of materials within the same material groups. We got a mean accuracy of 98.92\% across woods (SD=1.66), 98.84\% across plastics (SD=2.36), 97.25\% across textiles (SD=2.50), 95.90\% (SD=2.94) across paper-based materials, and 97.00\% (SD=2.16) across metals. The fact that paper-based sheets had the lowest rate is expected as they share the most similarities in their surface structures among different subtypes (e.g., cardstock vs. cardboard have a similar surface texture).

\subsection{Effect of Illumination and Sheet Orientation on Detection}
\label{analysis-three}
To understand how detection accuracy is affected under different illumination (ambient light) conditions and sheet orientations, we ran additional tests.

\begin{figure*}[h]
  \centering
  \includegraphics[width=0.9\linewidth]{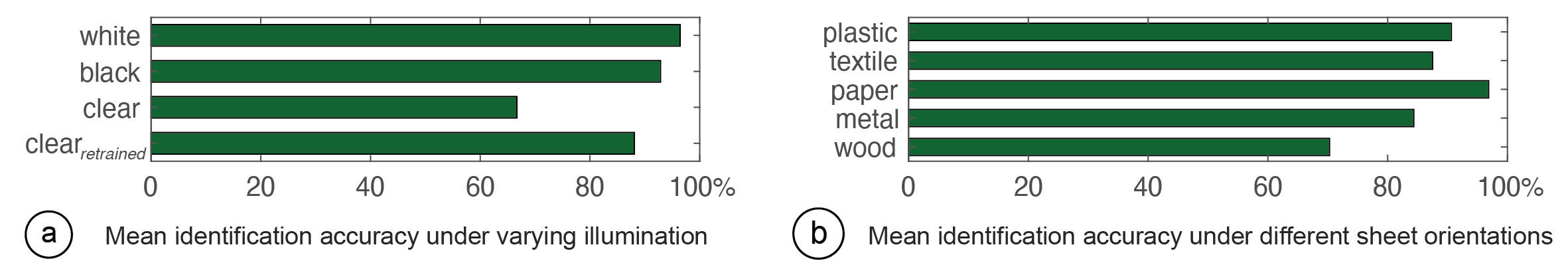}
  \caption{Effect of (a)~varying illumination and (b)~sheet orientation on detection. }
  \label{fig:Illumination-Sheet-Orientation}
\end{figure*}

\vspace{0.2cm}
\noindent\textit{Ambient Light:} 
When we captured the images for our main dataset, we kept illumination in the workshop low (i.e., all lights turned off).
To evaluate if the trained model can distinguish between materials even when the ambient light varies, we created an additional test set of images under different lighting conditions.
For this, we used two lamps, one on the left and one on the right corner of the room, resulting in three conditions 
(light1 on, light2 on, both lights on) that cover an illumination range of up to 80 lux, in addition to the initial data with all lights off. We tested this on different black and white sheets, representing the two ends of the light reflectance spectrum, as well as clear (transparent) sheets. 
We compare the accuracy in the following three scenarios: 8 white and 8 black sheets (for each color 2 sheets per type: plastic, paper, textile, wood), and 6 transparent sheets (all plastic).

The results are shown in Figure~\ref{fig:Illumination-Sheet-Orientation}a.
We found that the increased brightness did not have a major impact on the detection of white and black sheets.
For clear sheets, however, the mean accuracy was lower (66.67\%). The reason for this is that while opaque sheets benefit from data augmentation, this is not the case for clear sheets. We found that the illumination increase in the room was not realistically simulated in the digitally generated images of clear sheets because such materials allow light to pass through in all directions. This can be overcome by capturing additional images of clear materials under the varying light conditions and then retraining the model. Indeed, such retraining resulted in an increased accuracy of 88.10\% (shown in the last bar). We also found that 
retraining the model on this augmented dataset 
did not have a major impact on other materials' detection (only by 0.41\% on average). 

\label{Orientation-Sheets}
\vspace{0.2cm}
\noindent\textit{Orientation of the Sheets:} The images we captured for our main dataset were all taken in a specific sheet orientation. We thus evaluated if the classifier is still accurate when the sheets are arbitrarily rotated  for materials with uniform (e.g., acrylic) or non-uniform/irregular surface structure (e.g., wood).
For this, we created an additional test set by capturing speckle images while rotating the material sheets at 45$^{\circ}$ increments. For materials with uniform surface (plastic, textile, paper, metal), we picked two subtypes each (cast acrylic, Delrin, cardboard, matboard, felt, leather, aluminium, carbon steel).
For materials with non-uniform surface, we tested eight subtypes of wood  (oak, maple, walnut, birch, MDF, veneer, bamboo, laminated).

The results are shown in Figure~\ref{fig:Illumination-Sheet-Orientation}b. The lowest average detection accuracy was for wood sheets (70.31\%), which also had a high standard deviation among the wood subtypes (24.94\%). This is due to the fact that wood sheets included both artificial ones with regular surface structure (e.g., MDF), which resulted in 100\% detection accuracy, and natural woods with irregular surface structure (e.g., oak), which resulted in lower accuracies. The misidentified images for those materials were all captured at the odd degrees (45$^{\circ}$, 135$^{\circ}$, etc.). 
We believe this is due to the cellular 3D microstructure of natural wood that has a 90$^{\circ}$ rotational symmetry at the microscopic level~\cite{ansell_wood_2015}. We can increase detection accuracy for natural woods by augmenting the training dataset with more pictures taken at different angles, at the expense of longer capture time.

\subsection{Generalization to Different Material Batches and Manufacturers}
\label{analysis-four}
In our main dataset, each set of samples came from one manufacturer. To ensure that our trained model can work robustly for sheets from different batches of the same manufacturer or different manufacturers, we conducted the following two tests. 

\vspace{0.2cm}
\noindent\textit{New sheets from the same manufacturer:} Two months after we purchased our samples, we ordered a second batch of sheets from \textit{Ponoko} (two subtypes per material: oak, maple, cast acrylic, Delrin, felt, leather, cardboard, cardstock) and placed them inside the laser cutter to test if our trained model can still identify them. We found that only the maple sheet was incorrectly classified. As explained in Section~\ref{Orientation-Sheets}, this is likely due to unique microstructural orientation of natural wooden sheets.
    
\vspace{0.2cm}
\noindent\textit{New sheets from different manufacturers:} To test if sheets from different manufacturers can be reliably identified, we ordered 8 different sheets from various vendors on \textit{Amazon.com} (two subtypes per material: birch, cork, cast acrylic, Lexan,  leather, felt, cardboard, matboard). Only one of them, leather, was incorrectly classified. We later found out that the new sample was synthetic leather, whereas the classifier was trained on natural leather. 

\vspace{0.2cm}
While the above evaluation demonstrated that our classification model can  detect material types across various conditions, more longitudinal tests are needed to further verify its applicability across various workshop settings.

\section{Software Implementation}

Our user interface is Web-based and implemented using JavaScript and the \textit{Paper.js} library. When users request a material identification, our system automatically moves the laser cutter head to the corresponding xy-coordinates on the physical cutting bed.
These coordinates are offset by the distance between SensiCut's laser pointer and the cutting laser. For the z-value, a fixed distance to the sheet is used for capturing the speckles (Section~\ref{DataCaptureExperiment}).
To input the coordinates and initiate the movement, our system uses the \textit{PyAutoGUI} library to interface with the ULS Universal Control Panel (UCP).
It then detects if the laser head stopped moving, i.e., is stable enough to take a picture, by checking if the laser is in \textit{idle} mode. This is indicated via a color change in UCP, which our system can detect via \textit{PyAutoGUI}'s \textit{screenshot()} and \textit{getcolors()} functions.

\begin{figure}[t]
  \centering
  \includegraphics[width=0.67\textwidth]{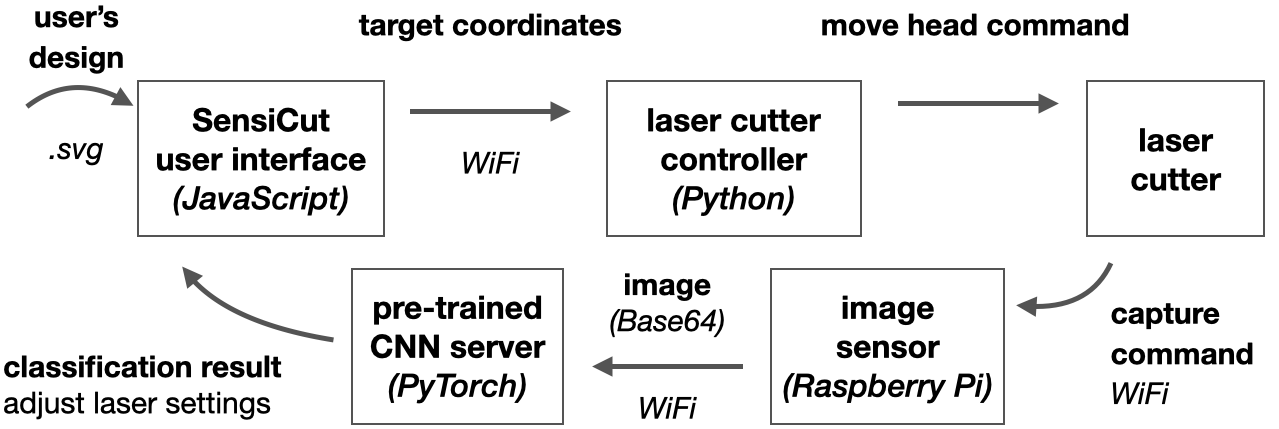}
  \caption{Our detection pipeline takes the user’s drawing as input and turns it into target points to capture speckle patterns. The captured images at those points are passed to the CNN to retrieve the material label.}
  \label{fig:Script-communication}
\end{figure}

Next, the captured image is wirelessly sent from the hardware add-on's Raspberry Pi board to the main server, which runs on an external computer. This Python server uses the image as input to the trained CNN model, which was implemented using \textit{PyTorch} and \textit{fast.ai} \cite{howard_fastai_2020}. The CNN returns the classification results, which are then displayed in the UI.
The communication between the JavaScript front-end and the Python back-end is handled by the \textit{Socket.IO} framework. The complete pipeline is shown in Figure~\ref{fig:Script-communication}.

After the user confirms the results, the laser power, speed, and PPI (pulses per inch) settings are retrieved from the ULS database based on the detected material. Because the UCP interface does not have an API, we extracted these values from the its back-end using a \textit{Firebird} server and \textit{Database Workbench 5 Pro} to create an interim datasheet from which we can look up values as needed.

To detect kerf-related issues, we  first dilate the drawing with a kernel of the size of the material-specific kerf. We check if two curves overlap or if the dilation results in extra blobs, i.e., the cut may lead to an undesired shape. The materials' kerf values are based on the \textit{"Minimum feature size"} values listed on \textit{Ponoko}.

For the multi-material object mode, our system samples points uniformly along the cutting path and processes the captured images according to the pipeline described previously. Our system then assigns the respective identified material to each part of the users' drawings.

\section{Discussion}
\label{Discussion}

In the next section, we discuss insights gained from our work, acknowledge the limitations of our approach, and propose future research.



\vspace{0.2cm}
\noindent\textit{Avoiding Dust on the Sensor:} In conventional cameras, lenses help prevent dust particles from landing on the sensor. Although we use a lensless image sensor to capture the speckle patterns, over the course of our research we did not observe the lack of the lens to interfere with classification results. We hypothesize that this is the case because (1)~the sensor is facing down, which prevents dust particles from reaching the surface of the sensor due to gravity, and (2)~the ventilation in the laser cutter bed sucks away particles from the image sensor.

\vspace{0.2cm}
\noindent
\textit{Confidence Scores for Misidentified and/or Unknown Materials}: 
The neural network's final layer outputs a vector for the confidence score of each material type. If multiple types have similar scores, the material is either misclassified or not included in the original training dataset. As part of our future work, we will extend the user interface to show bars to visualize the confidence scores and inform the user to act with caution when confidence scores of multiple materials are similar. Additionally, a confidence threshold for when a material is safely classified could be set by the workshop manager for all users of the workshop. 

\vspace{0.2cm}
\noindent
\textit{Effect of Scratches on Sheets}: 
Scratches on sheets are often local, i.e., they occur when a sheet's sharp corner abrades a spot on another sheet's surface. We picked 2 cast acrylic, 2 birch, 2 cardboard sheets with the most scratches from the material pile in our workshop and captured speckle patterns at 30 uniformly distributed points across the surface of each material sample. We found that the majority of points were correctly classified, i.e. 90\% for acrylic, 91.7\%  for birch, 86.7\% for cardboard. In future work, to make material detection robust to local scratches, SensiCut could take more than one image and cross-check the classification result at the expense of longer detection times.

\vspace{0.2cm}
\noindent\textit{Materials with Protective Cover:} Some material sheets come with a protective plastic/paper cover to avoid scratches during transportation and some users may prefer to leave it on during laser cutting. Since SensiCut needs access to the material's surface, users can peel a small section from the corner and use our interface's \textit{Pinpoint} function to detect the material type from that corner.

\vspace{0.2cm}
\noindent\textit{Estimating the Sheet Thickness from Speckles:} As the distance between the image sensor and the material surface increases, the speckles appear larger in the image~\cite{smith_colux_2017}. If the pictures are taken at a fixed height (i.e., a fixed distance of the laser head to the cutting bed), then the surface of thicker sheets is closer to the laser head, which results in smaller speckles, while thinner sheets are further away from the laser head, resulting in larger speckles. We tested if this can be used to detect the sheet thickness by using the same dataset and CNN structure as the material type classifier (ResNet). However, as this is a regression problem, we used mean squared error instead of cross-entropy as our loss function. An initial test across 14 material sheets gave us a mean error of 0.55mm (SD=0.68mm). For the ULS laser lens we have, the depth of focus (i.e., tolerance to deviations from the laser's focus) is 2.54 mm, which is larger than this detection rate. Thus, for future versions of SensiCut, we can also include thickness detection.

\vspace{0.2cm}
\noindent\textit{Labeling Workflows:} While some users may prefer to keep material sheets unlabelled and launch SensiCut every time they use the laser cutter, SensiCut can also support hybrid workflows, such as printing a sticker tag after identifying a sheet, which can then subsequently be attached to the material sheet. Similarly, the software interface could remind users to label the material sheet with a pen after use as a courtesy to the next maker.

\vspace{0.2cm}
\noindent\textit{Material Identification for Other Fabrication Tools:} For future work, we plan to explore how SensiCut's material identification method can be used for other personal fabrication machines as well. For example, in 3D printing, some manufacturers, such as \textit{Ultimaker}, add NFC chips into filament spools to allow the chip reader integrated in the 3D printer to automatically detect them. However, not all spools come with such chips. To address this issue, we plan to investigate how speckle sensing can be integrated into filament feeder systems to detect the filament type when a new spool is loaded onto the 3D printer.

\section{Conclusion}

In this chapter, we presented SensiCut, a material sensing platform that helps laser cutter users to identify visually similar materials commonly found in workshops. We demonstrated how this can be achieved with speckle sensing by adding a compact and low-cost hardware add-on to existing laser cutters. We showed how the material type detection can be used to create a user interface that can warn users of hazardous materials, show material-relevant information, and suggest kerf adjustments. Our applications demonstrated how SensiCut can help users identify unlabeled sheets, test various materials at once, and engrave onto multi-material objects. We discussed how we chose the materials in our dataset and how we trained the convolutional neural network for their classification. We reported on the detection accuracy for different material types and evaluated the impact of varying the room illumination, rotating the sheets, and using sheets purchased from different manufacturers. We then highlighted how our system can be extended to also detect the thickness of sheets. For future work, we plan to investigate how speckle sensing can be used to detect materials in other fabrication tools. Furthermore, we plan to collaborate with laser cutter manufacturers to integrate our material sensing approach into future commercial products, which only requires adding the lensless image sensor and adjusting the power of the existing visible laser pointer.

\chapter{StructCodes: Leveraging Fabrication Artifacts to Store Data in Laser-Cut Objects}
\label{thesis-StructCodes}

\section{Introduction}

\begin{figure}[h]
  \centering
  \vspace{0.33cm}
  \includegraphics[width=0.52\textwidth]{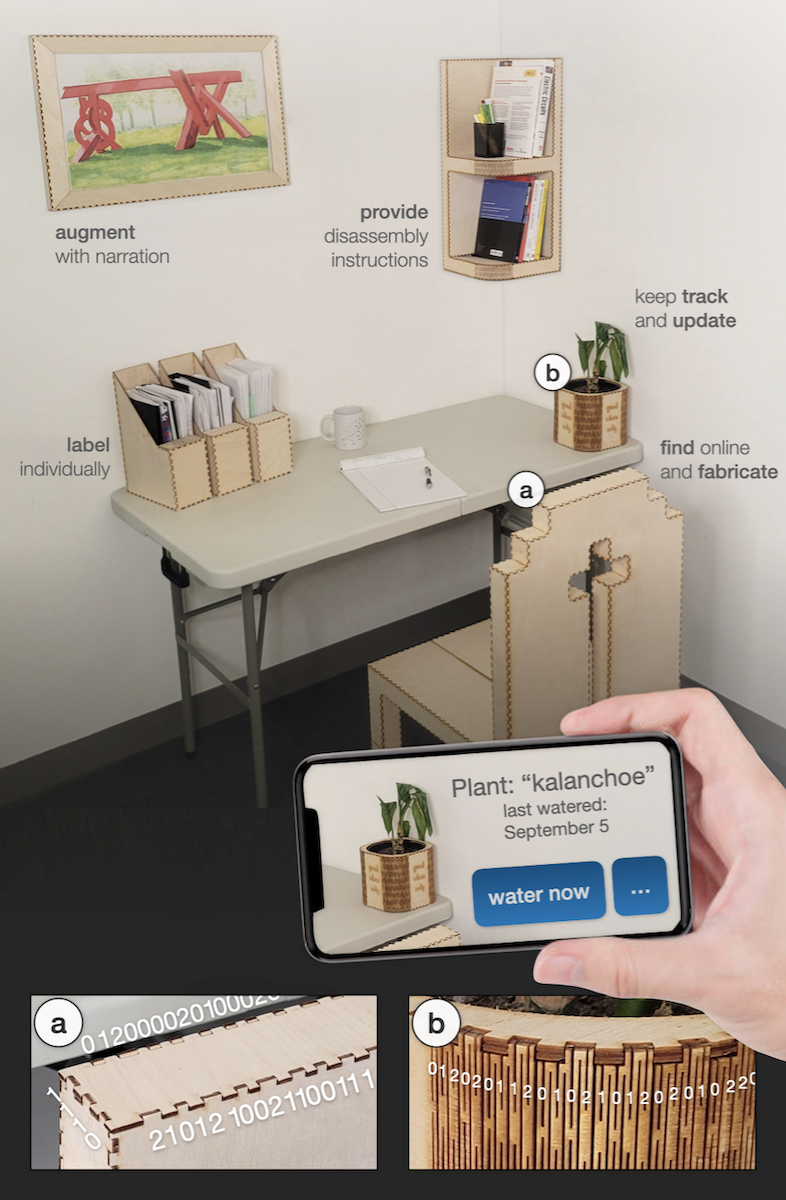}
  \caption{StructCode embeds data in the fabrication artifacts of laser-cut objects, such as the patterns of (a) finger joints and (b) living hinges, to augment objects with data.
  Here, the embedded StructCodes allow narration for a painting and status updates for a potted plant, among others.}
  \vspace{-0.2cm}
  \label{fig:teaser}
\end{figure}


To achieve unobtrusive passive tags that do not require object modification, 
researchers investigated ways to use the inherent characteristics of objects, which may manifest in the form of either naturally occurring or engineered features.
As an example of natural unobtrusive tags, \textit{SensiCut}~\cite{dogan_sensicut_2021} senses the micron-scale surface structure of materials using speckle sensing to identify the material's type. 
Similarly, \textit{Verifiable Smart Packaging}~\cite{wang_verifiable_2016} uses radio-frequency signals that penetrate through an object's exterior to identify the object by its internal structure.
The second approach, engineered unobtrusive tags, offers more flexibility to define how the tag should by embedded with the object.
For example, in agriculture, researchers developed tamper-evident tags using biodegradable silk particles, allowing for seed traceability and anti-counterfeiting~\cite{sun_integrating_2023}.

In this project, the question that prompted our research was:
Can we make use of existing fabrication artifacts of objects to store metadata in them?
Our goal is two-fold: (1) not introducing additional materials, parts, or features to the object, but simply \textit{make use of its existing visual structures that arise as a byproduct of the fabrication process}. (2) The added data should be \textit{decoded from regular camera images} so the meta-information can be easily used to augment the objects, e.g., in augmented reality (AR) applications.


Our proposed solution, \textit{StructCode}, slightly varies fabrication artifacts, i.e., the patterns resulting from joints, which does not change the main object geometry, but can still be detected from camera images. Thus, \textit{StructCode} does not add any new features to the object, but instead exploits the structures that are inevitable artifacts of the fabrication process.
While this idea can be applied using many different manufacturing processes and materials that allow for the customization of individual tangibles (to embed individual tags), for the scope of this project we focus on \textit{laser cutting}. Laser cutting is particularly suitable because it allows the rapid fabrication of sturdy, functional, and large-scale objects~\cite{baudisch_kyub_2019, dogan_sensicut_2021}. 
These typically contain many visual artifacts across their surfaces due to their finger joints and living hinges. 
\textit{StructCode} subtly changes the widths of these elements to represent machine-readable information.
The \textit{StructCodes} can be read as part of existing AR pipelines to 
augment the user's view with information relevant to the tangible objects that they are interacting with.


Figure~\ref{fig:teaser} shows how laser-cut objects use their existing joints to store meta-information, such as labels (e.g., used to retrieve watering updates for the plant), 
context (e.g., to access the digital model of the chair),
disassembly instructions (e.g., to recycle the bookshelf), and
narration (e.g., explanatory video for the artwork on the wall).
\textit{StructCode} achieves this by modifying the lengths of laser-cut finger joints and living hinges, which are inherent artifacts of commonly laser-cut objects. The modified lengths represent different bits while ensuring the codes are easy to capture with a mobile camera.
Because \textit{StructCodes} are an inherent part of the objects, they cannot be easily removed 
without causing damage to the object. 
Compared to other unobtrusive tagging methods in HCI, which use additional equipment such as infrared cameras~\cite{dogan_infraredtags_2022, willis_infrastructs_2013}, \textit{StructCode} only requires a conventional RGB camera for detection, and thus can be used on off-the-shelf mobile devices and headsets. 

While this project demonstrates the idea of opportunistic data embedding for the example of laser-cut objects, we discuss how it can be generalized to other types of fabricated objects in the future in Section~\ref{Discussion}.
We note that this chapter was originally published at \textit{ACM SCF 2023}~\cite{dogan_structcode_2023}. 
Our contributions can be listed as follows:

\begin{itemize}[leftmargin=0.5cm, noitemsep, topsep=2pt]
    \item An unobtrusive and integrated tagging method that embeds \textit{data} in the existing \textit{fabrication artifacts} of objects, as demonstrated for laser cutting, and only requires \textit{a standard camera} for detection.
    \item  A tag decoding pipeline that is robust to various backgrounds, viewing angles, and wood types.
    \item  A set of applications showing how this opportunistic embedding can enrich tangible interactions with laser-cut objects.
    \item A mechanical evaluation illustrating that the addition of StructCodes maintains the overall integrity of objects.
\end{itemize}

\section{Method: Embedding Data in StructCodes}
\label{Embedding-Information}

In this section, we discuss how to physically embed the code into existing laser-cut structures and which encoding scheme we use for StructCodes. Our goal is to be able to store custom text (e.g., sentences or website URLs).

\subsection{Identifying Features Suitable for StructCodes}
\label{Survey}

To better understand how to embed information in laser-cut objects' artifacts, we surveyed what joints are used in their structure and which of these are suitable to be leveraged as a StructCode.
We considered the top 100 laser cutting projects on the popular online repository \textit{Instructables}\footnote{\url{https://www.instructables.com/}} and classified the identified joints in objects. If an object contained more than one joint type, we took into account all of them.

In total, 58\% of projects had at least one type of joint. We found four main types of joints contained in the laser-cut objects: finger joints (26\%), mortise and tenon joints (16\%), slot joints (8\%), and living hinges (8\%).
We count living hinges as a type of joint because under \textit{joints}, we consider any connection between two planes in the laser-cut model. 
These joint types can also be commonly found in the research community, e.g., finger joints in \textit{Enclosed}~\cite{weichel_enclosed_2013}, \textit{CutCAD}~\cite{heller_cutcad_2018}, and \textit{Fresh Press Modeler}~\cite{chen_fresh_2016}, slot joints in \textit{FlatFitFab}~\cite{mccrae_flatfitfab_2014}, \textit{SketchChair}~\cite{saul_sketchchair_2010}, and \textit{Planar Pieces}~\cite{schwartzburg_fabrication-aware_2013}.

Out of these four joint types, we identify \textbf{finger joints} and \textbf{living hinges} as most suitable for illustrating the concept of StructCodes. 
Both contain repeating elements, i.e., finger joints have individual fingers repeated along the edges of a plate and living hinges are made up of repeated individual lines. 
They are both on the outside surface of the object, i.e., visible to the camera. Living hinges form a well-defined shape that can be tightly enclosed by a bounding rectangle.
While finger joints can also be applied to non-rectangular parts, they are most common on rectangular shapes 
(73.1\% of the joint projects had joints on rectangular plates).
Both structures offer large encoding capacity; particularly, living hinges have many line cuts that could be used as bits.

\subsection{Structural Embedding of Data Bits}

\begin{figure}[t]
  \centering
  \includegraphics[width=0.6\textwidth]{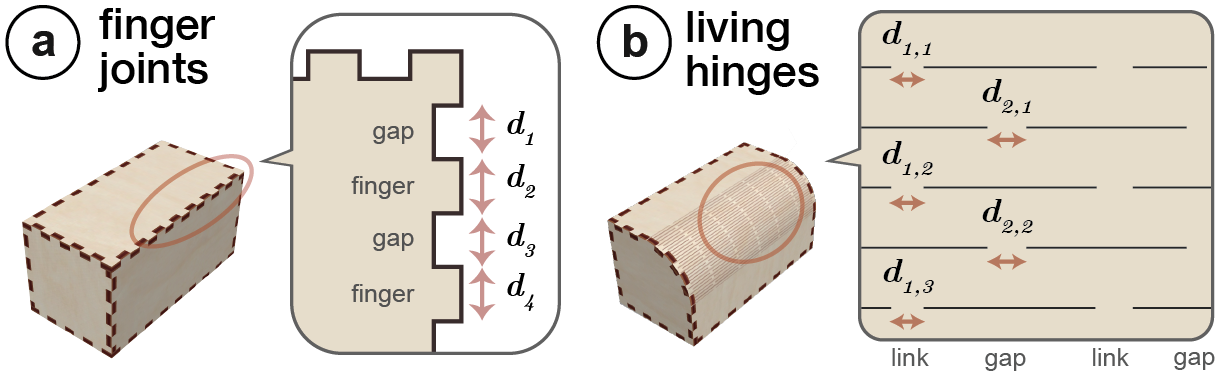}
  \caption{General design of (a) finger joints and (b) living hinges. To embed bits, we resize both the fingers and gaps in the finger joints, but only the links in the living hinges.}
  \label{fig:Structures}
\end{figure}

As shown in Figure~\ref{fig:Structures}, the identified structures typically contain material interleaved with cuts, i.e., gaps. For instance, in finger joints, the fingers are interleaved with cut outs that match the fingers of the neighboring plate, and in living hinges, the sheet is cut at specific distances to make the material bendable. 

To encode a message, we can either vary the length of the \textit{material} (\textit{fingers} of joints or \textit{links} of hinges) or the length of the \textit{gaps} between the fingers or between the links. Below we outline how we embed the codes for each of the two structural element types.

\paragraph{Finger joints}
To embed StructCodes into finger joints, we modify both the widths of the individual fingers and the widths of the gaps between the fingers, as labeled in Figure~\ref{fig:Structures}a (i.e., $d_1, d_2, d_3, ...$). We did not consider modifying the depth of the fingers because the depth has to be the same as the sheet thickness to ensure properly interlocking plates. To maintain the structural integrity of the joints, modifications to the widths of the elements (i.e., fingers or gaps) should be minimized to be as close to the original width as possible while being distinguishable by the camera and distributing evenly over the plate length (see Section~\ref{Detection-FingerJoints}). 


\paragraph{Living hinges}
There are two ways to embed StructCodes into living hinges, either by modifying the size of the individual links, the widths of the hinge gaps (i.e., the vertical cuts between the links), or both as shown in Figure~\ref{fig:Structures}b. To ensure that the hinge can be bent into the desired curved configuration after laser cutting, the links have to be vertically aligned. Therefore, we only modify the lengths of the \textit{links}. 
Furthermore, since the links are typically shorter than the gaps, they are less prone to warping when comparing the lengths against each other and thus more suitable for our purposes. This is because the links are situated in a smaller local region, which allows the projection to the image plane to maintain the ratio of distances~\cite{maia_layercode_2019}.
To ensure the structural integrity, 
we chose lengths that are as close to the original as possible while being distinguishable by the camera (see Section~\ref{Detection-LivingHinges}). 



\subsection{Encoding Scheme}
StructCode employs a base 3 (ternary) encoding scheme\footnote{In ternary logic, the digits are called \textit{trits} (\textbf{tri}nary digi\textbf{t}). However, for familiarity, we use the term \textit{bits} in this chapter although it is normally used for the binary system.}, i.e., it uses the bits
\texttt{0}, \texttt{1}, and \texttt{2}. This is represented in the physical code as a narrower element ("\texttt{0}"), a medium element ("\texttt{1}"), and a wider element ("\texttt{2}"). 
Four elements, i.e., four bits, represent one character. The different combinations of element widths that make up the four bits can generate up to 81 different variations, which allows to embed 81 different character types, including 62 alphanumeric characters and 18 special characters for strings that can include URLs. The remaining bit combination is used as the start/end sequence, which is asymmetric ("\texttt{0120}"). This allows us to identify where the code starts and thus helps to find the location in the captured image where we need to begin decoding. In addition, since the start/end sequence contains the \texttt{0}, \texttt{1}, and \texttt{2} bits, it makes it easier for us to detect the widths of the narrow, medium, and wide elements. We chose the base 3 encoding scheme rather than a base~2 scheme (only \texttt{0} and \texttt{1}, such as in conventional barcodes) as it allows us to increase the amount of information we can embed in the same number of elements.
We did not go up to base 4 or higher in order not to sacrifice subtlety (Section~\ref{discussionInconspicuousness}).
If a bit string is relatively short and the plate or hinge has a large area, our tool automatically repeats the string throughout the structure to achieve a more even look.


\paragraph{Finger joints} 
The bit sequence is encoded in a circular fashion beginning at one of the four corners of a rectangular plate so that it covers its whole perimeter.
When investigating the average number of joints used in rectangular plates on \textit{Instructables}, we found that the ones that have joints on all four sides had on average a total of 50.3 fingers and gaps (std=33.6). 
By varying both the finger and the gap, this allows us to embed 7 characters with a base 2 scheme, 
or 12 characters with a base 3 scheme, a capacity sufficient to store longer URLs using shortened URLs (similar to regular QR codes).
For instance, 
\href{http://t.ly/aNPf}{\textit{t.ly/aNPf}}
with 9 characters is the shorted version of a 39-character \textit{kyub} link to a 3D model.


\paragraph{Living hinges} 
A row of columns is encoded from left to right, and an individual column goes top to bottom. As an example, in Figure~\ref{fig:Structures}b, the first column is read from top to bottom ($d_{1,1}$, $d_{1,2}$, $d_{1,3}$), and then the second column is read ($d_{2,1}$, $d_{2,2}$, ...). We found that the average number of links in two-dimensional hinges from the  \textit{Instructables} dataset is 85.7 (std=43.1), excluding the hinges that are completely circular and thus would not fit in a single camera frame. This allows for a data capacity of  21 characters using our base 3 scheme. Thus, we conclude that compared to finger joints, living hinges usually fit longer StructCodes because they have more repeating elements.

\section{End-to-End Workflow}

We next describe how a user can embed StructCodes into their laser cut object. Our workflow is designed to work with laser-cut objects made in the 3D editor \textit{kyub}~\cite{baudisch_kyub_2019} 
or 2D models (.svg) from other sources imported into \textit{kyub} using \textit{assembler$^3$}~\cite{roumen_assembler3_2021}.
Users import the project from \textit{kyub} as a 2D file into our Web-based tool to embed StructCodes in it. After fabricating the object in the laser cutter, the user can detect the code using our mobile application.

\subsection{User Interface for Embedding StructCodes}
\label{User-Interface}


Once the user has imported their \textit{kyub} object into our tool, they click on the \textit{Identify compatible structures} button in the tool (Figure~\ref{fig:UI}), which marks the jointed plates and hinges that allow embedding codes. In our walkthrough, the user is making a box and the StructCode tool shows that the two rectangular plates of the box and the one with the living hinge can be used to embed a code. It also reminds the designer to choose a structure that is facing towards the user when they take a photo of it. Since the hinge pattern on the top part of the object is most visible, the designer chooses it for the StructCode.

\begin{figure}[h]
  \centering
  \includegraphics[width=1\linewidth]{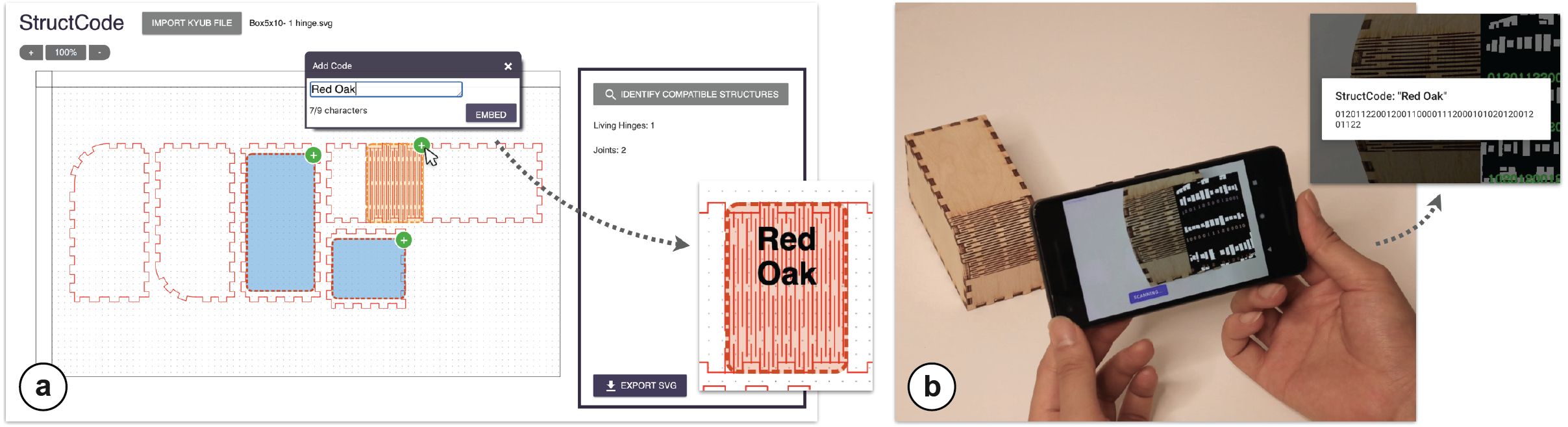}
  \caption{Workflow for encoding and decoding information. (a) After the tool highlights the compatible structures, the user selects the hinge and encodes \textit{"Red Oak"}. (b) They use the StructCode mobile application to decode the message.}
  \label{fig:UI}
\end{figure}

Once we select one of the structures, the tool shows how many characters can be embedded in it based on its number of joints or lines in the hinge pattern, e.g., that the selected hinge can embed up to 15 characters in it. In this example, we want to embed a StructCode that can later help us identify which material the object was laser cut from, i.e., red oak plywood. We type in "Red Oak" which is within our character limit (Figure~\ref{fig:UI}a) and as a result, the tool modifies the hinge cuts. After embedding this code, users have the option to also add more codes into the other structural elements of the object. In our case, we do not need to embed additional messages, so we leave all other structural elements as they are. 

Finally, the user exports the laser-cut design with the embedded StructCodes as an SVG file and send it to the laser cutter's software. After fabrication, the user assembles the parts using the instructions generated by \textit{kyub}. Our tool preserves the plate numbers from \textit{kyub} to facilitate the assembly process.

\subsection{Mobile Interface for Reading StructCodes}
The StructCode mobile application is used for decoding data embedded in objects. To do so, the user points their phone at the structure as shown in Figure~\ref{fig:UI}b. The mobile application automatically starts capturing images and processes them to decode the message. After enough images are taken to ensure correct decoding, the app displays the encoded message, i.e., in our case "Red Oak". The decoded information can be used to augment various mobile experiences, which are illustrated in Section~\ref{Applications}.

\section{Applications}
\label{Applications}

We demonstrate how StructCodes enriches object interactions with data, including identifiers for labels, object context, such as instructions, as well as overlaid media, such as narration.

\subsection{Embedding Identifiers for Static or Dynamic Labels}
\label{FileFolderApplication}
Users can add identifiers to objects which allow static or dynamic labels for the specific use cases of the objects.

\begin{figure}[h]
  \centering
  \includegraphics[width=0.95\linewidth]{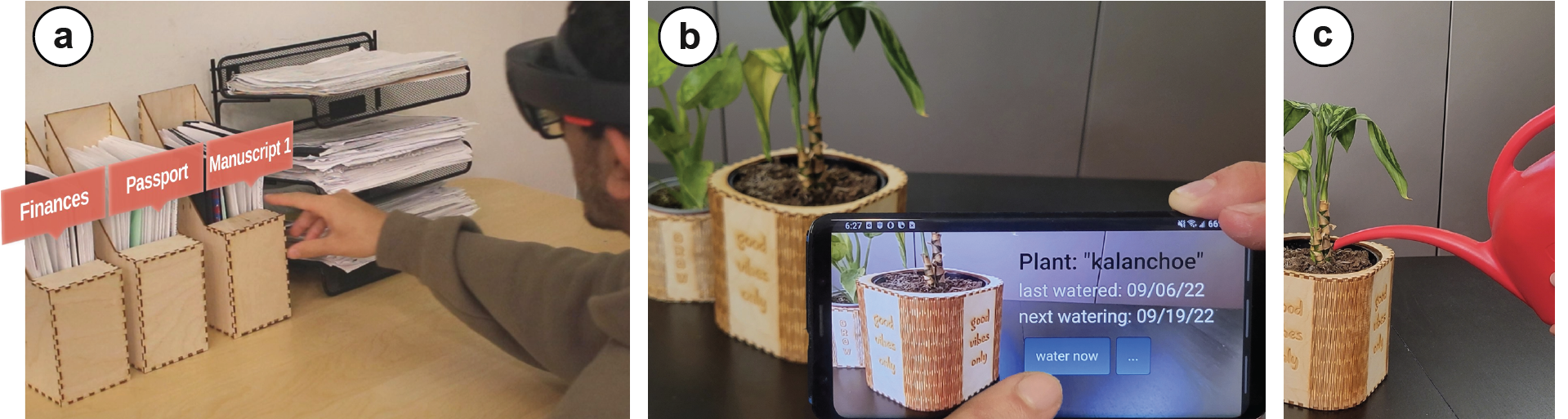}
  \caption{(a) Document folders with encoded personal labels, which the user views in AR to retrieve the right one. (b, c) Status updates inform the user about when the plant needs watering.}
  \label{fig:IDsLabels}
\end{figure}


\paragraph{Static labels}
As shown in Figure~\ref{fig:IDsLabels}a, a set of office folders contain various personal documents (e.g., \textit{personal finances} or \textit{visa documentation}) that the owner would rather not have visibly shown.
The user embeds a StructCode into the finger joints of the folder's front plates. Thus, the information is only visible to the user in AR and otherwise the folder looks unlabeled. For the most sensitive tags, the mapping of code to label can be encrypted and available to that user only.

\paragraph{Dynamic labels}
The embedded identifiers can be also used for dynamic labels, i.e., status updates.
In the example shown in Figure~\ref{fig:IDsLabels}b, the designer is creating a plant pot that lets the user pull up an online dashboard to keep track of the plant's health (i.e., watering, trimming the branches, fertilizing). The pot has an identifier embedded into its living hinges as StructCodes. When the designer leaves for vacation, they ask their friends to take care of the plants.
StructCode allows the friend to identify the plant and retrieve its recent status, e.g., that it needs to be watered today.

\subsection{Embedding Context: Resources and Instructions}
\label{EmbeddingContext}

StructCodes allow embedding references to the object's context, 
such as accessing the digital model, on-demand renewal, and (dis)-assembly instructions.

\begin{figure}[h]
  \centering
  \includegraphics[width=0.65\textwidth]{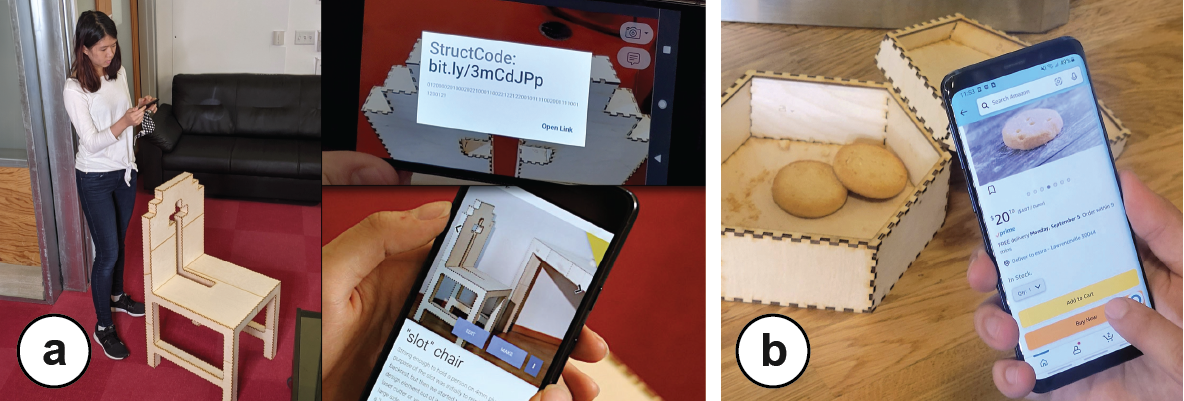}
  \caption{Providing context. (a) The user can access the fabrication files of the furniture.
  (b) Running low on cookies, the user adds a new batch to the online shopping cart through the embedded reorder link.}
  \label{fig:MetadataApplication}
\end{figure}

\paragraph{Accessing the digital model}
As shown in Figure~\ref{fig:MetadataApplication}a, the laser-cut chair contains a shortened link to its \textit{kyub} page. 
Users who see the chair and like the design can scan the StructCode to download its files online and make a copy for themselves.


\paragraph{On-demand renewal}
As shown in Figure~\ref{fig:MetadataApplication}b, the user reorders cookies by simply capturing its container, which has the reorder link encoded as a StructCode. This allows users to access and renew supplies whenever needed.

\begin{figure}[h]
  \vspace{5pt}
  \centering
  \includegraphics[width=0.7\textwidth]{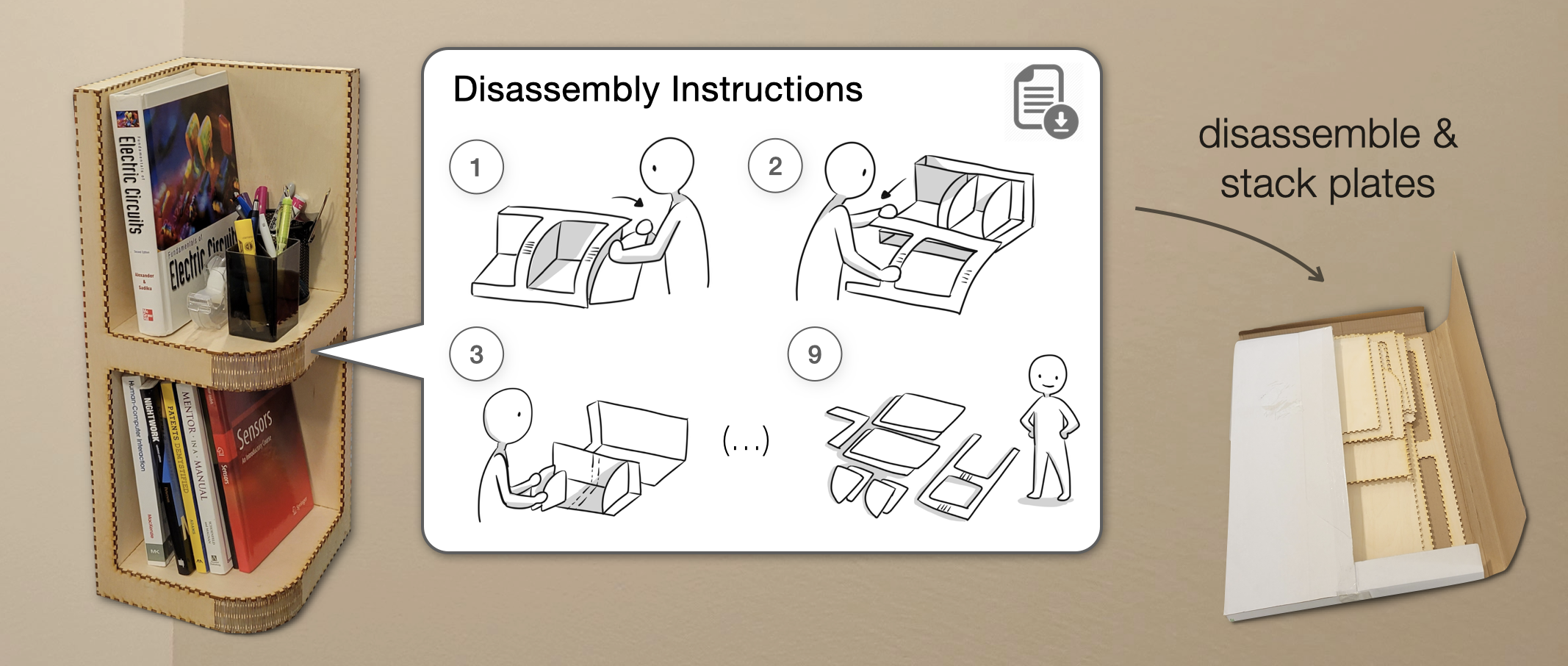}
  \caption{Disassembly instructions are linked to the shelf in case the user needs to take it apart for recycling or transport.}
  \label{fig:DisasemmblyFurniture}
\end{figure}

\paragraph{Assembly / disassembly instructions}
Furniture manufacturers such as \textit{IKEA} provide not only the assembly instructions for their products, but also the \textit{disassembly} instructions to improve recycling for better sustainability~\cite{allen_ikea_2021}.
However, users frequently misplace instruction manuals over time.
Similarly, for fabricated objects, users may struggle to locate the related instructions that originally came with the digital file.
As shown in Figure~\ref{fig:DisasemmblyFurniture}, the owner retrieves the disassembly manual through the StructCode when they need to discard of the piece, or transport and reconstruct it elsewhere.

While we showed the use case of StructCodes for linking \textit{disassembly} instructions, our method can also be integrated into existing systems for helping users with the \textit{assembly} of laser-cut objects~\cite{abdullah_roadkill_2021, park_foolproofjoint_2022}.
For instance, the part numbers could be embedded into individual plates so that our app guides the user to pick the right one in each step.

\subsection{Embedding Overlaid Media}
\label{EmbeddingOverlaidMedia}
StructCodes enable users to overlay objects with related media such as narrative videos or illustrations using AR.

\begin{figure}[h]
  \vspace{5pt}
  \centering
  \includegraphics[width=0.73\linewidth]{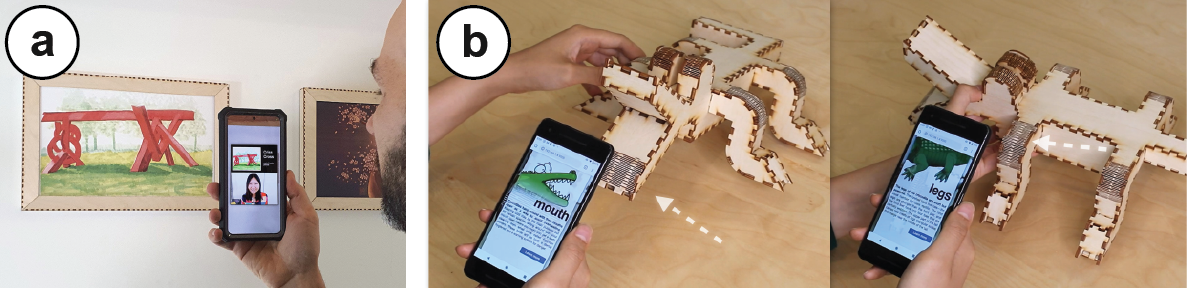}
  \caption{Overlaying media. (a) In an exhibit, visitors  use an AR app to view narrative videos by the artists. 
  (b) The crocodile has StructCodes that describe what part they are located on, which is used for educational applications.}
  \label{fig:EmbeddingMedia}
\end{figure}

\paragraph{Narrative media}
As shown in Figure ~\ref{fig:EmbeddingMedia}a, StructCode links a video of the original artist narrating their artwork. Using StructCode, students quickly set up a temporary gallery by laser cutting wooden frames for their own works.
Visitors can use their phones to view the art in AR, which overlays the video on the related artwork. 
The app can tailor the content and duration to the individual viewer's level of interest.

\paragraph{Illustrative sublabels}
Individual parts of an object can have different functions. StructCodes are used to associate relevant information with each part using illustrative sublabels. One example of this is the educational toy model shown in Figure~\ref{fig:EmbeddingMedia}b that has different sublabels embedded into different limbs of the crocodile. Thus, users access educational materials for each limb. This allows teachers to interactively introduce new concepts in classrooms.
For such use cases, we expect that users are made aware of the existence of StructCodes through 
the toy description or manual.

\section{Detection of StructCodes}
\label{ImageProcessing}

The image processing pipeline of StructCodes is implemented 
using 
\textit{OpenCV}~\cite{bradski_opencv_2000}.
As shown in Figure~\ref{fig:ImageProcessing}, StructCodes are detected by locating the structure of interest, detecting the modifications in it to extract the bits, and finally decoding the message.

\begin{figure}[t]
  \vspace{5pt}
  \centering
  \includegraphics[width=1\linewidth]{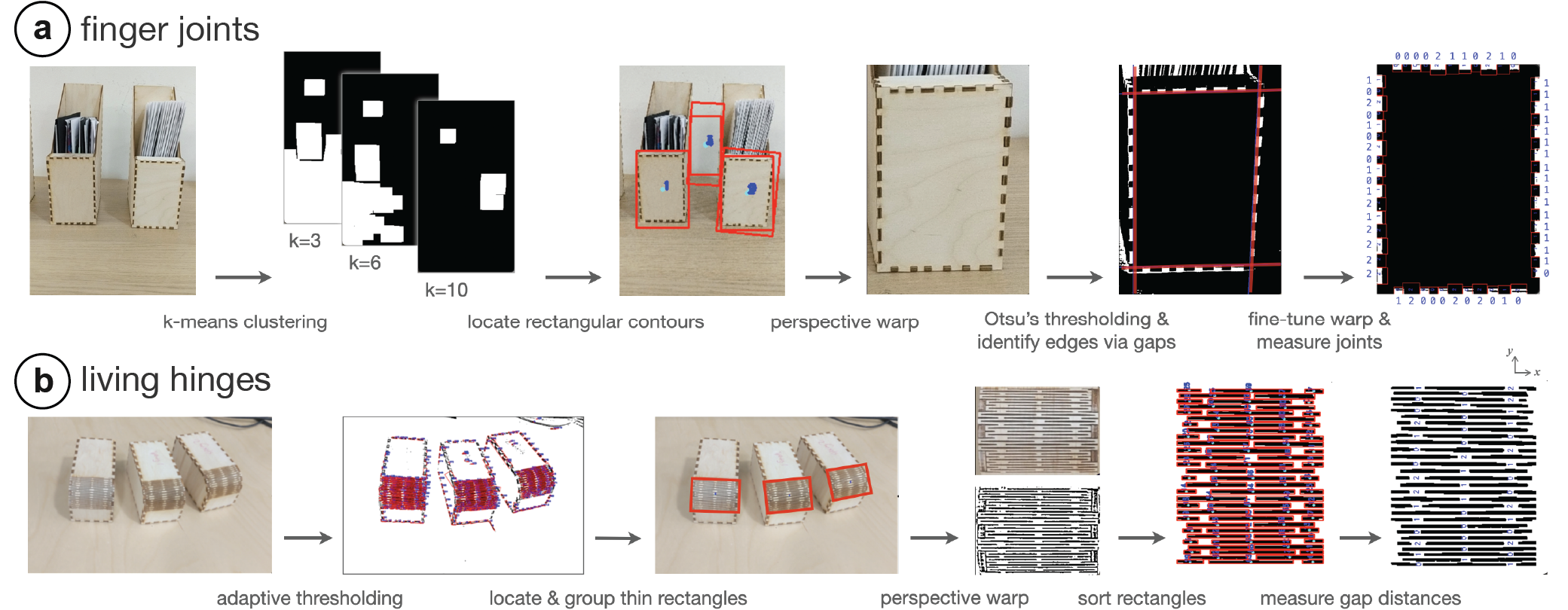}
  \caption{Image processing steps for (a) finger joints and (b) living hinges.}
  \label{fig:ImageProcessing}
\end{figure}

This section describes the decoding pipeline for finger joints and living hinges, which use slightly different image processing techniques.
If the user selects what type of feature to decode when launching the application, StructCodes can run the corresponding pipeline directly on the phone's processor. However, running both pipelines at the same time 
requires higher processing power and is thus recommended to run on a server, which in our applications takes less than a second.
We implemented both standalone phone detection for simpler applications and the server-based approach for other use cases for which joints and hinges are expected to be identified at the same time.

\subsection{Detection of Finger Joints}
\label{Detection-FingerJoints}
We explain the steps to read the code from the raw image of the finger joints as shown in Figure~\ref{fig:ImageProcessing}a.


\paragraph{Isolate the plate of interest}
StructCode's finger joint detection pipeline starts by grouping the pixels  using k-means based on each pixel’s HSV values. This creates multiple black and white masks 
such that a plate of interest is white in at least one of the masks. 
It runs this process for \textit{k=3, 6, 10}. It creates a different mask for each \textit{k}, \textit{k} (i.e., a total of 3+6+10 masks).
The algorithm then applies a morphological opening on each mask to disconnect the plates from the background.
It takes the larger rectangles found over all masks (with a padding of 25\% to account for any error margin) and applies a 4-point perspective transformation to isolate the plate.


\paragraph{Correct for perspective}
The next step in the pipeline applies Gaussian blur to reduce noise and use Otsu's thresholding to turn the gaps between fingers into contours. 
For each edge of a plate, it samples points on the gaps' contours, and draws a line of best fit through the points closest to the center,
rather than using the details closer to the sides. 
This allows the the algorithm to segment the interior plate even if the background has a similar color. 
It uses these four fitted lines to find the bounding quadrilateral for the plate interior.
It then applies another perspective warp that maps the four corners of the quadrilateral to an axis-aligned rectangle. 
And finally, it slightly extends the bounding rectangle 
to include the fingers but exclude the background.


\paragraph{Decode the message}
The gaps of interest are the white rectangles in the final image, and the width of the fingers can be calculated by the distance between consecutive gaps along each edge. The pipeline then runs k-means with \textit{k=3} to classify the gap and finger lengths as \texttt{0}, \texttt{1}, or \texttt{2} bits. It converts the entire ternary string into characters by grouping bits into blocks of 4. 
To identify the correct reading direction irrespective of the plate orientation, the start/end sequence, which has an asymmetric order (i.e., "\texttt{0120}", see Section~\ref{Embedding-Information}) is detected from one of the corners and used as an anchor.

\subsection{Detection of Living Hinges}
\label{Detection-LivingHinges}
We next explain the pipeline to detect living hinges, as demonstrated in Figure~\ref{fig:ImageProcessing}b.

\paragraph{Identify individual hinge gaps} 
The first step of the pipeline uses adaptive thresholding to turn the image into black and white. A morphological opening and closing are applied to eliminate small isolated patches of black pixel noise.
Because the living hinge consists of many thin and long gaps (i.e., the cuts between links) that are parallel and close together, the algorithm first searches for thin rectangles. It measures each contour's similarity to a rectangle by comparing its enclosed area to the minimum-area rectangle that contains it. 

\paragraph{Group gaps to isolate the hinge} 
The next step groups rectangles based on proximity, as living hinges consist of many parallel nearby gaps with similar dimensions and orientation. For this, it creates a graph with edges between similar rectangles and find the bounding boxes around the largest connected components. 
For each bounding box, it applies a 4-point perspective transform and crop out the rest so that the hinge is axis-aligned and takes up the whole image.

\paragraph{Sort the gaps}
This step removes noise due to the side joints 
via morphological opening, resizes the image to thicken the rectangular gaps,
and iterates through the gaps to assign them to rows and columns by their coordinates. 

\paragraph{Decode the message}
Since every row of \textit{gaps} is determined, the final step calculates the position and lengths of the \textit{links} (i.e., the bits) by measuring the distances between the gaps within each row. It scans the links from top to bottom, where it groups each link together with the previous link that has approximately equal x-coordinate. StructCode runs k-means with \textit{k=3} to classify each link length as a \texttt{}{0}, \texttt{1}, or \texttt{2} bit. The entire ternary string is then converted into characters by grouping bits into blocks of 4. The correct reading direction is determined by identifying the start/end sequence “\texttt{0120}”.

\subsection{Evaluation of the Detection Pipelines}
\label{Detection-Evaluation}
We evaluated what the smallest detectable length difference $\Delta d$ is between individual bit categories (\texttt{0}, \texttt{1}, or \texttt{2}) when processing the camera images with our detection pipeline. If \texttt{0} is represented by distance $d$, then \texttt{1} is represented by $d+\Delta d$, and \texttt{2} is represented by $d+2\Delta d$. Our goal is to only use the smallest possible length difference $\Delta d$ to maintain the mechanical integrity of the object. 

The smallest possible length difference $\Delta d$ is related to the length of the captured joint plate $w_{plate}$ and the camera distance since larger plates require the the camera to be held further to capture all features (joints or hinge links), which results in them appearing smaller in the image.
Further, in certain applications, 
the user may want to identify more than one object at the same time and hold the phone further away, thus differences may become even subtler.
We formalize this relationship as $\Delta d_{min} = \frac{w_{plate}}{\alpha} $, where $\alpha$ is the camera distance scaling factor and $w_{plate}$ is the longest plate dimension or the longest dimension in the bounding box around living hinge regions.

To obtain a conservative bound for $\alpha$, we conduct the following test in our workshop (80-150 lux), which is in line with regular indoor lighting conditions~\cite{national_optical_astronomy_observatory_recommended_2016}.  We first cut multiple joint plates of a fixed size (15cm x 10cm, from the folder application in Section~\ref{FileFolderApplication}) with 6 different $\alpha$ values with increments of 5 (from $\alpha = 75$ to $\alpha = 100$),  which corresponds to a range of 1.5-2mm for $\Delta d$.  Next, we captured the plate with a phone (12.2MP on \textit{Pixel 2}) and downscaled the images to 2048x1536 for fast processing. 
To ensure that at least two objects can be identified from a single image as shown in the use cases, we held the camera far enough (45cm) so that at least three plates can fit into the frame. 

We then ran our image processing pipeline on the resulting images and found that it was able to distinguish between bits with $\alpha <=  80$. Thus, with the given $\alpha$, a 15x10 cm plate requires a difference $\Delta d$ = 1.88 mm, allowing up to 26 fingers, which can store up to 14 characters. 
By contrast, a larger plate of 20x20 cm (60 fingers) can store up to 31 characters but requires a larger difference $\Delta d$ = 2.5 mm since the camera has to be held further away.

We repeated this experiment for living hinges. We first cut multiple copies of a hinge (5cm x 3.7cm, the model from Section~\ref{User-Interface}), with 6 different $\alpha$ values with increments of 5 (from $\alpha = 25$ to $\alpha = 50$),  which corresponds to a range of 1-2mm for $\Delta d$.
We held the camera at a distance far enough (20.7cm) so that at least three of these hinges can be detected from the captured shot.

The processing pipeline managed to distinguish between bits with $\alpha <= 45 $. Thus, with the given $\alpha$, a hinge of a size of 5 cm x 3.7 cm requires a difference $\Delta d$ = 1.11 mm, resulting in 72 hinge cuts in the area, which can store up to 9 characters. In contrast, a larger hinge of 10 cm x 7.4 cm (193 cuts) can store up to 36 characters but requires a larger difference $\Delta d$ = 2.22 mm since the camera has to be held further away.

\paragraph{Evaluation of the viewing angle}
\label{EvaluationViewingAngles}
Using the above results, we evaluated the maximum camera capture angle at which the message can still be decoded relative to the plate normal.
To do this, we fixed the model with finger joints on a surface and rotated the camera around the plate of interest until the code is no longer detectable, 
while keeping track of the angle using a protractor attached onto the surface. We did this for three different codes (Figure~\ref{fig:IDsLabels}) and three different backgrounds (black, white, wood). The maximum viewing angle was 25.28$^{\circ}$
(std=2.81) across the nine resulting conditions. When repeated with the living hinge samples, we found that the maximum angle was 37.19$^{\circ}$ (std=4.52) around the axis perpendicular to the hinge gaps (y-axis in Figure~\ref{fig:ImageProcessing}b), and  19.08$^{\circ}$ (std=2.95) around the axis along the gaps (x-axis) across the nine conditions.
The reason the second value is smaller is that due to the curvature, the outermost gaps become more easily occluded with deviations away from the center.

\begin{figure}[h]
  \centering
  \vspace{0.2cm}
  \includegraphics[width=0.47\textwidth]{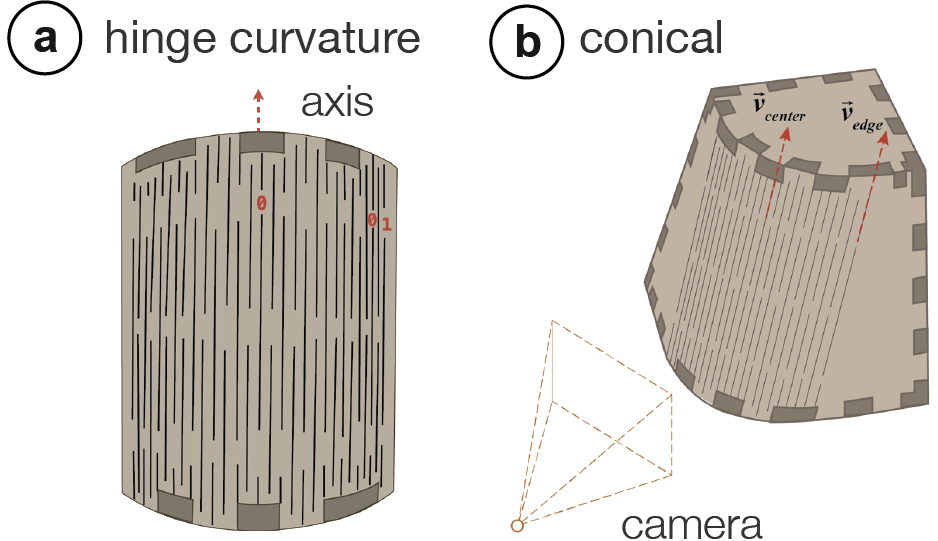}
  \caption{Curvature of living hinges. (a) As bit lengths may be distorted due to curvature, StructCode encodes them with sufficient tolerance to ensure correct detection. (b) StructCodes can be encoded on various shapes, e.g., conical hinges. 
  }
  \label{fig:Curvature}
\end{figure}

\paragraph{Evaluation of hinge curvature} 
For reliable detection, we need to ensure that different bits can be correctly distinguished even though their lengths may be distorted as a result of hinge curvature. For instance, the more curved an (outward) hinge is, the more likely it is that a \texttt{0} bit at the center of the hinge appears longer than a \texttt{0} bit at the edge of the hinge. Figure~\ref{fig:Curvature}a shows an exaggerated case of this where the camera is very close the the hinge (4cm).
As length is the differentiating factor between bits, this distortion creates a risk of incorrect detection (i.e., mistaking a \texttt{0} bit for a \texttt{1} bit as a result of length distortion).
Based on the pinhole camera model~\cite{sturm_pinhole_2014}, we formalize the condition to avoid this risk using the expression
$\Delta d > d\left [ \frac{dist_{edge}}{dist_{center}} -1 \right ]$, where $dist_{center}$ is the camera distance from the bit at the center and $dist_{edge}$ is the camera distance from the bit at the edge.
The difference between these distance values is proportional to the curvature. However, since the camera is sufficiently far from the object, 
we have $dist_{center} \approx dist_{edge}$. Thus, in practice the right-hand term is smaller than the $\Delta d $ values used.

For instance, for the typical viewing distance of the model evaluated in Section~\ref{Detection-Evaluation}, we need $\Delta d > 2.1mm \left [ \frac{21.3cm}{20.7cm} -1 \right ]$ $ \approx 0.06mm$, which is satisfied since the $\Delta d$ used is 1.11 mm. This gives enough legroom to ensure correct detection in various conditions, such as different angles as mentioned in the earlier section. Another example is conical hinges (Figure~\ref{fig:Curvature}b), here the lines on which the hinge cuts lie have different angles. The legroom allows the detection of the correct measurements, however, the maximum viewing angles reported in Section~\ref{EvaluationViewingAngles} are reduced by the angle difference between $\vec{v}_{center}$ and $\vec{v}_{edge}$ on which the central and the edge cuts lie.

\section{Mechanical Evaluation}
\label{MechanicalEvaluation}

When embedding StructCodes, the individual elements of mechanical structures are slightly modified, i.e., the width of the finger joints and the distance between the living hinge cuts are adjusted by multiples of $\Delta d$. To evaluate how much this change affects the mechanical integrity, we conducted several tests with modified finger joints and living hinges.

\subsection{Compression Evaluation of Finger Joints}
\label{Sturdiness_FingerJoints}

To compare how the addition of StructCodes affects sturdiness, we compared the ultimate compressive strength of box structures before and after embedding the code (i.e., unmodified vs. modified).


\paragraph{Experiment setup}
We evaluated two different box sizes because each size results in a different resizing difference $\Delta d$ based on the camera scaling factor $\alpha =  80$ for finger joints as explained in Section~\ref{Detection-Evaluation}. The smaller box, which can carry up to 5 characters, had a plate size of 5cm x 5cm and required a $\Delta d$ of 0.625mm. The larger box, which can carry up to 10 characters, had a plate size of 12cm x 12cm and required a $\Delta d$ of 1.5mm. The dimensions for the large box were chosen since they represent the largest size that can fit into the measurement machine. We laser cut 4 unmodified and 4 modified boxes of each box size (total of 16 boxes) from 3mm birch plywood sheets.
The messages embedded in the modified boxes were produced using a random string generator.
Similar to the technical evaluation for \textit{kyub}~\cite{baudisch_kyub_2019}, we used a common low-cost material to obtain a conservative lower bound for the sturdiness of the tested objects. All boxes were held together solely press fitting their joints together, i.e., without glue.

\begin{figure}[h]
  \centering
  \includegraphics[width=0.7\textwidth]{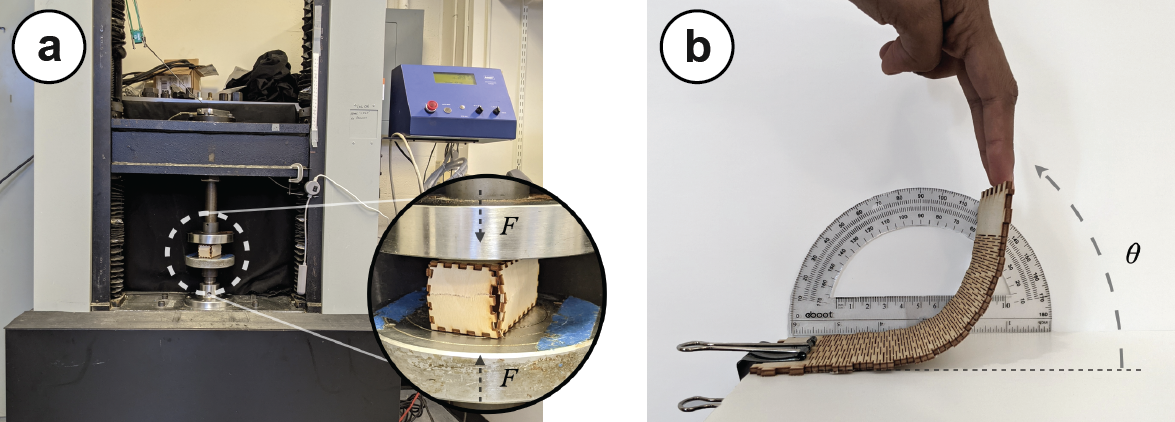}
  \caption{Mechanical evaluation of StructCodes: (a) Compressive strength of finger joints and (b) bending of hinges.}
  \label{fig:Evaluation}
\end{figure}


\paragraph{Experiment procedure}
We used an \textit{Instron} universal testing machine (UTM) as shown in Figure~\ref{fig:Evaluation}a. We placed the modified boxes such that the side with the StructCode faced upward. The Instron increased the compression force on the box and we measured the ultimate compression strength to the point where the box failed under crush loading. The ultimate compression strength represents the maximum stress the structure can sustain.


\paragraph{Results}
We found that the width modification of joints required to embed StructCodes did not strongly affect the sturdiness when compared to the values reported in previous literature. In their tests with \textit{kyub} objects,  Baudisch et al. ~\cite{baudisch_kyub_2019} have reported that (unmodified) objects were still intact when they exceeded the 500kg (4,903 N) value range of their measuring device. We were able to confirm this as shown in Table~\ref{table:JointPeakLoad}. Both unmodified and modified objects go well beyond the reported 4,903 N value. Even though the peak load the box can handle decreased on average by 14.5\% for the 5cm box and increased on average by 0.16\% for the 12cm box after modifying the joints, the force that each of these boxes can withstand is still larger than what is required of most objects used in daily life. The high standard deviation, specifically for the large boxes, is likely due to the fact that there is a large variation in the composition of plywood sheets even though they all came from the same batch.

\begin{table}[h]

\centering
\begin{tabular}{llll}
 &
  \textbf{\begin{tabular}[c]{@{}l@{}}\small Unmodified\\ \end{tabular}} &
  \textbf{\begin{tabular}[c]{@{}l@{}}{\small Modified with}\\ {\small StructCode}\end{tabular}} &
  \textbf{\begin{tabular}[c]{@{}l@{}}\small Relative change\\ \end{tabular}} \\ \hline
5 cm &
  \begin{tabular}[c]{@{}l@{}} 15,939.7 N\\ \small (std=621)\end{tabular} &
  \begin{tabular}[c]{@{}l@{}} 13,630.6 N\\ \small (std=641)\end{tabular} &
  Decrease by 14.5\% \\ \hline
12 cm &
  \begin{tabular}[c]{@{}l@{}}12,578.6 N\\ \small (std=2,020)\end{tabular} &
  \begin{tabular}[c]{@{}l@{}}12,599.3 N\\ \small (std=1,760)\end{tabular} &
  Increase by 0.16\% \\
\end{tabular}%
\caption{Average peak load comparison of the finger joints.}

\label{table:JointPeakLoad}
\end{table}

\subsection{Bending Evaluation of Living Hinges}

To evaluate if the introduction of StructCodes into living hinges changes their flexibility, we measured up to what angle the living hinge can bend before fracturing and compared the results for both modified and unmodified living hinges. 


\paragraph{Experiment setup}
Similar to the setup from  Section~\ref{Sturdiness_FingerJoints}, we created 4 pairs of hinges based on the camera scaling factor for living hinges $\alpha =  45$, which was explained in Section~\ref{Detection-Evaluation}.
We repeated this for hinges of two sizes: a smaller one that can carry 10 characters (4.7cm x 3.64cm, $\Delta d = 1.04$mm) and a larger one that can carry 50 characters (12.2cm x 9.6cm, $\Delta d = 2.71$mm). 
We laser cut  the 4 unmodified and 4 modified hinges of each size (16 hinges in total) from 3mm birch plywood sheets. The messages embedded in the modified ones were produced using a random string generator.
While different hinge designs may exhibit different flexibility~\cite{fenner_lattice_2012-1}, we use the default pattern generated by \textit{kyub} as this is the tool we used for fabricating our application samples.


\paragraph{Experiment procedure}
To evaluate the bending angle, we bent the living hinge pattern either up to 180$^{\circ}$ (maximum) or up to the point where it started cracking, and then read the corresponding maximum angle from a protractor as shown in Figure~\ref{fig:Evaluation}b.


\paragraph{Results}
For the larger hinge, we found that both the unmodified and modified hinge were able to bend to 180$^{\circ}$  (Table~\ref{table:LivingHingeMaximumRotation}). Thus, the change did not impact the performance. 
This is likely because the cut distance is not a main factor impacting the maximum hinge bend angle, which is rather directly linked to the the sheet thickness and the number of links in series~\cite{fenner_lattice_2012-1}.
For the smaller hinge, the unmodified hinge was also able to bend to the maximum of 180$^{\circ}$, whereas the modified hinge was still able to bend to 175$^{\circ}$ (i.e., a decrease only 2.8\% after the insertion of the code). 
We believe this should not affect the use of hinges as they are typically bent to 90$^{\circ}$ for most objects (see Section~\ref{Discussion}).

\begin{table}[h]
\centering
\begin{tabular}{llll}
 & \textbf{\begin{tabular}[c]{@{}l@{}}\small Unmodified\\ \end{tabular}} &
  \textbf{\begin{tabular}[c]{@{}l@{}}{\small Modified with}\\ \small StructCode\end{tabular}} &
  \textbf{\begin{tabular}[c]{@{}l@{}}\small Relative change\\ \end{tabular}} \\ \hline
\multicolumn{1}{l}{smaller}  & 180$^{\circ}$  & 175$^{\circ}$ \small (std=5)          & Decrease by 2.8\% \\ \hline
\multicolumn{1}{l}{larger} & 180$^{\circ}$      & 180$^{\circ}$       & No change       \\
\end{tabular}
\caption{Average bend angle of the living hinges.}
\label{table:LivingHingeMaximumRotation}
\end{table}

\vspace{0.1cm} 
The above evaluation demonstrated that the addition of StructCodes preserves the large compression strength of laser-cut objects held together with joints and curvatures achieved via hinges. However, more longitudinal tests might be needed to further examine the joints' usage in diverse applications where, for instance, the effect of shearing (i.e., when the direction of the force is parallel to the plane of the object) is more important than the effect of compression. Similarly, while the hinge applications presented were for static objects, further analysis could be conducted to determine the long-term impact of repeated bending on hinges with StructCodes.

\section{Software Implementation}

Our software tool for embedding StructCodes is Web-based and uses \textit{JavaScript} as well as the \textit{Paper.js} canvas library. 


\paragraph{Extracting laser-cut plates from SVG}
To extract the laser-cut structures from the user's \textit{kyub} design file (.svg), our software first parses through its layers that contain the individual path segments of the drawing. For each plate, our tool utilizes the annotations in the \textit{kyub} file to create a data structure instance that contains the ID of the plate and the IDs of interlocking plates on the plate's sides.


\paragraph{Identifying structures in each plate}
Once the plates are extracted from the SVG file, the software interface identifies the finger joints and living hinges in each plate using our algorithm and adds the line segments that represent them to the data structure.
To identify joints, it detects parallel line segments whose endpoints match in either the x- or y-axis and stores these segments in an array inside the data structure. To identify hinges, it detects groups of adjacent parallel lines with tiny distances between them. For hinges, once the last line is found, a bounding box is created to encompass all segments within that hinge.


\paragraph{Converting text into bit sequences}
The user-inputted characters are transformed into 4-bit, base 3 sequences to form a bit string according to a pre-defined dictionary.

\paragraph{Modifying structures to embed bits}
The fingers joints and living hinges are then manipulated to encode the computed bit sequence in the following manner:

\vspace{0.1cm} 
\textit{For finger joints}, we first compute $\Delta d_{min}$ based on the formula described in Section~\ref{Detection-Evaluation} ($\Delta d_{min} = \frac{w_{plate}}{\alpha } $) for the selected plate. 
While modifying the fingers and gaps of the joints, we readjust their widths to ensure that the joints cover the whole side of the plate based on the computed $\Delta d_{min}$ value as well as the number of \texttt{0}s, \texttt{1}s, and \texttt{2}s needed for the specific message.
We modify the widths by shifting each parallel line from the line before it by the required calculated difference, and this shifting continues along each of the sides in a counter-clockwise manner until complete.
When a side is done, the neighboring plate and its side are also shifted with the corresponding bit subarray to ensure interlocking. 

\vspace{0.1cm} 
\textit{For living hinges}, we similarly first compute $\Delta d_{min}$ based on the formula described in Section~\ref{Detection-Evaluation} ($\Delta d_{min} = \frac{w_{hinge}}{\alpha } $) for the hinge selected by the user, and use this value to find the lengths corresponding to the different bits.
These length values are used to modify the endpoints of the two adjacent line segments that represent a gap between hinge cuts.

\vspace{-0.2cm}

\section{Discussion}
\label{Discussion}
In the next section, we discuss insights gained from our work, acknowledge its limitations, and propose future research.

\paragraph{Aesthetics vs. code capacity}
\label{discussionInconspicuousness}
Even though varying structural elements keeps the surface and the main geometry of the object intact, the varying patterns may come across as unfamiliar or less smooth to users that are used to the standard joint or hinge patterns. Thus, similar to \textit{FoolProofJoint}~\cite{park_foolproofjoint_2022}, which varies the joints to facilitate assembly, the use of StructCodes comes at an aesthetic cost based on the user's familiarity and experience with these patterns.
StructCode aims to minimize this by choosing $\Delta d$ values as small as possible while also ensuring machine-readability.
Future work can explore how to optimize the look of joints for being completely unnoticeable to humans by aiming to go under
the \textit{Just Noticeable Difference}, i.e., the minimum level a stimulus that needs to be changed for humans to perceive it~\cite{hecht_visual_1924}. Psychologists show this difference is proportional to 
the original length~\cite{boring_sensation_1942}. Therefore, distinguishing between two rectangles (e.g., joints) with different lengths is more difficult when their \textit{average length} increases while the difference between them is kept constant~\cite{wadhwa_series_2020}.

\vspace{0.1cm} 
The current version of StructCode uses a base 3 encoding scheme, which allows us to embed a variety of characters sufficient to represent, e.g., a URL. Other numeral systems (e.g., base 5) can embed more characters with fewer bits, but they require more variation in the joint lengths. Thus, the difference between individual joints may be more visible and have a stronger impact on the object integrity and look.
A future version of StructCode can offer multiple encoding schemes and allow users to decide on the best trade-off between data capacity, subtlety, and mechanical performance.

\paragraph{Error correction}
Error correction codes (ECCs) such as Reed-Solomon~\cite{reed_polynomial_1960} or Hamming code~\cite{hamming_error_1950} could be added to StructCodes to further increase detection robustness.
By adding redundancy, 
these may help detect and correct errors that may occur in  particularly noisy or blurry images. However, this comes at the expense of reduced code capacity.

\paragraph{The effect of post-hoc polishing}
After laser cutting an object, some users prefer smoothing the finger joints depending on intended use and time availability. However, sanding finger joints may impact the size of the detected joints. We evaluated how much sanding our image processing pipeline can endure for different wood materials, i.e., birch plywood, walnut, and medium-density fiberboard (MDF).
We cut the same sample with these materials and sanded them with sandpaper of grit sizes 1000, 600, 220, and 110 (fine to coarse), 30 passes for each. We could decode each sample before sanding. 
For birch plywood and walnut, we found that sanding with grit size 1000 and 600 preserved decodability. For MDF, only grit size 1000 preserved decodability.  
This is likely because plywood sheets consist of multiple stacks of veneers, one of which naturally has a darker color that remains intact even after sanding.
In the future, our detection algorithm could be further optimized to increase recognition under tougher conditions by enhancing contrast using methods such as CLAHE~\cite{yadav_contrast_2014}.
Future research may also consider using specialized cameras~\cite{dogan_infraredtags_2022} for cases where fabrication artifacts like joints are intentionally occluded by the designer.

\paragraph{Extending StructCode to other types of shapes and objects} 
Our initial survey 
showed that finger joints and living hinges are most suitable for StructCodes.
We therefore optimized our image processing pipeline to detect these structures.
However, the StructCode concept is not limited to these.
We plan to embed codes into more freeform shapes in the future, such as joints placed along circular plates.
Future work can investigate ways to embed StructCodes into joints used in other fabrication processes, e.g.,
stitching in \textit{Joinery}~\cite{zheng_joinery_2017},
3D printing textiles in \textit{DefeXtiles}~\cite{forman_defextiles_2020} or joints in \textit{Hybrid Carpentry}~\cite{magrisso_digital_2018}, 
and machining traditional or modern woodworking joints in \textit{Tsugite}~\cite{larsson_tsugite_2020}, \textit{JigFab}~\cite{leen_jigfab_2019} and \textit{MatchSticks}~\cite{tian_matchsticks_2018}.

\section{Conclusion}
In this chapter, we presented StructCode, a technique to embed data into laser-cut fabrication artifacts. 
By modifying the patterns of laser-cut joints while maintaining their functionality, StructCode enables the embedding of data that can be decoded using a mobile phone camera.
We explained the encoding scheme used to embed data into finger joints and living hinges.
We then presented a software tool for embedding codes into existing 3D models, and a mobile application to decode them.
Our applications illustrated how fabrication artifacts can be leveraged to augment laser-cut objects with data such as labels, instructions, and narration. We explained our image processing pipeline which extracts the data from camera images. Finally, we evaluated the mechanical integrity of the fabricated objects to ensure that they are stable after StructCodes are embedded, and discussed how StructCodes can be further developed to make them fully inconspicuous to humans and more robust using  error correction codes.

This work aims to bring us one step closer to the vision of embedding data for augmented objects as an inherent part of the fabrication process~\cite{dogan_fabricate_2022}.

\chapter{G-ID: Identifying 3D Prints Using Slicing Parameters}
\label{thesis-G-ID}

\begin{figure}[t]
  \centering
  \includegraphics[width=0.6\columnwidth]{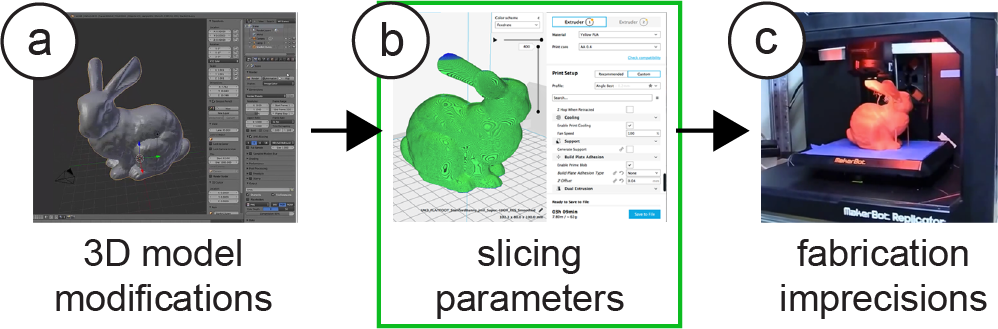}
  \caption{Different ways to embed tags into 3D models while leaving the surface intact:
(a) changing the internal geometry (\textit{InfraStructs}),
(b) varying slicing parameters (\textit{G-ID}),
(c) relying on fabrication imprecisions (\textit{PrinTracker}).}
  \label{fig:gid-2-RelatedWork}
\end{figure}


\section{Introduction}

As identified in the earlier chapters, a key challenge when using machine-readable tags is how to make them unobtrusive. For this purpose, researchers have investigated methods to leave the surface intact and instead change the inside geometry of a model.
\textit{InfraStructs}~\cite{willis_infrastructs_2013}, for instance, scans the object’s interior with a terahertz scanner, while \textit{AirCode}~\cite{li_aircode_2017} requires a projector and camera setup to detect internal air pockets using subsurface scattering.
Even though these approaches leave the object’s surface intact, both need large equipment, which prevents these solutions from being used in everyday scenarios.

To be able to use regular scanning equipment, such as a mobile phone camera, researchers proposed to analyze small imprecisions on the object’s surface that are created during the fabrication process. Those imprecisions are unobtrusive yet machine readable and can therefore be used as tags. Such imperfections make it possible to identify which fused deposition modeling (FDM) printer was used to create an object~\cite{li_printracker_2018}.


In this work, we propose a different approach to identification. When a 3D model is prepared for 3D printing, it first undergoes \textit{slicing}, a computational process which converts a 3D model into layers and their corresponding print path (a \textbf{G}-code file), which the extruder then follows to create the 3D object. The parameters for the slicing process can be modified for each individual instance, which allows \textbf{G}-ID to create unique textures on the surface of objects that can be detected with a commodity camera, such as those available on a mobile phone. Since our approach allows us to be in control over which printed instance has been modified with which slicer settings, we can identify each instance and retrieve associated labels previously assigned by a user.
We note that this chapter was originally published at \textit{ACM CHI 2020}~\cite{dogan_g-id_2020}.

\vspace{0.2cm}
The contributions of G-ID can be summarized as follows:

\begin{itemize}[leftmargin=0.5cm, noitemsep, topsep=2pt]
    \item  A method to utilize the subtle patterns left as an inevitable byproduct of the conventional 3D printing process to identify objects without the need to embed an additional tag.
    \item  A tool for users who want to create multiple instances of an object but intend to give each one a unique identifier.
    \item A mobile app that helps users take pictures of objects to be identified, as well as a stationary setup to detect finer variations in slicing parameters using image processing.
    \item  An evaluation of the space of slicing parameters that can be varied to generate unique instances of a 3D model and the corresponding detection accuracy under different environmental and hardware-related conditions.
\end{itemize}
 
We demonstrate these contributions with a diverse set of interactive applications.

\begin{figure}[t]
  \centering
  \includegraphics[width=1\columnwidth]{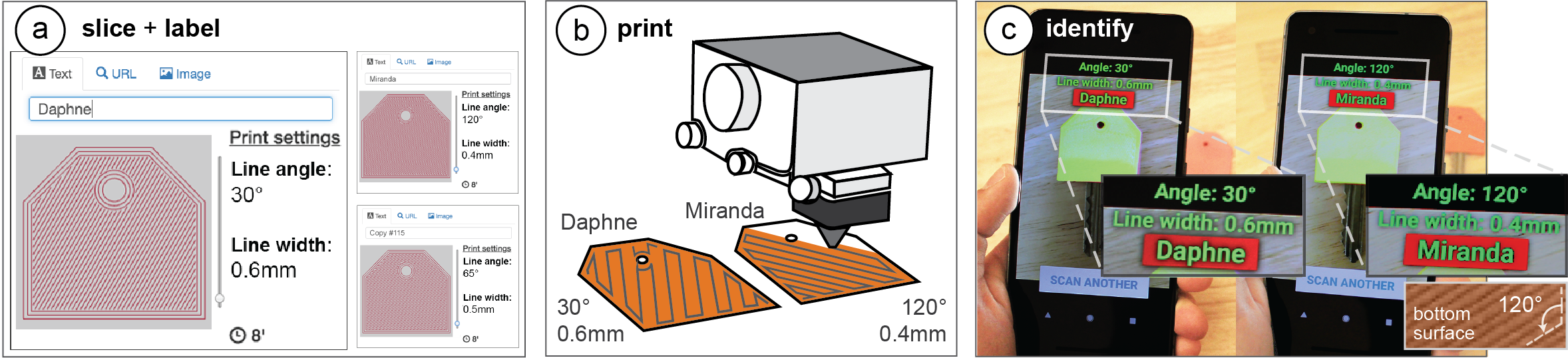}
  \caption{3D printed objects inherently possess surface patterns due to the angle of the print path and the thickness of the trace the 3D printer lays down. G-ID exploits such features 
  to identify unique instances of an object without the need to embed an obtrusive, additional tag. G-ID provides (a) a user interface for slicing individual instances of the same object with different settings and assigning labels to them. After (b) 3D printing, users can (c) identify each instance using the G-ID mobile app that uses image processing techniques to detect the slicing parameters and retrieve the associated labels.}
  \label{fig:gid-1-Teaser}
\end{figure}

\section{Method: Labeling and Identifying Objects By Their Slicing Parameters}

The main contribution of G-ID is a framework to \textit{label} and \textit{identify} 3D printed objects by their distinct slicer settings.

G-ID \textit{labels} 3D printed objects by intentionally varying the slicing settings of an unmodified 3D model which determine the path the extruder will follow. This allows G-ID to produce multiple instances that all have a unique artifact, e.g., the small grooves on the surfaces of the object that can be shaped differently when the print path is laid down. 

G-ID then \textit{identifies} the 3D printed object by such textures, i.e. after users take a picture of the object with a commodity camera, G-ID applies image processing techniques to first extract and then correlate the features with their initial slicing settings to retrieve the identifying label.  

\subsection{Main Benefits of Using Different Slicing Parameters }
Slicing parameters reveal themselves on any printed object as a fabrication byproduct that is normally ignored. One may make use of these inevitable textures that come for free due to 3D printing. G-ID combines a wide range of slicing parameters to create a sufficiently large parameter space. For each slicing parameter for surface and infill, there is a variety of values available (see section “Spacing of Slicing Parameters”). The use of so many values is enabled by G-ID’s recognition algorithm, which uses a Fourier-based method for precise measurements. Detecting these values precisely in turn enables new applications such as “Finding optimal print settings” (see “Application Scenarios”).

\subsection{G-ID Workflow for an Identification Application}
In the following section, we describe how we use (1) the G-ID labeling interface to assign each instance of a 3D printed object a unique tag prior to 3D printing, and how we use (2) the G-ID identification app that runs on a mobile device to detect each object’s tag after 3D printing.

We explain the workflow of G-ID using an application scenario, in which we will create a set of forty key covers—each with an unobtrusive feature that identifies its owner. We use these key covers in our research lab: At the end of the semester when members depart, we often forget to write down who has returned their keys. Using G-ID, we can quickly identify whom the previously returned keys used to belong to and then send a reminder to those who have outstanding keys.

\subsection{Labeling Interface (Slicer)}
\begin{figure}[t]
  \centering
  \includegraphics[width=0.7\columnwidth]{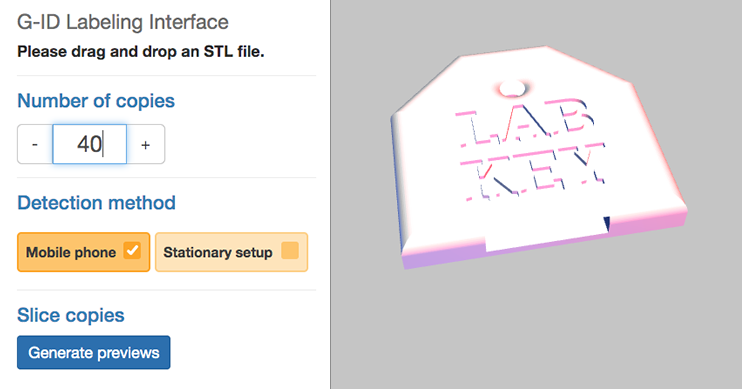}
  \caption{G-ID labeling interface: load a 3D model and enter number of unique instances needed.}
  \label{fig:gid-3-UI}
\end{figure}

To assign each key cover a unique tag, we open G-ID’s labeling interface (Figure~\ref{fig:gid-3-UI}) on our computer and load the 3D model of the key cover by dragging it onto the canvas. Since we want to create 40 key covers, we enter 40 instances into the left-hand panel of the interface.

\textit{\#1 Generate Instances:} We select “Mobile phone” as the desired detection setup. Next, we click the “Generate previews” button, which slices each instance of the key cover with a unique set of slicing settings. In this case, since we only request 40 instances, G-ID only varies the slicing parameters initial bottom line angle and initial bottom line width (more on this in section “Slicing parameters used for labeling”). After slicing is completed, G-ID previews each instance as shown in Figure~\ref{fig:gid-1-Teaser}a.

\textit{\#2 Enter Labels:} We can now enter a label in the form of a text, an image, or a URL next to the preview of each instance. Since we want to give each key cover the name of one of our lab members, we enter one name per instance. We can update these labels later any time, for instance, when a new team member joins our lab and we need to reassign the key.

\textit{\#3 3D Printing:} Upon clicking the “Export” button, each instance’s G-code file is saved, as well as a digital file (XML) that stores the object information and the entered label corresponding to each instance. We can now send the G-code files to our FDM printer to obtain the printed instances. We also transfer the digital file that stores the object information to our smartphone to be used later for identification.

\subsection{Identification Interface (Mobile App with Object Alignment)}

\begin{figure}[t]
  \centering
  \includegraphics[width=0.55\columnwidth]{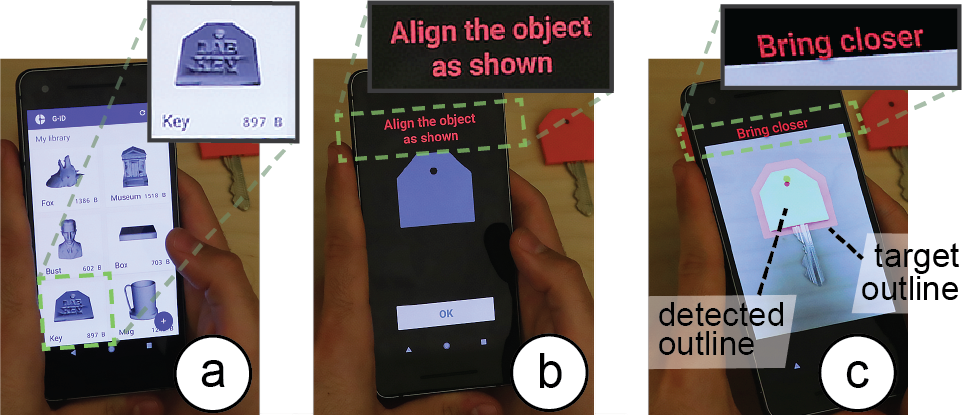}
  \caption{G-ID mobile app for identifying instances: (a) select model from library, (b) face the object, (c) once outlines are aligned, the app automatically takes the image.}
  \label{fig:gid-4-App}
\end{figure}

At the end of the semester when we update our key inventory, we use the G-ID mobile app on our phone to identify which of the returned keys belonged to whom. After launching the app, we first select the model we would like to scan, i.e., the key cover, from our object library (Figure~\ref{fig:gid-4-App}a). The app then helps us to align the camera image with the object by showing an outline of the object on the screen, similar to how check cashing or document scanning apps work (Figure~\ref{fig:gid-4-App}b). When the outlines are aligned in this human-in-the-loop setting, the app automatically captures and processes the image (Figure~\ref{fig:gid-4-App}c). It then identifies the features in the photo associated with the surface-related slicing parameter settings, retrieves the user-assigned label, and shows it on the screen (Figure~\ref{fig:gid-1-Teaser}c). We check off the lab members who returned their keys and send a reminder to everyone else.

\subsection{Surface and Interior: Detecting Infill Using a Light Source}

In the above described scenario, G-ID was able to successfully label each instance using only slicing parameters that affect the object’s surface, such as the initial bottom line width and angle because the number of instances was small. However, for scenarios requiring more instances, G-ID can also sense the interior of objects (infill) at the expense of adding a small light source as described in the next scenario.

\begin{figure}[t]
  \centering
  \includegraphics[width=0.57\columnwidth]{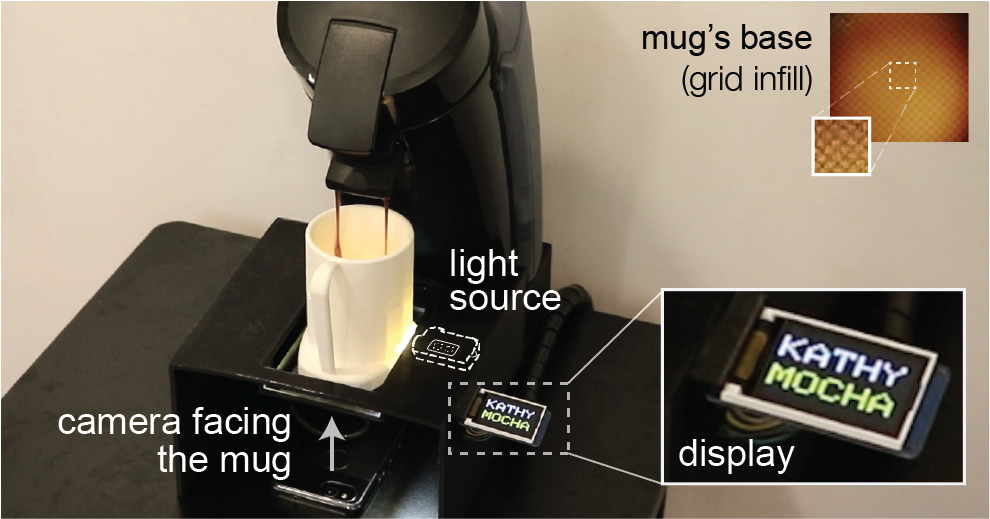}
  \caption{By adding a small light source, we can also detect variations in infill, such as different infill angles, patterns, and densities, which allow for a larger number of instances. Here, the coffee maker recognizes the mug’s infill and pours the user’s preferred drink.}
  \label{fig:gid-5-Infill}
\end{figure}

For our department’s annual celebration, we are asked to print a set of 300 coffee mugs as a giveaway. Each coffee mug, when inserted into a smart coffee machine (camera and light source below the tray table), automatically fills the mug with the user’s preferred drink. Similar to the previous scenario, we use G-ID’s labeling interface to generate the instances, but this time G-ID also varies the parameters infill angle, infill pattern, and infill density once it used up the parameter combinations available for the surface. As users insert their mug into the smart coffee machine, which has a stationary setup, the integrated light makes the infill visible due to the translucent nature of regular PLA 3D printing filament (Figure~\ref{fig:gid-5-Infill}). G-ID takes a picture, extracts the infill angle, pattern, and density in addition to the previously mentioned bottom surface parameters, and after identification, pours the user’s favorite drink. 

\section{Slicing Parameters Used for Labeling}
In the next section, we report on the types of slicing parameters for surface and infill that can be used for creating unique identifiers. 

\subsection{Surface Parameters}

\begin{figure}[h]
  \centering
  \includegraphics[width=0.55\columnwidth]{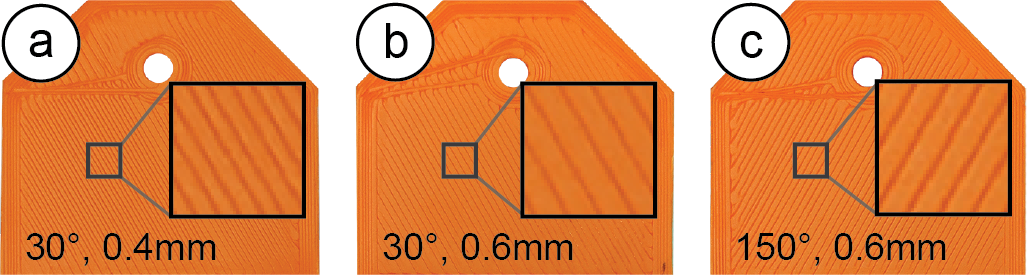}
  \caption{Combinations of different line widths and angles.}
  \label{fig:gid-6-Bottom}
\end{figure}

Bottom Surface: Resolution and Angle
When the bottom layer is printed by moving the print head along a path, two parameters influence how the path on this first layer is laid out. Initial bottom line width defines the width of a single line on the bottom surface and thus the resulting resolution. Initial bottom line angle sets the direction when drawing the lines to construct the surface. Combinations of these two parameters are shown in Figure~\ref{fig:gid-6-Bottom}. 

\begin{figure}[h]
  \centering
  \includegraphics[width=0.55\columnwidth]{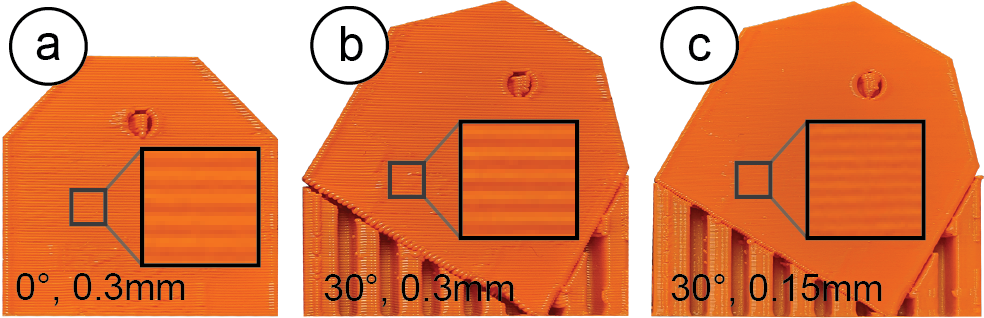}
  \caption{Different layer angles achieved by rotating the model at the expense of additional support material and print time (a vs. b), and various print qualities (b vs. c).}
  \label{fig:gid-7-LayerAnglesRotation}
\end{figure}

\vspace{0.2cm}
\textit{Intermediate Layers: Resolution and Angle}
It is possible to vary the slicing parameters for the intermediate layers in the same way as for the bottom surface. Layer height when varied leads to different layer thicknesses across the printed object and thus affects the overall print resolution. Rotating the 3D model on the build plate leads to different layer angles across the side surface. Combinations of these two parameters can be seen in Figure~\ref{fig:gid-7-LayerAnglesRotation}.

However, using the slicing parameters for layers comes at several drawbacks. As can be seen, changing the layer orientation results in a significant increase in print time due to the extra support material required. Further, changing the layer resolution can result in a notable difference in print quality across different instances. We still include it here to provide a complete overview of all available parameters.

\subsection{Infill Parameters}
Next, we review slicing parameters that change an object’s internal print path.

\begin{figure}[t]
  \centering
  \includegraphics[width=0.525\columnwidth]{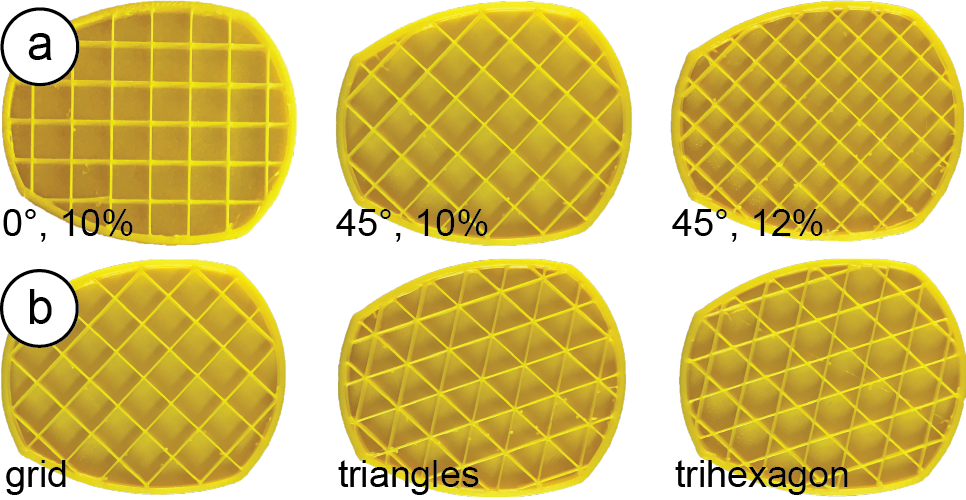}
  \caption{Cross-sections of the mug model show (a) different infill angles and densities, (b) different infill patterns.}
  \label{fig:gid-8-Infill}
\end{figure}

\vspace{0.2cm}
\textit{Infill: Resolution, Angle, and Pattern}:
Three parameters influence how the infill is laid out. \textit{Infill line distance} determines how much the lines of the infill are spaced out and thus determines the internal resolution.
The denser the infill lines, the higher the infill density.
\textit{Infill angle} rotates the infill lines according to the direction specified in degrees. Different combinations of these two parameters are shown in Figure~\ref{fig:gid-8-Infill}a. \textit{Infill pattern} allows for different layouts of the print path (Figure~\ref{fig:gid-8-Infill}b), such as grid or triangle shapes.

\subsection{Selecting Parameters that Minimize Print Time and Material}

When the user enters the number of instances, G-ID varies those slicing settings first that have the least amount of impact on print time and required material. For instance, the bottom line angle does not add any print time, does not change the resolution, and does not require additional material, whereas changing the bottom width changes the resolution slightly. Infill parameters are varied afterwards since they affect most layers of the model, starting with the infill angle, followed by infill pattern and width. 

\section{Detecting and Identifying Slicing Parameters}

To detect these slicing parameters on a 3D print, we apply common image processing techniques. Our pipeline is implemented using \textit{OpenCV} and uses \textit{SimpleElastix}, a state-of-the-art image registration library [25].

\subsection{Aligning the Object’s Base in Handheld Camera Images}
In the first processing step, G-ID needs to further refine the position and orientation of the object in the photo the user has taken to match the outline of the 3D model that was shown on the screen. For such alignment, most existing tagging approaches include specific shapes with the tags. For example, QR codes [17] have three square finder patterns and AirCodes [22] have four circles that are used to align the image. We did not want to add such markers and therefore decided to infer the position and orientation of the object based on the contour of its surface, which G-ID can extract from the 3D model.

\begin{figure}[t]
  \centering
  \includegraphics[width=1\columnwidth]{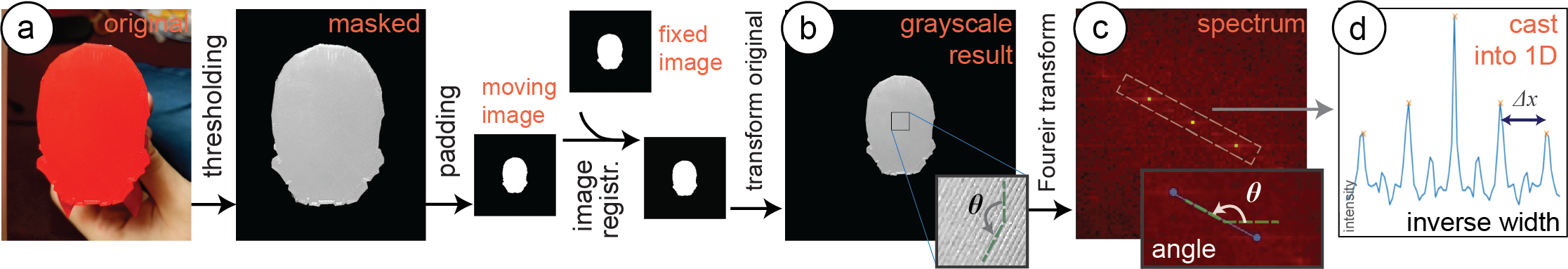}
  \caption{Image registration \& processing pipeline. (a) The captured outline is registered with that of the 3D model model for (b) improved alignment. (c) Its Fourier transform is used to infer line angle $\delta$ and width $d$. (d) The distance $\Delta x$ between intensity peaks on the inclined line is inversely proportional to $d$.}
  \label{fig:gid-9-ImageRegistrationProcessing}
\end{figure}

\vspace{0.2cm}
\textit{Images Used for Alignment}:
When the user processes the 3D model in the G-ID labeling interface, it automatically saves the outline of its base in the XML file as a binary image (stored as a Base64 string). When the user loads the XML file in the app, the object appears in the user’s model library. After choosing the desired model, the app shows the user the stored outline to assist them with facing the object from the right angle. The app automatically captures the image when the contours are matched (i.e., the bitwise XOR error between the detected contour and target contour is below an acceptable value). G-ID then applies several pre-processing steps, such as applying a bilateral filter to smooth different color and shade regions without blurring their boundaries.

\vspace{0.2cm}
\textit{Removing Overhangs}:
In the next processing step, G-ID removes overhangs, i.e., filters out object geometry that is not part of the bottom surface but appears in the image due to the camera angle. G-ID filters these parts out based on their shading in the camera image: Since the bottom surface is flat and located at a different height than an overhang, the two surfaces have different reflection properties, and thus appear brighter or darker in the image (Figure~\ref{fig:gid-9-ImageRegistrationProcessing}a). To find the shading of the bottom surface, G-ID samples a small region of interest to extract the HSV values to get the corresponding threshold range. G-ID then uses the threshold to mask the contour. 

\vspace{0.2cm}
\textit{Reducing Perspective from the Image (Undistorting)}:
After extracting the contour, G-ID applies an affine transformation (translation, rotation, scaling, shearing) to compute the deformation field required for alignment. As input to this image registration, we use the masked image from the previous step (converted to a binary outline) as the moving image. The fixed image is the outline of the 3D model at a constant scale. While a projective transformation would best rectify the perspective, it can be approximated by an affine transform since the perspective deviation is minimized due to human-in-the-loop camera image capturing. 

To find the best affine transformation, we use adaptive stochastic gradient descent as the optimizer. As the objective function, we use mean squares since we work with binary images, which have little gradient information. The computed parameter map of the affine transformation is then applied to the image the user has taken to align it with the digital 3D model outline (Figure~\ref{fig:gid-9-ImageRegistrationProcessing}b). 

\subsection{Detecting Bottom Line Angle and Width}

Since the traces of the 3D printed surface have a periodic layout, we are able to detect their orientation and widths by looking at the frequency spectrum, i.e., we take the 2D Fourier transform of the image. From this spectrum, we can determine the bottom line angle $\theta$ by extracting the slope of the line on which the periodic peaks lie (peaks are marked yellow in Figure~\ref{fig:gid-9-ImageRegistrationProcessing}c). We can determine the bottom line width d by casting the intensity values on this inclined line into a 1D array and computing the distance $\Delta x$ between the maxima, which is inversely proportional to d. This approach is more robust than looking at the original image itself because in case the lines have irregularities, their distances may be inconsistent, whereas the Fourier transform acts as a smoothing filter and provides an averaged value.

\subsection{Error Checking}

If the picture the user has taken is of poor quality (i.e., out of focus, poor lighting, or accidental shaking of camera), the lines on the object surface will not be clear enough to extract correct measurements of parameters. Fortunately, these false readings can be avoided due to the nature of the 2D Fourier transform. In the Fourier spectra of digital photos, there is a strong intensity along the x and y-axis since real-world objects have many horizontal or vertical features and symmetries [38, 48]. If surface lines are not distinguishable, peak intensities appear onthe x and y-axis and therefore erroneously result in detection of 0° or 90°. Thus, the detection of either of these two angles indicates a false reading. Therefore, we exclude these two values from our allowed slicing parameter space. Whenever our algorithm detects these two angles, we notify the user that the image has to be retaken for correct measurement.

\subsection{Detecting Infill Angle, Width, and Pattern}

\begin{figure}[t]
  \centering
  \includegraphics[width=0.575\columnwidth]{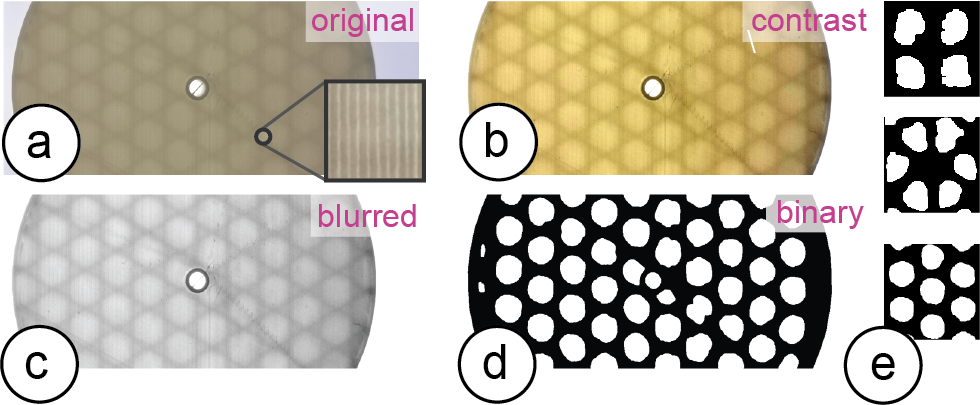}
  \caption{Image processing to extract the components of the infill pattern: (a) photo, (b) contrast increased, (c) blurred, (d) binarized, and (e) matched to the template of the respective infill type: grid, triangles, trihexagon from top to bottom.}
  \label{fig:gid-10-ImageProcessingInfill}
\end{figure}

To detect the infill parameters, we first remove noise from the image that is caused by the bottom lines on the surface (Figure~\ref{fig:gid-10-ImageProcessingInfill}). We remove them by (a) increasing the contrast of the image, (b) blurring the image with a 2D Gaussian smoothing kernel, and (c) applying adaptive thresholding. 

To detect the infill pattern, we compare the resulting shapes after thresholding to the known infill pattern templates shown in Figure~\ref{fig:gid-10-ImageProcessingInfill}d/e. To determine infill density, we compare the infill templates at different scales to the size of the shapes in the image, the matching template then indicates the size of the pattern. Similarly, the infill angle is detected by rotating the template and finding the angle that gives the smallest sum of squared difference to shapes in the image. Since infill is detected in the stationary setup, alignment of the base is less of a concern and image registration can either be simplified or ignored for shorter processing times.

\section{Spacing of Slicing Parameters}

We conducted an experiment to determine which parameter spacing can be reliably identified using our detection method. For the experiment, we printed a number of instances with different slicing settings and used our detection method to identify each pattern. Table 1 summarizes the results of this experiment under regular office light conditions. We focus this analysis on the parameters related to bottom surface and infill, which are seen from the object’s base, and do not consider those related to the side (intermediate layers).

\begin{table}[]
\centering
\begin{tabular}{lllll}
                 & Min  & Max     & Spacing & Variations        \\ \hline
\textbf{Bottom:} &      &         &         &                   \\ \hline
Angle (°)        & 0°   & 180°    & 5°      & 36-2=34*          \\
Width (mm)       & 0.35 & 0.6     & 0.05    & 6                 \\
Bottom Total:    &      &         &         & = 204             \\ \hline
\textbf{Infill:} &      &         &         &                   \\ \hline
Angle (°)        & 0°   & 60°/90° & 5°      & 12 / 18           \\
Width (mm)       & 2.6  & 3.2     & 0.6     & 2                 \\
Pattern (type)   & -    & -       & -       & 3                 \\
Infill Total:    &      &         &         & = 84              \\ \hline
\textbf{Total:}  &      &         &         & \textbf{= 17,136}
\end{tabular}
\caption{Each slicing parameter’s range (min, max) and incremental spacing values as determined by our experiments. The last column shows the number of variations that can be realized. (*2 angles reserved for error checking).}
\label{tab:gid-slicing-results}
\end{table}

Using the object’s base for identification has many advantages: (1) it is easier for users to take aligned pictures of the base since it is flat, (2) it is more time and material efficient to manipulate base-related features since they only affect a single layer and have no influence on the print quality of the main surface of the object, (3) there are more combinations of identifiable features since both bottom surface and infill can be seen just from the base, (4) it is convenient to computationally process a flat surface. 
We therefore decided to first focus on the bottom layer and infill parameters, however, further analysis can be done concerning side surfaces by repeating our experiment.

\subsection{Selecting 3D Models to Evaluate Parameter Spacings}

How finely differences in slicing parameters can be detected depends on the size of the area of the bottom surface. The larger the surface, the more features can be used by the algorithm for classification. To determine a spacing of parameters that works well across different 3D models, we used objects with varying surface areas for our experiment.

To select these objects, we downloaded the top 50 3D models from \textit{Thingiverse} [48] and ranked them by their bottom surface area (i.e., the largest square one may inscribe in the contour of the first slice, determined by an automated \textit{MATLAB} script). We found that 25 models had a large surface area (>6cm$^{2}$), 7 models had a medium surface area (1.2-6cm$^{2}$), and 18 models had small surface areas (<1.2cm$^{2}$). These ranges were determined empirically based on our initial tests. We randomly picked one object representing each of these three categories and printed multiple instances using the parameters below.

\subsection{Determining the Range for Each Slicer Setting}

Before slicing each of the models with different settings, we first determined the min and max values for each setting.

\vspace{0.2cm}
\textit{Bottom:} For the bottom angle, we can use  0°-180°. Going beyond 180° would cause instances to be non-distinguishable (i.e., 90° looks the same as 270°). We took the min value for initial bottom line width as 0.35mm, the default value recommended in the slicer \textit{Cura}. Although this parameter can be as large as twice the nozzle size (2*0.4mm), we limit the max value to 0.6mm to avoid disconnected lines. For the pattern settings, which do not have min and max values, we considered the “line” pattern for the bottom surface. 

\vspace{0.2cm}
\textit{Infill:} As for the infill angle, we can use a range of 0°-60° for the trihexagon and triangular patterns, and 0°-90° for the grid pattern, as their layouts are periodic with period 60° and 90°, respectively. For infill line distance (density), we determined that having infill units smaller than 2.6mm makes the pattern unrecognizable — we thus used it as the min value. The max value is 3.2mm for objects with medium base area but may go up to 8.0mm for larger objects. Going beyond this value would imply an infill density of less than 10\%, and thus fragile, less stable objects. The three infill pattern (type) settings do not have min or max values.

\subsection{Slicing with Different Spacings and Capturing Photos}

Next, we used our three selected objects (small, medium, large), and printed them with different slicing settings. Our goal was to determine how finely we can subdivide the given parameter ranges for accurate detection. To find the optimal spacing for each parameter, we made pairwise comparisons of two values (e.g., for angles, instance \#1: 8° - instance \#2: 5°; difference: 3°), while keeping all other parameters constant. Based on 16 pictures taken for each pairwise comparison, we report the accuracy at which we can distinguish the two instances. For the prints with infill variations, we held a small light source (\textit{Nitecore Tini} [34]) against the side of the 3D printed object before taking the image.

\subsection{Results of the Experiments}

As expected, objects from the “small” category did not have sufficient base area to fit in enough infill units and thus not give satisfactory results. We therefore conclude that G-ID cannot be used for very small bases and excluded them from the rest of the analysis. We next discuss the results for each slicing parameter for the medium and large object category.

\begin{figure}[t]
  \centering
  \includegraphics[width=1\columnwidth]{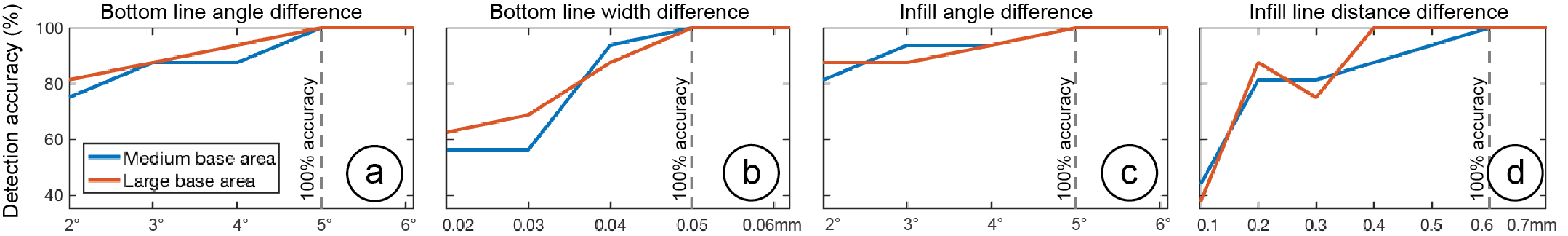}
  \caption{Detection accuracies vs. parameter spacing between pairs of instances for the four slicing parameters (bottom line angle and width, as well as infill angle and width). All plots have the same y-range.}
  \label{fig:gid-11-DetectionCombinedResults}
\end{figure}

\vspace{0.2cm}
\textit{Bottom Line Angle and Width}:
The dashed lines in Figure~\ref{fig:gid-11-DetectionCombinedResults}a,b indicate that a spacing of 5° and 0.05mm provides a classification accuracy of 100\% for both the medium and large base area categories, respectively. Thus, a range of 0°-180° would give us 36 variations for the angle. However, we exclude the two degrees 0° and 90° from the bottom line angle range since these are reserved for error checking (as described in section “Detecting Slicing Parameters”), therefore we have 34 variations. A range of 0.35-0.6mm allows 36 variations for the width.

\vspace{0.2cm}
\textit{Infill Angle and Line Distance (Width)}:
The dashed lines in Figure~\ref{fig:gid-11-DetectionCombinedResults}c,d show that a spacing of 5° and 0.6mm provides a 100\% detection accuracy for the two categories, respectively. Thus, for the ranges of 0°-60° and 0°-90°, we can use 12 or 18 variations, respectively. For the width, we can use, for medium objects, a range of 2.6-3.2mm (2 variations); and for large objects 2.6-8.0mm (10 variations). We report the smaller number in Table~\ref{tab:gid-slicing-results}.

\begin{figure}[t]
  \centering
  \includegraphics[width=0.7\columnwidth]{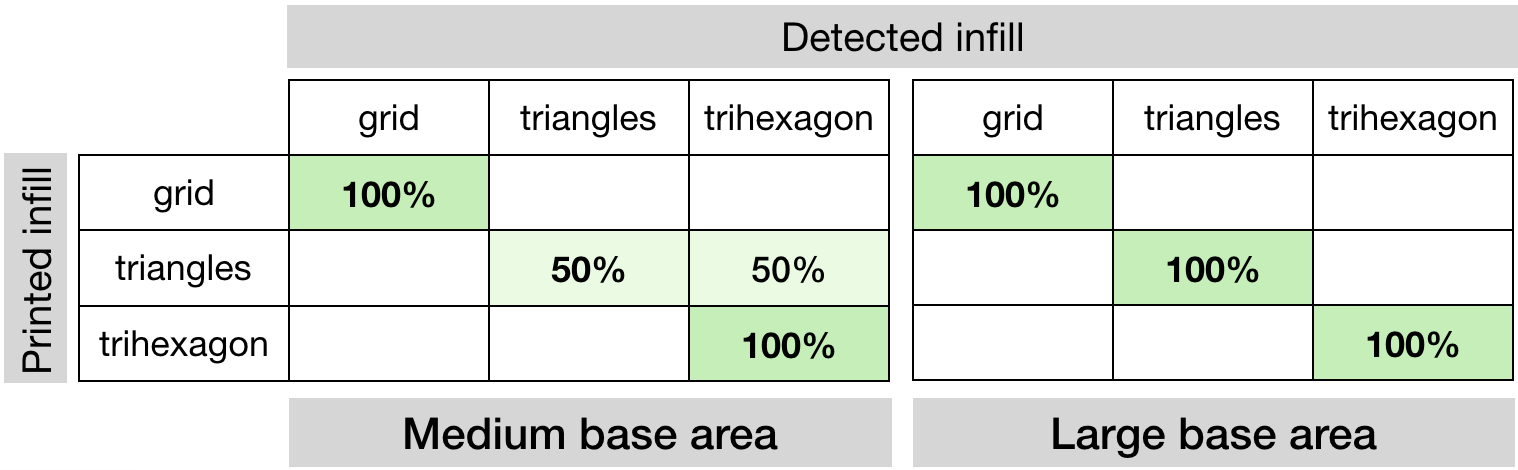}
  \caption{Detection accuracies vs. parameter spacing between pairs of instances for the four slicing parameters (bottom line angle \& width, as well as infill angle \& width). All plots have the same y-range.}
  \label{fig:gid-12-InfillConfusionMatrix}
\end{figure}

\vspace{0.2cm}
\textit{Infill Pattern}:
The confusion matrix in Figure~\ref{fig:gid-12-InfillConfusionMatrix} shows that the “grid” and “trihexagon” work for both medium and large classes. For large objects, we can also use all three different patterns.

\subsection{Total Number of Instances Possible}

Based on the results, using a parameter spacing that works for both medium and large objects, we can achieve a total of 204 instances if we only use the camera, or 17,136 instances if we use both the camera and light (see  Table~\ref{tab:gid-slicing-results}). In comparison to other types of tags, we can say that these parameter spaces have a larger code capacity than a 1D barcode with 7 or 14 bits, respectively.

\subsection{Cross-Validation}

To cross-validate our parameter spacing, we printed another random set of 16 objects (10 large, 6 medium) from the same top 50 model list from \textit{Thingiverse} (excluding the ones we used in the previous experiment) with white filament. Before printing, we randomly assigned each model a combination of slicing settings from the available parameter space and then tested if G-ID can identify them. Our second goal was to see if the parameter spacing still applies when multiple parameters are varied at the same time (our previous experiment only varied one slicing parameter per instance at a time). Among the 10 objects with large bottom area, all slicing parameters were correctly identified. Among the 6 objects from the medium category, 5 were correctly identified. In total, the detection accuracy is 93.75\%.
The falsely identified model had the smallest bottom surface area (1.4cm$^{2}$), which confirms the fact that objects without sufficient surface area cannot be recognized.

\begin{figure}[t]
  \centering
  \includegraphics[width=0.7\columnwidth]{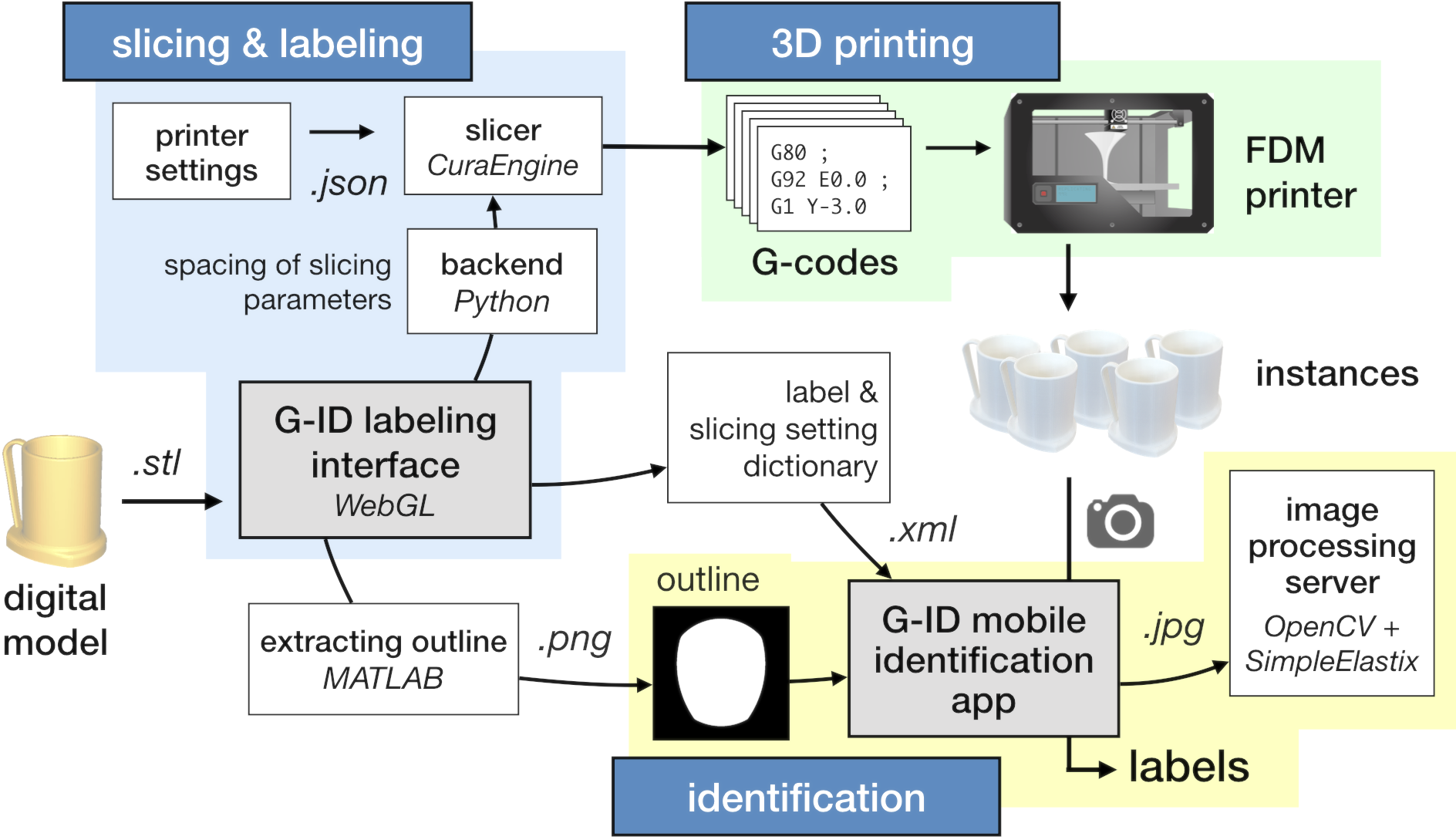}
  \caption{System overview.}
  \label{fig:gid-13-Implementation}
\end{figure}

\section{System Implementation}

An overview of the system is shown in Figure~\ref{fig:gid-13-Implementation}. G-ID’s labeling interface for creating multiple instances runs on the browser and is based on \textit{WebGL}. Once the user drags a 3D model (.stl) onto the canvas and enters the number of instances, the interface calls its backend written in \textit{Python}, which is responsible for the distribution of the slicing parameters. Once the slicing parameters are determined for each individual instance, the backend calls the slicer \textit{CuraEngine} [43] to compute the G-code for each instance.

After the instances are sliced with their individual slicer settings, the user is shown the 2D sliced layers as well as the 3D model. For rendering these previews, we use the \textit{JavaScript} library \textit{Three.js}. Finally, G-ID saves an .XML file with the slicing parameters, labels and the contour of the object’s base as an image (created using an automated \textit{MATLAB} script) for future identification with the G-ID mobile app.

\section{Evaluation}

\subsection{Different Materials, Lighting Conditions, Thicknesses}

\begin{figure}[t]
  \centering
  \includegraphics[width=0.7\columnwidth]{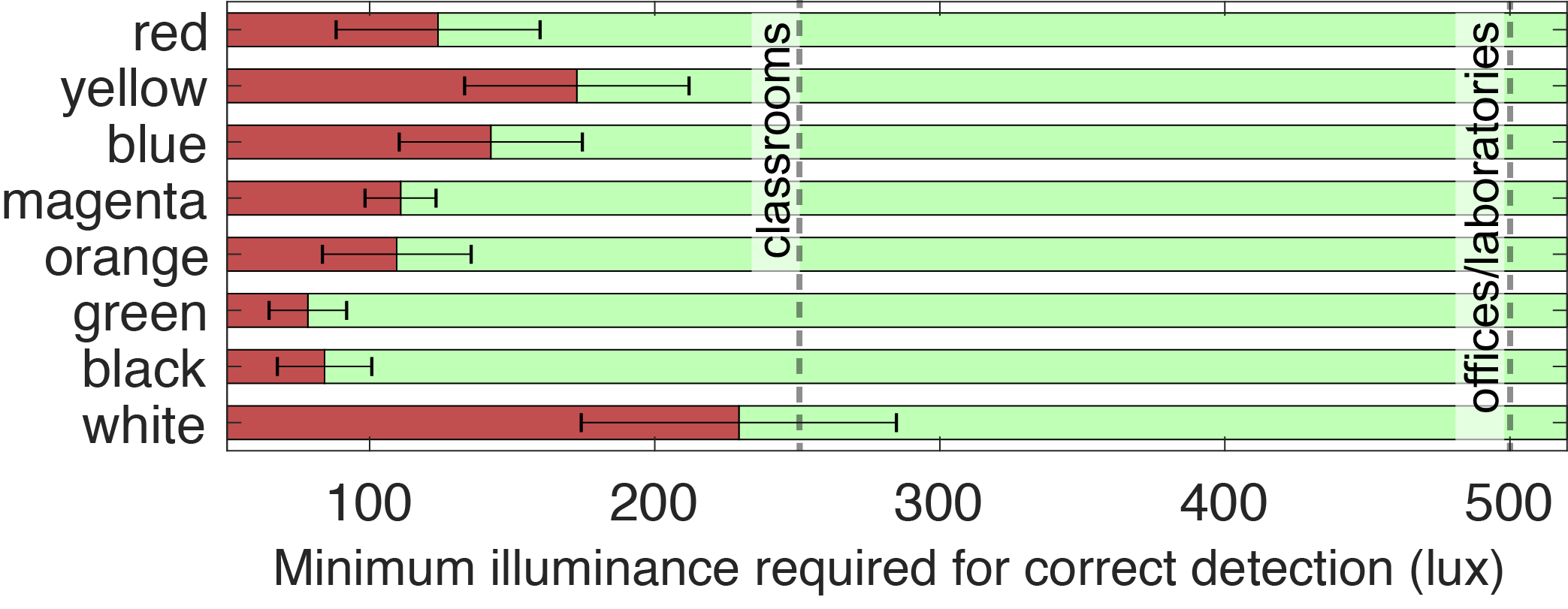}
  \caption{Different filament colors vs. minimum illuminance required to correctly detect the traces on the bottom surface.}
  \label{fig:gid-14-FilamentColorLighting}
\end{figure}

Surface: To see how the filament’s color and different lighting affect the detection of surface parameters, we printed six instances of the key cover with eight different colors of \textit{Ultimaker} PLA filaments (total of 48 instances) and using a dimmable LED lamp varied the light in the room from 0 – 500 lux (measured with lux meter \textit{Tacklife LM01}). We picked the filament colors to be the primary and secondary colors of the RYB model, as well as white and black filament. We selected the light intensities to represent different low-light conditions. For slicing, we used surface parameters distributed evenly along bottom line angle and width within the allowed range from Table~\ref{tab:gid-slicing-results}. The results are shown in Figure~\ref{fig:gid-14-FilamentColorLighting}. All colors worked well for lighting conditions above 250 lux. This shows that our approach works well in classrooms (recommended light level: 250 lux) and laboratories (recommended light level: 500 lux)~\cite{national_optical_astronomy_observatory_recommended_2016}. The results also show that the camera needs more light to resolve the lines for lighter colors (i.e., white and yellow) than for darker colors. Since we used white filament for our parameter spacing evaluation, the results in Table~\ref{tab:gid-slicing-results} work even for worst-case scenarios. This also means that for other filament colors, an even smaller surface area would suffice for correct detection since the surface details can be better resolved.

\textit{Infill:} We were able to detect the patterns for all of the aforementioned filament colors except for black due to its opaque nature. Further, the brighter the light, the thicker the object base may be: 145 lumens suffice for a base of 1mm (suggested thickness in \textit{Cura}), 380 lumens suffice for 1.75 mm.

\subsection{Different 3D Printers}

Since G-ID takes as input universal units like millimeters (width) and degrees (angle), our method extends to FDM printers other than the one we used for the experiments. To confirm this, we fabricated six test prints on the 3D printers \textit{Prusa i3 MK3S} and \textit{Creality CR-10S Pro} in addition to the \textit{Ultimaker 3} that we used for our experiments, and inspected the traces laid down with a microscope. The line widths of the prints had an average deviation of 9.6$\mu$m (\textit{Prusa}) and 10.7$\mu$m (\textit{Creality}) from the Ultimaker 3 prints. The line angles had an average deviation of 0.5° (\textit{Prusa}) and 0.25° (\textit{Creality}). These deviations are insignificant for our detection method since the spacing values chosen for the parameters are much larger than these values. To verify this, we used our mobile app to take pictures of these samples and ran our algorithm, which correctly detected the unique identifiers.

\subsection{Camera Distance and Angle}

\vspace{0.2cm}
\textit{Distance:} If the phone is held too far away from the object, the camera cannot detect the detailed grooves on the surface. To determine how far the camera can be held from the object, we took pictures with smartphones of different camera resolutions. We found that the \textit{iPhone 5s} (8MP) can resolve the slicing parameter features up to 26cm, \textit{Pixel 2} (12.2MP) up to 36cm, and \textit{OnePlus 6} (16MP) up to 40cm. Taking into account the cameras’ field of view, this means that users can fit in an object with one dimension as large as 29cm, 45cm, and 52cm, respectively. G-ID guides users into the correct camera distance by varying the size of the object outline displayed on the user’s phone for alignment.

\vspace{0.2cm}
\textit{Angle:} Since our image registration technique uses affine transformation, it is not able to fully remove the perspective distortion if the camera angle varies strongly. However, since our app guides users to align the object, the distortion is negligible. To show our algorithm can robustly read bottom surface parameters on a variety of shapes, we created a virtual dataset similar to [28]. We downloaded the first 600 models from the \textit{Thingi10K} database [57] that have a rotationally non-symmetric base of appropriate size, sliced them with a random set of slicing settings, and rendered the G-codes using 3D editor \textit{Blender}. We placed the virtual camera at points located on a spherical cap above the object base, with 8 evenly sampled azimuthal angles for each of the 5 polar angles. The percentage of shapes read correctly for each polar angle value $\theta$ is given in  Table~\ref{tab:gid-angle-detectionAccuracy}. The spacing values chosen for the parameters act as a buffer to prevent false readings. The objects for which detection failed at small angles had rather rounded bases, which makes alignment challenging.

\begin{table}[]
\centering
\begin{tabular}{l|lllll}
Camera angle \textbf{$\theta$} & 4°      & 6°      & 8°      & 10°     & 12°     \\ \hline
Accuracy   & 98.50\% & 94.50\% & 86.67\% & 75.00\% & 64.50\%
\end{tabular}
\caption{Polar angle $\theta$ vs. the percentage of identified objects.}
\label{tab:gid-angle-detectionAccuracy}
\end{table}

\section{Applications}

Below, we outline three further application examples in addition to the two previously explained use cases.

\begin{figure}[h]
  \centering
  \includegraphics[width=0.7\columnwidth]{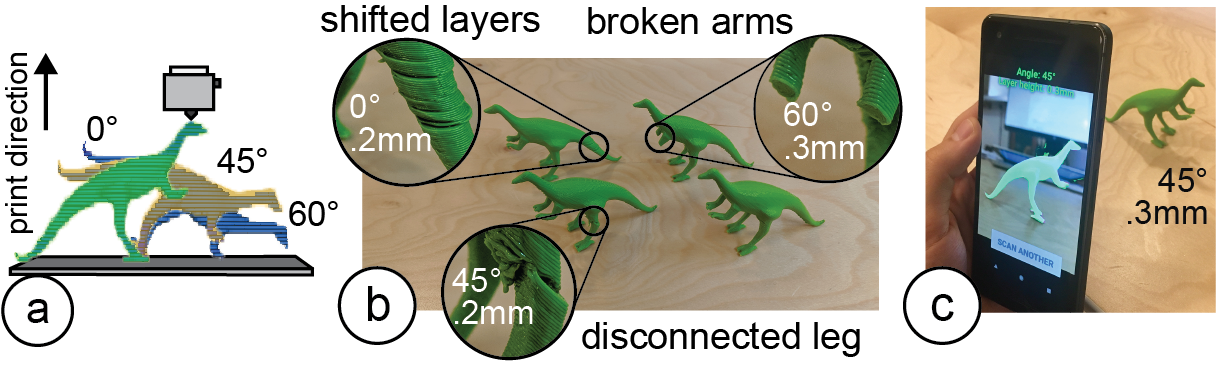}
  \caption{Identifying particular print settings with G-ID: (a) exploring different slicing parameters and (b) printing them, (c) retrieving the best settings using G-ID.}
  \label{fig:gid-15-DinoApplication}
\end{figure}

\vspace{0.2cm}
\textit{Finding Optimal Print Settings:} G-ID can be used to identify which slicing parameters a particular 3D print was created with. In Figure~\ref{fig:gid-15-DinoApplication}, a maker is trying to find the best angle for optimizing mechanical strength, and prints the model multiple times with varying settings. Rather than carefully writing down which settings were used for which one, the maker can retrieve a particular setting from a print simply by taking a picture of it. It is unlikely that they users estimate these settings by eye correctly as seeing tiny differences is not trivial.

\begin{figure}[h]
  \centering
  \includegraphics[width=0.6\columnwidth]{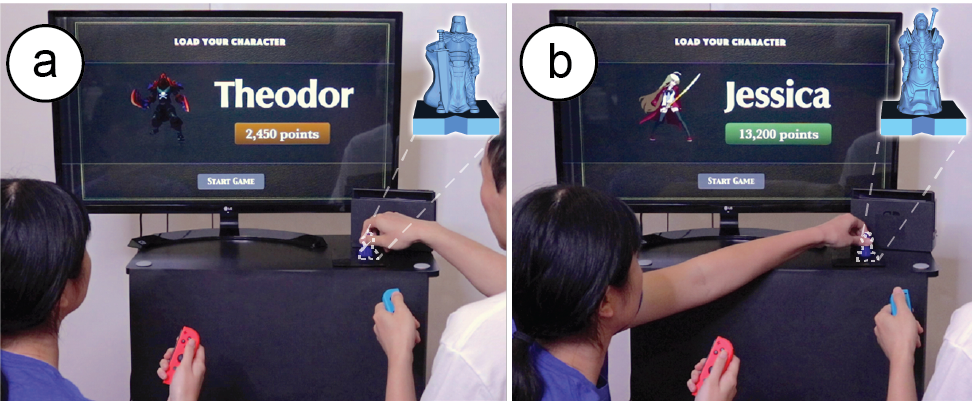}
  \caption{Toys-to-life figurines are used to identify the player and display their info in the video game.}
  \label{fig:gid-16-GamingApplication}
\end{figure}

\vspace{0.2cm}
\textit{User Identification:} G-ID can be used to create identifiable objects that belong to a specific user cheaply and rapidly as it doesn’t require the user to embed, e.g., an RFID or NFC tag in a separate step. For instance, in toys-to-life video games, physical character figurines that carry a G-ID label can be used to load a player’s identity and score. When users insert their figurine into the G-ID reader, it communicates the user’s ID to the game to display their name and score (Figure~\ref{fig:gid-16-GamingApplication}).

\vspace{0.2cm}
\textit{Labeling a Commercial 3D Print for Anti-Counterfeiting:}
Online 3D printing services such as \textit{3D Hubs}~\cite{hubs_3d_2020} or \textit{makexyz}~\cite{makexyz_llc_makexyz_2020} ensure to refund customers if they can show a 3D model was not printed according to the user’s specifications. By slicing a model using certain settings and storing this information, they can verify that a returned object was indeed fabricated by them before a refund is approved. Let us assume a 3D model is leaked and frauds attempt to print a copy themselves and then return that to get a refund (although they never bought it in the first place). Businesses can use G-ID to cross-verify the print setting used to create the original.

\section{Discussion and Limitations}

Next, we discuss limitations and future work for labeling and identifying objects by their slicing parameters.

\subsection{Other Slicing Parameters}
In this work, we focused on surface and infill parameters, which offer a large number of unique identifiable features. Other slicing parameters, such as those that create geometry that is removed after fabrication (e.g., those related to support material or build plate adhesion) are less suitable. 

\begin{figure}[h]
  \centering
  \includegraphics[width=0.6\columnwidth]{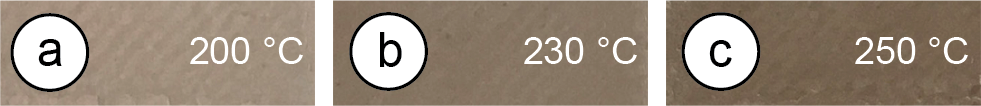}
  \caption{The hotter the nozzle, the darker the print’s finish.}
  \label{fig:gid-17-DiscussionPrintingTemperature}
\end{figure}

However, for special materials, such as wood filaments, the shade of the object’s color can be altered by varying the speed and temperature, which could be used to create differences among instances: the hotter or slower the extruder, the more the wood particles burn and the darker the resulting surface (Figure~\ref{fig:gid-17-DiscussionPrintingTemperature}). However, since the changes affect the objects appearance, we do not consider them for our work.

\subsection{Rotational Symmetry of Outlines}
Since we use the outline of objects to extract the orientation of features, objects whose bases are rotationally symmetric are less suitable for our approach. Thus, the number of identifiable angles is reduced for certain shapes, e.g., for a square base, the range narrows down from 0°-180° to 0°-90°.

\subsection{Non-Flat Side Surfaces}
\begin{figure}[h]
  \centering
  \includegraphics[width=0.55\columnwidth]{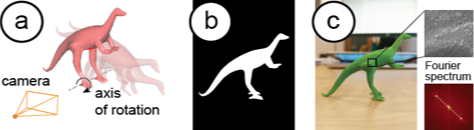}
  \caption{(a) Model’s axis of rotation is orthogonal to the image plane of the virtual camera, (b) rendered target outline for the user to align model, (c) image has parallel traces.}
  \label{fig:gid-17-DiscussionRotation}
\end{figure}

When a camera’s optical axis is parallel to the object’s axis of rotation on the build plate, the layer traces on a curved surface appear as straight, parallel lines and the features can be extracted similarly using Fourier transforms. For that, we instruct the user to take a picture from the correct viewpoint (as shown in Figure~\ref{fig:gid-17-DiscussionRotation}) by generating a silhouette guide of the object taking into account the camera’s focal length.

\subsection{Applicability Beyond FDM Printing}
\begin{figure}[h]
  \centering
  \includegraphics[width=0.3\columnwidth]{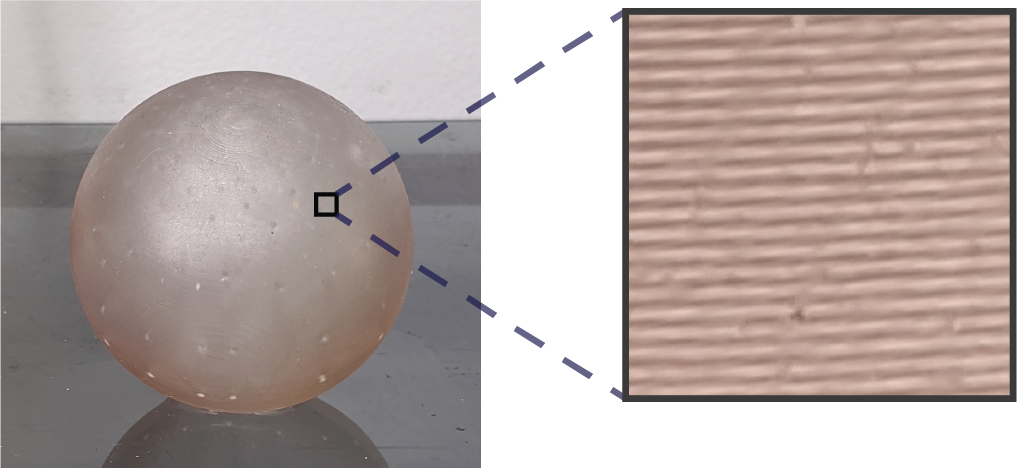}
  \caption{A sample from a \textit{Formlabs} SLA printer captured by a conventional mobile phone camera.}
  \label{fig:gid-17-DiscussionSLAPrinting}
\end{figure}

FDM and SLA printing have been the most accessible consumer techniques for the last decade. Although SLA achieves better resolutions, individual layers on objects can still be distinct (Figure~\ref{fig:gid-17-DiscussionSLAPrinting}). As for DLP printing, the projected pattern creates rectangular voxels that cause the edges to look stepped; different voxel sizes thus lead to different appearances~\cite{formlabs_sla_2020}. Most printing methods use infill, so the general idea of varying infill still applies. We thus think our method will stay relevant. Even if printing imperfections become smaller in the future, there are two factors to consider: As 3D printers improve in resolution, camera resolution improves over time (see \textit{Samsung}'s latest 200MP sensor~\cite{samsung_isocell_2021}). Computer vision gets better too: Neural networks can now pick up details beyond what the human eye or traditional image processing can detect, e.g., they have been used to identify tiny defects in solar cells~\cite{chen_solar_2018} or dents in cars~\cite{zhou_automatic_2019}.

\section{Conclusion}
We presented G-ID, a method that utilizes the subtle patterns left by the 3D printing process to identify objects and that provides both a labeling and identification interface for creating and detecting surface and infill parameters using only a commodity camera. We reported on the code capacity enabled by an off-the-shelf slicer. In the future, this can be scaled up by building a custom slicer for creating unique print paths that go beyond what current slicers offer, e.g., spatial pattern tiling could expand the number of encodings~\cite{ma_anisotag_2023}. Also, our current implementation uses optimization-based image registration, which takes a few seconds. In the future, we can enable continuous detection for faster image capturing using optimization-free contour matching methods~\cite{domokos_parametric_2010, domokos_simultaneous_2012}.
\chapter{InfraredTags: Embedding Invisible Markers Using Infrared-Translucent 3D~Printing Filaments}
\label{thesis-InfraredTags}



\section{Introduction}
In the last decade, researchers have investigated several ways to embed tags \textit{inside} 3D objects to make them unobtrusive. One method to accomplish this is to leave air gaps inside the object that represent the bits of a tag. For instance, \textit{AirCode}~\cite{li_aircode_2017} embeds air gaps underneath the surface of 3D printed objects and uses scattering of projected structured light through the material to detect where the air gaps are located. \textit{InfraStructs}~\cite{willis_infrastructs_2013} also embeds air gaps into the object but scans it in 3D using terahertz imaging, which can penetrate better through material than visible light. 

While both of these methods can embed tags inside 3D objects, they require complex hardware setups (e.g., a projector-camera setup as in \textit{AirCode}), expensive equipment (e.g., a terahertz scanner as in \textit{InfraStructs}), and long imaging time on the order of minutes. To address these issues, we propose a new method that combines air gaps inside the 3D printed structure with infrared transmitting filament. This makes the object semitransparent, and the air gaps are detectable when viewed with an infrared camera. Thus, our method only requires a low-cost infrared imaging module, and because the tag is detected from a single frame, scanning can be achieved much faster. 

One method that has used infrared-based 3D printing materials is \textit{LayerCode}~\cite{maia_layercode_2019}, which creates 1D barcodes by printing objects from regular resin and resin mixed with near-infrared dye. Thus, while the printed objects look unmodified to humans, infrared cameras can read the codes. However, this method required a modified SLA printer with two tanks, custom firmware, and custom printing material.
In contrast, our method uses more readily available low-cost materials.

In this chapter, we present InfraredTags, a method to embed markers and barcodes in the geometry of the object that does not require complex fabrication or high-cost imaging equipment. We accomplish this by using off-the-shelf fused deposition modeling (FDM) 3D printers and a commercially available infrared (IR) transmitting filament~\cite{kunststoffe_fur_3d-drucker_pla_2021} for fabrication, and an off-the-shelf near-infrared camera for detection. 
The main geometry of the object is 3D printed using the IR filament, while the tag itself is created by leaving air gaps for the bits. Because the main geometry is semitransparent in the IR region, the near-infrared camera can see through it and capture the air gaps, i.e., the marker, which shows up at a different intensity in the image. The contrast in the image can be further improved by dual-material 3D printing the bits from an infrared-opaque filament instead of leaving them as air gaps. Our method can embed 2D tags, such as QR codes and ArUco markers, and can embed multiple tags within the object, which allows for scanning from multiple angles while tolerating partial occlusion. To be able to detect InfraredTags with conventional smartphones, we added near-infrared imaging functionality by building a compact module that can be attached to existing mobile devices.

To enable users to embed the tags into 3D objects, we created a user interface that allows users to load tags into the editor and position them at the desired location. The editor then projects the tags into the 3D geometry to embed them with the object geometry. After fabrication, when the user is taking a photo with our imaging module, our custom image processing pipeline detects the tag by increasing the contrast to binarize it accurately. This enables new applications for interacting with 3D objects, such as remotely controlling appliances and devices in an augmented reality (AR) environment, as well as using existing passive objects as tangible game controllers. 
We note that this chapter was originally published at \textit{ACM CHI 2022}~\cite{dogan_infraredtags_2022}.

\vspace{0.2cm}
In summary, our contributions are as follows:

\begin{itemize}[leftmargin=0.5cm, noitemsep, topsep=2pt]
    \item A method for embedding invisible tags into physical objects by 3D printing them on an off-the-shelf FDM 3D printer using an infrared transmitting filament.

    \item A user interface that allows users to embed the tags into the interior geometry of the object.
    
    \item An image processing pipeline for identifying the tags embedded inside 3D prints.
    
    \item A low-cost and compact infrared imaging module that augments existing mobile devices.
    
    \item An evaluation of InfraredTags detection accuracy based on 3D printing and imaging constraints.
    
\end{itemize}

\section{The InfraredTags System}

InfraredTags are embedded such that the objects appear opaque and unmodified under visible light but reveal the tag under near-infrared light. We accomplish this by 3D printing the main geometry of the object using an infrared-transmitting filament, while the tag itself is created by leaving air gaps for the bits. Because the main geometry is semitransparent in the infrared region, the near-infrared camera can see through it and capture the air gaps, i.e., tag, which shows up at a different intensity in the image. 
We refer to the infrared-transmitting filament as \textit{infrared filament} or \textit{IR filament} in the remainder of the chapter.

In the next sections, we describe the properties of the IR filament and the appropriate infrared camera, and then discuss how the IR filament can be used either as a standalone single-material print or together with another filament to create markers inside the object. 

%




\subsection{Infrared Filament}
\label{Infrared_Filament}

We acquired the IR filament from manufacturer 3dk.berlin~\cite{kunststoffe_fur_3d-drucker_pla_2021} 
(ca. \$5.86/100g). It is made out of polylactic acid (PLA), the most common FDM printing filament, and can be used at regular 3D printing extrusion temperatures. To the naked eye, the filament has a slightly translucent black color, however, when 3D printed in multiple layers it looks opaque. 

\begin{figure*}[h]
  \centering
  \includegraphics[width=1\textwidth]{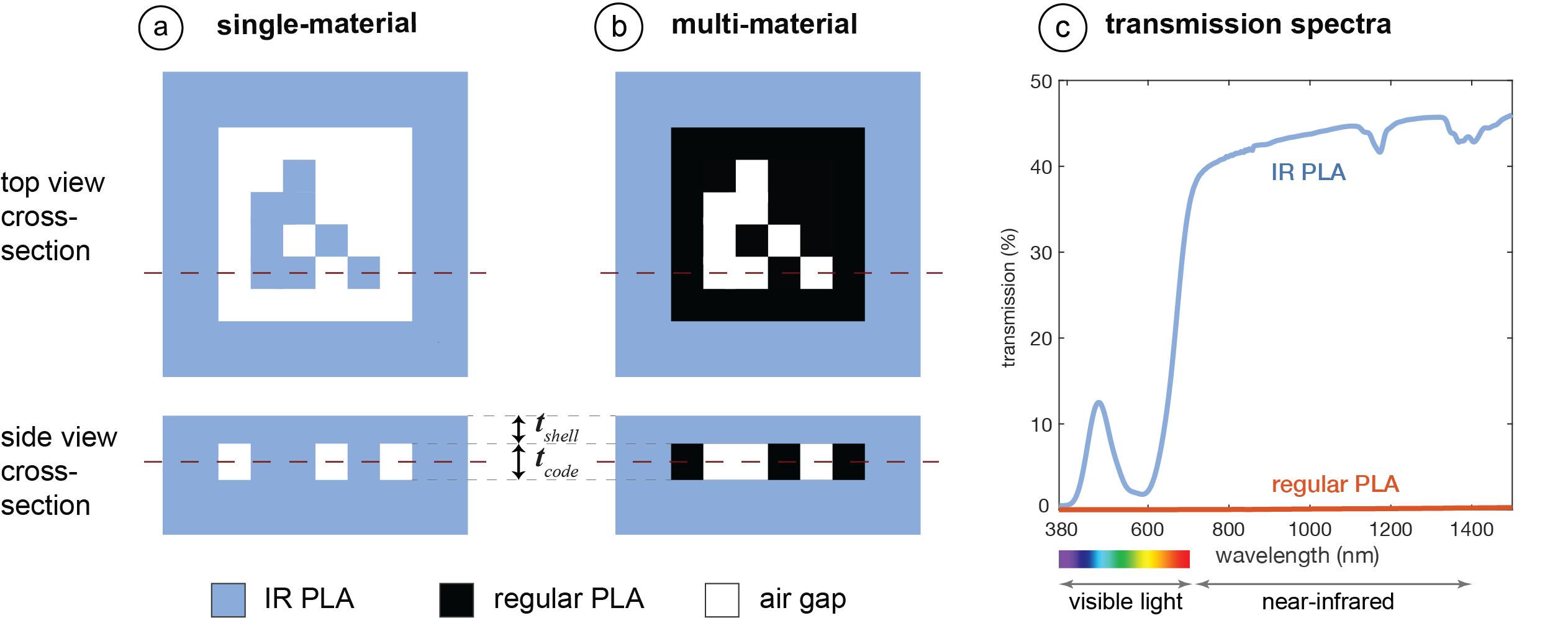}
  \caption{Material composition of the tags for a sample ArUco marker. We modify the interior of the object to embed the tag based on (a) single- or (b) multi-material printing. (c) The transmission spectrum of the IR PLA and regular PLA.}
  \label{fig:ModelInterior}
\end{figure*}

\vspace{5pt}
\noindent\textbf{IR Translucency:} Since the manufacturer does not provide data on the light transmission characteristics for different wavelengths, we manually measured it using a UV/VIS/NIR spectrophotometer (\textit{PerkinElmer Lambda 1050}). The transmission spectra for both the IR PLA and comparable regular black PLA filament are given in  Figure~\ref{fig:ModelInterior}c. Both spectra are for 1mm thick 3D printed samples.
Because the regular PLA has close to 0\% transmission in both visible and near-infrared regions, it always appears opaque. In contrast, the IR PLA transmits near-infrared at a much higher rate ($\sim$45\%) compared to visible light (0\%-15\%), and thus appears translucent in the IR region and mostly opaque in the visible light region.

\subsection{Choosing an Infrared Camera}

To choose the image sensor and filter that can see infrared light and thus can read the tag, we considered the following:

\vspace{5pt}
\noindent\textbf{Filter:} Almost all commercial cameras have an infrared cut-off filter to make camera images look similar to human vision. This filter thus prevents near-infrared light from reaching the image sensor. Since for our purposes, we want to capture the infrared light, we can either buy a camera that has this filter already removed, e.g., the \textit{Raspberry Pi NoIR} camera module, or remove the embedded filter from a camera manually.

\vspace{5pt}
\noindent\textbf{Image Sensor:}  Different cameras' sensors have different sensitivity for different parts of the light spectrum.  
To best detect the markers, the sensor should have a high sensitivity in the maximum peak of the material's near-infrared transmission. However, as can be seen in Figure 2c, since the transmission is similar across the entire infrared-region, all cameras that can detect light in the IR region would work for our purposes. For instance, off-the-shelf cheap cameras, such as the \textit{Raspberry Pi NoIR} (\$20), can detect up to 800-850nm in the near-infrared range according to several vendors\footnote{\scriptsize This module has an \textit{Omnivision 5647} sensor. \url{https://www.arducam.com/product/arducam-ov5647-noir-m12x0-5-mount-camera-board-w-raspberry-pi/} \\ \url{https://lilliputdirect.com/pinoir-raspberry-pi-infrared-camera}}. More expensive IR cameras that have sensitivity beyond the near-infrared range, such as \textit{FLIR ONE Pro}\footnote{\scriptsize \url{https://www.flir.com/products/flir-one-pro/}}, can detect up to 14,000nm but may cost more than \$400. However, since the infrared transmission does not increase much with higher infrared wavelengths, the low-cost camera is sufficient for our purposes.


\subsection{Composition of the Tags and Materials}

To create InfraredTags, we need to create two geometries with different IR transmission properties that form the object. The different IR transmission properties will cause the two geometries to appear with different intensities in the resulting infrared image. We found that there are two ways to accomplish this. 


\vspace{5pt}
\noindent\textbf{Single-Material Print (IR PLA):} Our first method uses the IR filament for the main geometry of the object, air gaps for the outside bits of the marker, and IR filament for the inside bits of the marker as shown in Figure~\ref{fig:ModelInterior}a.
The contrast between the bits arises from the fact that the IR light transmission reduces by $\sim$45\% per mm of IR filament (Section~\ref{Infrared_Filament}).
Under IR illumination, the light rays first penetrate the IR filament walls of the 3D printed object and then hit the air gap inside the object or the filled interior area. 
When the object is imaged by an IR camera, the light intensity reduces for each pixel differently depending on whether it is located on an air gap or not.
The rays that go through the air gaps lead to brighter pixels since they penetrate through less material than the other rays.
This intensity difference in the resulting image is sufficient to convert the detected air gaps and filled areas into the original tag.

\vspace{5pt}
\noindent\textbf{Multi-Material Print (IR PLA + Regular PLA):} We explored multi-material 3D printing to further improve the contrast of the marker in the image. This second approach uses IR PLA for the main geometry of the object, regular PLA for the outside bits of the marker, and air gaps for the inside bits of the marker as shown in Figure~\ref{fig:ModelInterior}b. When the user takes an image, the IR rays penetrate the IR filament walls of the 3D printed object, and then either hit the air gap inside the object or the regular PLA.
The air gaps will appear as brighter pixels since they transmit IR light, whereas the regular PLA filament will appear as darker pixels since it is nearly completely opaque in the IR region (Figure~\ref{fig:ModelInterior}c). This leads to a higher contrast than the previously discussed single-material prints.



We also considered filling the air gaps with IR filament to avoid empty spaces inside the object geometry. However, this requires frequent switches between the two material nozzles for regular PLA and IR filament within short time frames, which can lead to smearing. We therefore kept the air gaps for objects that we printed with the dual-material approach (Figure~\ref{fig:ModelInterior}b).




\vspace{0.2cm}
\noindent
\textbf{Code Geometry}: When embedding the code (i.e., the 2D tag) into the geometry of the object, the code and the geometry surrounding it (i.e., the shell) need to have a certain thickness. 

\vspace{0.2cm}
\noindent
\textit{Shell Thickness}: The shell thickness $t_{shell}$ should be large enough to create sufficient opaqueness so that the user cannot see the code with their eyes, but small enough to ensure detectability of the code with the IR camera.

\begin{figure}[h]
  \centering
  \includegraphics[width=0.7\columnwidth]{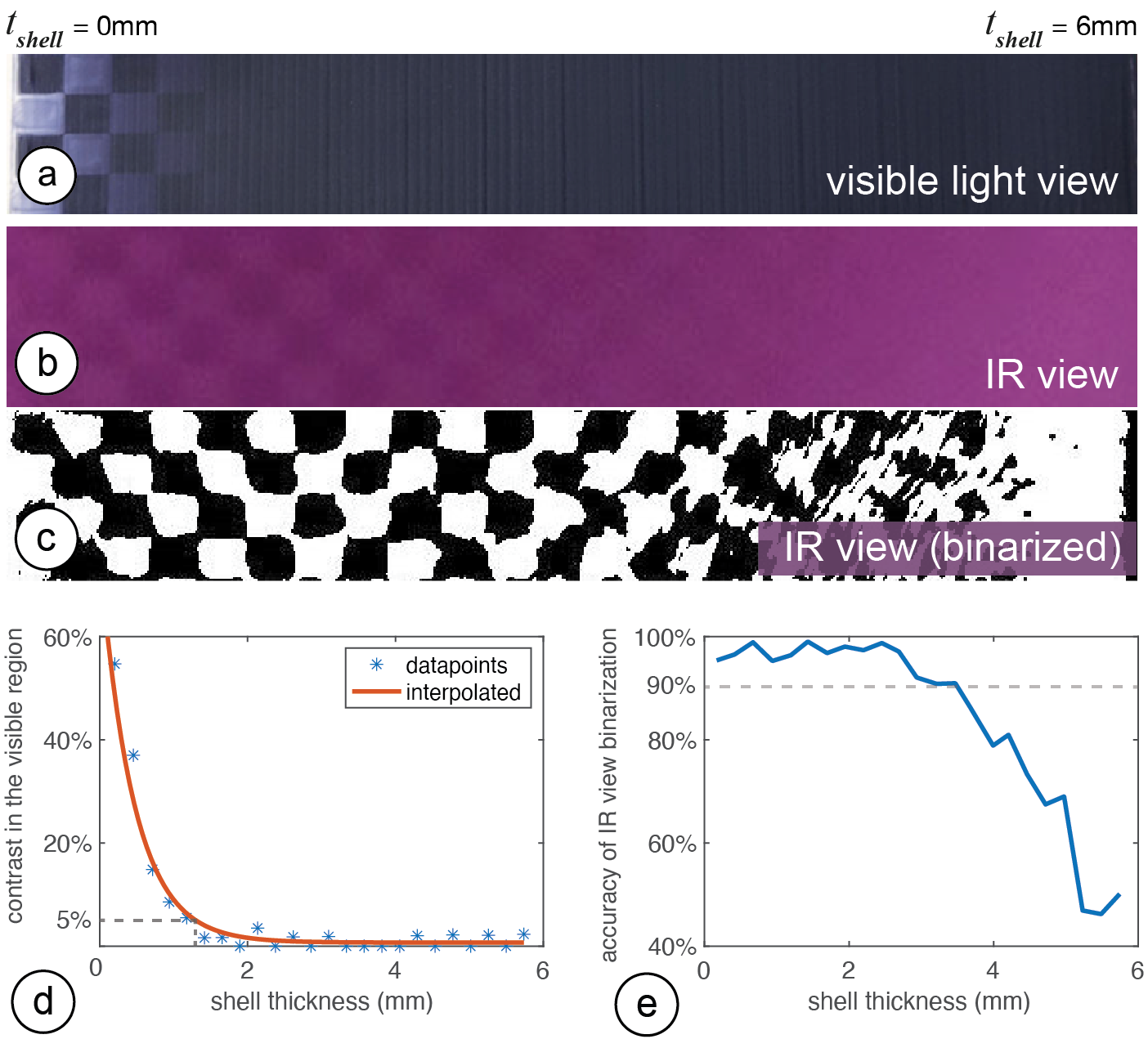}
  \caption{Determining the shell thickness for a multi-material print with white PLA. As $t_{shell}$ increases, the checkerboard pattern becomes less visible in both (a) the visible camera and (b) IR camera image. Thus, it gets more challenging to (d) identify the contrast in the pattern for humans, and to (c, e) binarize it correctly from the IR view.}
  \label{fig:ShellThicknessValues}
\end{figure}

Since the IR filament is slightly translucent to the naked eye, with small $t_{shell}$, it becomes possible for the user to see the code inside the object (Figure~\ref{fig:ShellThicknessValues}a).
Thus, for the lower bound of $t_{shell}$, our goal is to find a value that achieves a contrast in the image smaller than 5\% when the image is taken with a regular camera (i.e., with an IR cut-off filter). The image taken with regular camera represents the visible light region sensitivity, i.e., that of human vision. We chose 5\% because this is the contrast value at which humans cannot differentiate contrast anymore~\cite{bijl_visibility_1989}.

On the opposite side, the larger $t_{shell}$ is, the more IR light it absorbs and thus the darker the overall image becomes, reducing the contrast of the code in the IR region (Figure~\ref{fig:ShellThicknessValues}b).
Thus, for the upper bound for $t_{shell}$, our goal is to find the value at which the code is no longer detectable in the IR camera image.

To determine these bounds, we 3D printed a checkerboard pattern as an InfraredTag with a shell of varying thickness (range: 0mm-6mm). As shown in Figure~\ref{fig:ShellThicknessValues}, we captured the 3D print with both a visible light camera and an IR camera. In  Figure~\ref{fig:ShellThicknessValues}d, we plot the contrast between the "white" and "black" parts of the checkerboard as a function of shell thickness in the visible light camera image. We see that the visible light camera contrast drops to 5\% at approximately 1.32mm thickness, which defines the lower bound, i.e., the minimum thickness needed so that the tag is invisible to humans.
On the other hand, Figure~\ref{fig:ShellThicknessValues}c shows the binarized version of the IR camera image (Figure~\ref{fig:ShellThicknessValues}b).
In  Figure~\ref{fig:ShellThicknessValues}e, we show how the binarization of the checkerboard deteriorates as shell thickness increases. 
The graph shows that a shell thickness of up to 3.5mm could be used to achieve 90\% binarization accuracy, which defines the upper bound. However, for the sake of maximum detectability, we use the lower bound values when fabricating the objects.

For multi-material 3D printing, different filament colors can be used for the regular PLA part (i.e., the code). Each color requires a different shell thickness to prevent users from seeing the code. For instance, because the IR filament appears \textit{black} in the visible light region, it blends more easily with \textit{black} or \textit{blue} PLA, thus requiring a thinner top layer to hide the resulting code than when the code is printed in white PLA. Table~\ref{tab:CodeThickness} shows the minimum shell thickness needed to make codes fabricated from different colors unobtrusive.




\vspace{0.2cm}
\noindent
\textit{Code Thickness}: While the shell thickness affects the overall contrast of the image in the visible region, the code thickness $t_{code}$ determines the contrast between the individual bits of the embedded code in the IR region. 
If the code layer is too thin, there might not be enough contrast between the "white" and "black" bits and thus the code will not be detectable.

We conducted a test similar to the one shown in Figure~\ref{fig:ShellThicknessValues}, where we, instead of varying the shell thickness, varied the code thickness to determine which values provide enough contrast.
The values are summarized in Table~\ref{tab:CodeThickness}. Going below the values listed makes the material too thin such that the IR light starts going through the code bits, which reduces the contrast in IR view and thus detectability.
Going above this value is possible but does not improve the contrast further.

\begin{table}[h]
\centering
\begin{tabular}{lll}
 & {\footnotesize\textbf{Shell thickness} $t_{shell}$} & {\footnotesize\textbf{Code thickness} $t_{code}$} \\
 \cline{2-3}
\multicolumn{1}{l|}{\begin{tabular}[c]{@{}l@{}}Single-material\\ {\footnotesize (IR PLA)}\end{tabular}}            & 1.08 mm & 2.00 mm \\ \hline
\multicolumn{1}{l|}{\begin{tabular}[c]{@{}l@{}}Multi-material\\ {\footnotesize (IR PLA + white PLA)}\end{tabular}} & 1.32 mm & 0.50 mm \\ \hline
\multicolumn{1}{l|}{\begin{tabular}[c]{@{}l@{}}Multi-material\\ {\footnotesize (IR PLA + black PLA)}\end{tabular}} & 1.08 mm & 0.50 mm \\ \hline
\multicolumn{1}{l|}{\begin{tabular}[c]{@{}l@{}}Multi-material\\ {\footnotesize (IR PLA + blue PLA)}\end{tabular}}  & 1.20 mm & 0.50 mm \\
\end{tabular}
\vspace{0.2cm}
\caption{Thickness values for the shell and code layers.}
\label{tab:CodeThickness}
\end{table}

Lastly, an important observation we made is that IR filament spools ordered from the same manufacturer~\cite{kunststoffe_fur_3d-drucker_pla_2021} at different times showed slightly different transmission characteristics. This is likely linked to the possibility that the manufacturer might have adjusted the amount of IR-translucent dye used to make the spools. 
We suggest that users conduct similar analysis with a test print as shown in Figure~\ref{fig:ShellThicknessValues} to determine the optimal values for each new IR spool.

\section{Embedding and Reading InfraredTags}

We next explain how users can embed codes into 3D objects using our user interface and then discuss our custom add-on for mobile devices and the corresponding image processing pipeline for tag detection.

\subsection{User Interface for Encoding InfraredTags}
\label{User-Interface}

\vspace{5pt}
\noindent\textbf{Import and Position Tags:} The user starts by loading the 3D model (.stl file) into our user interface, which is a custom add-on to an existing 3D editor (\textit{Rhinoceros 3D}). Next, users import the tag as a 2D drawing (.svg) into the editor, which loads the marker into the 3D viewport. The marker is then automatically projected onto the surface of the 3D geometry (Figure 1a). Users can move the code around in the viewport and scale it to place it in the desired location on the 3D object. 

\vspace{5pt}
\noindent\textbf{Select Printing Method:} In the user interface, users can then select the printing method, i.e., if they want to fabricate the object with single material (IR-PLA only) or dual-material printing (IR-PLA + regular PLA). As a result, the user interface generates the geometry to accommodate the selected printing method. For example, for dual-material printing, it generates two .stl files, one for the main geometry and one for the embedded tag. The UI ensures that the tag is accurately spaced from the surface of the object (Table~\ref{tab:CodeThickness}). The user can then slice both files with the 3D printer's slicing software and print the object. 


\begin{figure}[t]
  \centering
  \includegraphics[width=0.67\columnwidth]{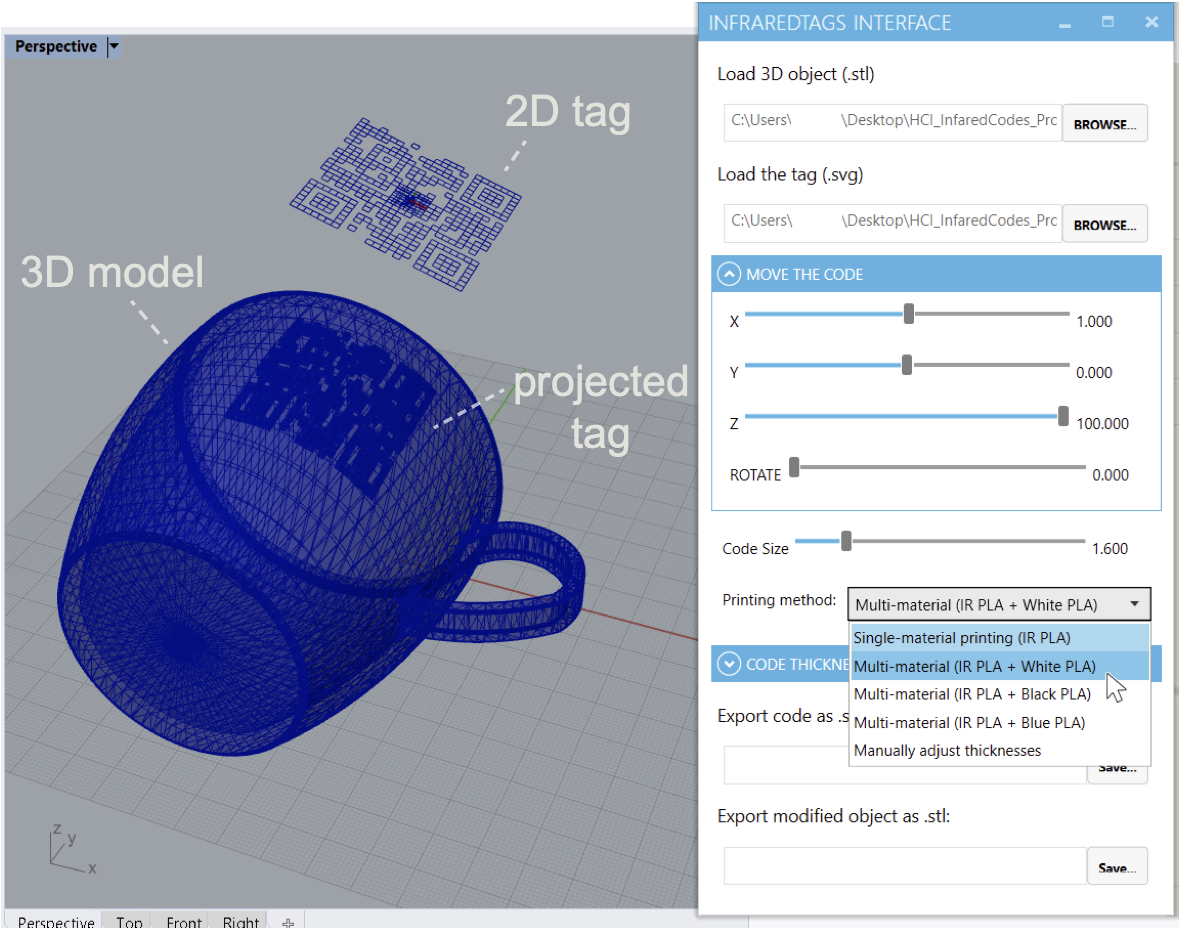}
  \caption{InfraredTags embedding interface.}
  \label{fig:UI}
\end{figure}


\subsection{IR Imaging Module for Reading the Tags}
\label{DetectionInterfaceIRModule}
InfraredTags can be read with digital devices that have an infrared camera attached to them. Even conventional USB webcams for personal computers can be used for this purpose by manually removing their infrared cut-off filter\footnote{\scriptsize A sample tutorial can be found on  \url{https://publiclab.org/wiki/webcam-filter-removal}.}.

Today, several recent smartphones already come with an IR camera either on the front (\textit{iPhone X} and newer models) or the rear (\textit{OnePlus 8 Pro}), however, the phones' APIs may not allow developers to access these for non-native applications. Furthermore, not all mobile phones contain such a camera at the moment. To make our method compatible independent of the platform, we built an additional imaging add-on that can easily be attached to existing mobile phones.

\begin{figure}[h]
  \centering
  \includegraphics[width=0.67\columnwidth]{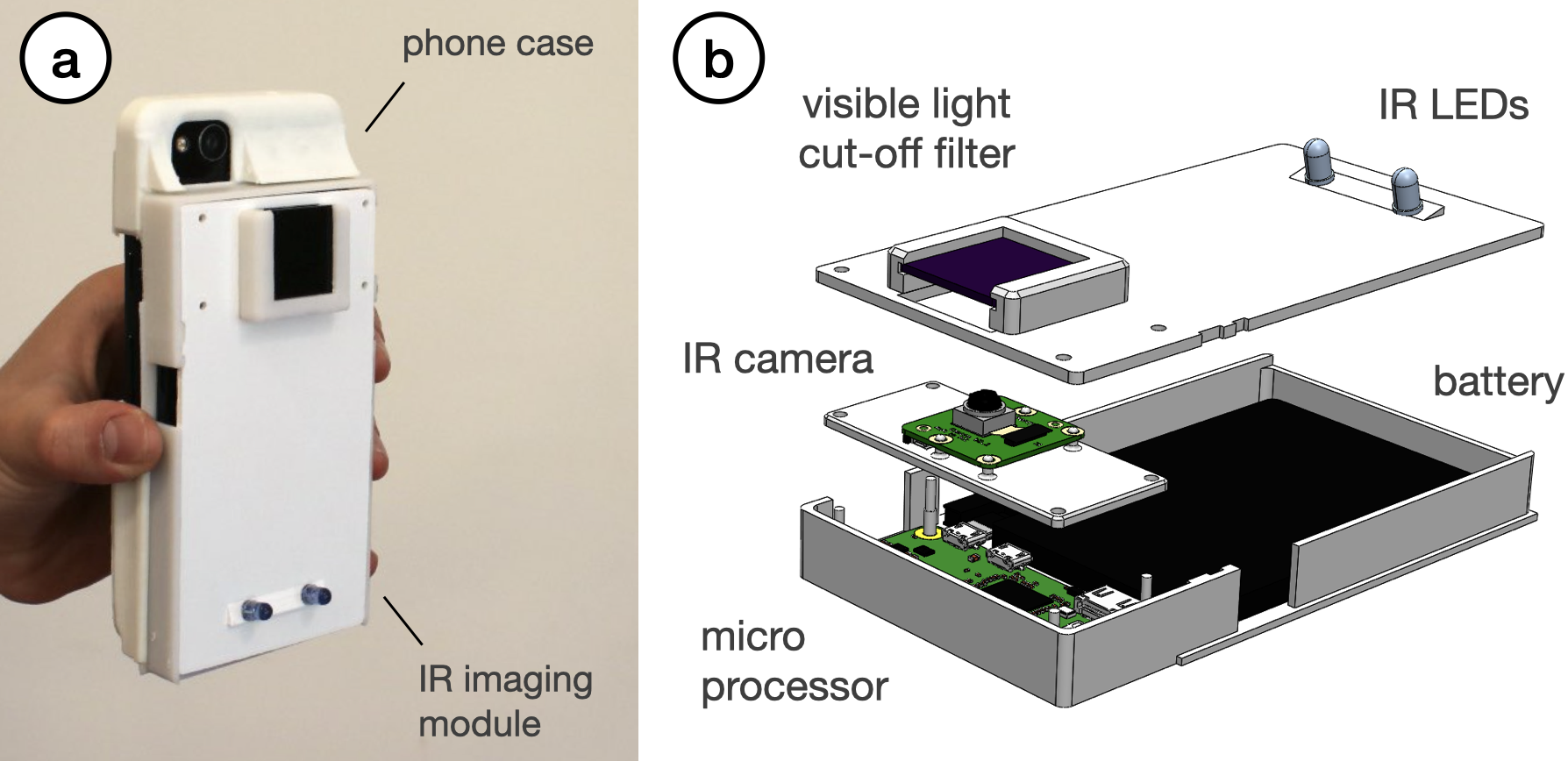}
  \caption{Infrared imaging module. (a) The module is attached onto a flexible case that can be 3D printed based on the user's mobile device. (b) The module's hardware components. }
  \label{fig:IRImagingModule}
\end{figure}

\vspace{5pt}
\noindent\textbf{Attaching the IR camera module:}  As shown in Figure~\ref{fig:IRImagingModule}, our add-on contains an infrared camera (model: \textit{Raspberry Pi NoIR}). This camera can see infrared light since it has the IR cut-off filter removed that normally blocks IR light in regular cameras. Additionally, to remove the noise from visible light and improve detection, we added a visible light cut-off filter\footnote{\scriptsize \url{https://www.edmundoptics.com/p/1quot-x-1quot-optical-cast-plastic-ir-longpass-filter/5421/}}, as well as 2 IR LEDs (940nm) which illuminate the object when it is dark.
This add-on has two 3D printed parts: a smartphone case from flexible TPU filament that can be reprinted based on the user's phone model, and the imaging module from rigid PLA filament that can be slid into this case. The imaging module has a \textit{Raspberry Pi Zero} board and a battery and weighs 132g.

\vspace{5pt}
\noindent\textbf{Detecting the Tag:} To detect the tag, users open the InfraredTags detection application on their mobile phones and point the camera to their object. The application shows the phone camera's view, which is what the user sees with their eyes instead of the IR view. 
This way, more information can be overlaid on top of the regular view for AR applications.
Under the hood, the imaging module continuously streams the images to our image processing server. If the server detects any tags, it sends the location and the encoded message to the smartphone app to show to the user.

\section{Software Implementation}

In this section, we explain how we implemented the code embedding user interface, as well as our infrared imaging module and image processing pipeline.

\subsection{UI Implementation}

Our embedding user interface is based on 
\textit{Rhinoceros 3D} CAD software\footnote{ \scriptsize \url{https://www.rhino3d.com/}} (referred to as \emph{Rhino}) and \emph{Grasshopper}\footnote{\scriptsize \url{https://www.grasshopper3d.com/}} which is a visual programming language that runs within \textit{Rhino}.

\vspace{5pt}
\noindent\textbf{Importing the Tag and the 3D Model:}
After the user loads an STL file representing the 3D object, our software converts it into a mesh utilizing a Python subprocess function call. The script then centers the mesh along its center of mass. When the user imports a tag as an SVG file, it creates a plane that contains the paths that represent its bits, i.e., the air gaps. While the user is positioning the code, our software always orients the plane of the code to face the mesh's surface. For this, it uses the normal on the mesh that is closest to the plane that holds the code.


\vspace{5pt}
\noindent\textbf{Embedding the Tag into the Object}:
Depending on the type of embedding selected (i.e., single-material or multi-material 3D printing), the tag is projected into the object in one of two ways:

\vspace{0.2cm}
\noindent
\textit{Single-Material:}
Our software first projects the tag onto the curved surface of the mesh and then translates it along the inverted closest mesh normal (i.e., pointing it towards the mesh) by the shell thickness ($t_{shell}$, see Table~\ref{tab:CodeThickness}). We then extrude the tag along the inverted normal by the code thickness ($t_{code}$), which creates a new mesh inside the object representing the air gaps inside the 3D geometry. To subtract the geometry that represents the air gaps from the overall geometry of the 3D object, we first invert the normals of the air gap mesh and then use a Boolean join mesh function to add the holes to the overall object geometry. This results in the completed mesh with the code, i.e., air gaps, embedded that the user can export as a single printable STL file.

\vspace{0.2cm}
\noindent
\textit{Multi-Material:} For multi-material prints, our software generates two meshes as illustrated in Figure~\ref{fig:ModelInterior}b: one for the tag (printed in regular PLA) and one for the shell (printed in IR PLA).
We start by following the steps described for the single-material approach, i.e., project the path representing the tag's bits onto the curved surface, translate it inwards, and extrude it to generate the tag mesh. Next, we find the bounding box of the this mesh, invert its normals, and join it with the main object's mesh. This creates a new mesh (i.e., the IR PLA shell), which once printed will have space inside where the regular PLA tag can sit.

\subsection{Mobile IR Imaging}
The mobile application used for capturing the tags is Web-based and has been developed using \textit{JavaScript}. It uses \textit{Socket.IO}\footnote{https://socket.io/} to communicate with a server that runs the image processing pipeline for tag detection explained in Section~\ref{ImageProcessing}.

The image processing server receives the images from the live stream shared by the microprocessor (\textit{Raspberry Pi Zero W}) on the imaging module and constantly runs the detection algorithm. If a tag is detected, the server sends the tag's location and the decoded message to the Web application, and shows it subsequently to the user. Because the imaging module does not use the resources of the user's personal device and is Web-based, it is platform-independent and can be used with different mobile devices.

\subsection{Image Processing Pipeline}
\label{ImageProcessing}

InfraredTags are identified from the images captured by the IR camera on the mobile phone or attached imaging module. Although the tags are visible in the captured images, they need further processing to increase the contrast to be robustly read. We use \textit{OpenCV}~\cite{bradski_opencv_2000} to perform these image processing steps as shown in Figure~\ref{fig:ImageProcessing}.

\begin{figure}[h]
  \centering
  \includegraphics[width=0.7\columnwidth]{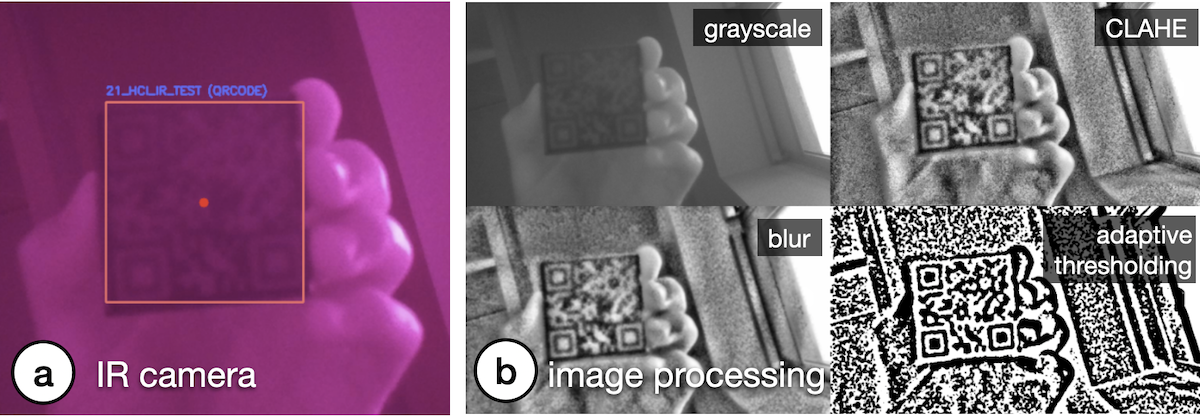}
  \caption{Image processing to read the tags. (a) Infrared camera view. (b)  Individual processing steps needed to decode the QR code message: \textit{"HCI\_IR\_TEST"}.}
  \label{fig:ImageProcessing}
\end{figure}

\vspace{0.2cm}
\noindent
\textbf{Pre-processing the Image:} We first convert the image to grayscale and apply a  contrast limited adaptive histogram equalization (CLAHE) filter \cite{pizer_adaptive_1987} to improve the local contrast (\texttt{clipLimit}=20, \texttt{tileGridSize}=(8,8)). 
For our pipeline, CLAHE is more appropriate than a standard histogram equalization as it redistributes the pixel intensity values based on distinct sections of the image~\cite{garcia-martin_vein_2020}.
To reduce the high-frequency noise that arises due to CLAHE, we smooth the image with a Gaussian blur filter. We then binarize the image using Gaussian adaptive thresholding to obtain black-and-white pixels that contain the code (\texttt{constantSubtracted}=4).

\vspace{0.2cm}
\noindent
\textbf{Code Extraction:} Once the binary image is generated, it can be used  to detect the respective code using existing libraries, such as \textit{ZBar}\footnote{\scriptsize  \url{http://zbar.sourceforge.net/}}. On average, it takes 6ms to decode a 4x4 ArUco marker and 14ms to decode a 21x21 QR code from a single original frame. The images we use as input are 
512x288 pixels; in the future, the detection could be made even faster by downsampling the image to a dimension optimal for both readability and speed.


\vspace{0.2cm}
\noindent
\textbf{The Effect of Tag Distance}:
The readability of the binarized tag depends on the parameters used for the pre-processing filters. More specifically, we found that a different Gaussian kernel for the blur (\texttt{ksize}) and block size for the adaptive threshold (\texttt{blockSize}) need to be used depending on the size of the tag in the captured image, i.e., the distance between the tag and the camera.
This is especially important for QR codes since they generally have more and smaller bits that need to be correctly segmented.


One strategy to increase detection accuracy is to iterate through different combinations of the filter parameters. To identify the effect of the number of filter parameter combinations on detection accuracy, we ran the following experiment: We captured 124 images of a 21x21 QR code from different distances (15-80cm from the camera). We then generated 200 different filter parameter combinations and used them separately to process the captured images. We then evaluated which filter parameters correctly binarized the QR code. We found that even with a small number of filter combinations, we can have sufficient detection results comparable to existing QR code detection algorithms. For instance, three different filter combinations (Table~\ref{tab:filter_combinations}) achieve an accuracy up to 79.03\% (existing QR code readers achieve <57\% for blurred tags\footnote{\scriptsize  Peter Abeles. 2019. Study of QR Code Scanning Performance in Different Environments. V3. \url{https://boofcv.org/index.php?title=Performance:QrCode}}). It is possible to further increase the number of filter parameter combinations to improve the accuracy further at the expense of detection time.



\begin{table}[]
\centering
\begin{tabular}{ll}
Filter combinations & Accuracy \\ \hline
\small (\texttt{ksize}=3, \texttt{blockSize}=23)    & 56.45\%  \\ \hline
\small (\texttt{ksize}=3, \texttt{blockSize}=23), \small(\texttt{ksize}=1, \small\texttt{blockSize}=37)   & 70.97\%   \\ \hline
\small(\texttt{ksize}=3, \texttt{blockSize}=23), \small(\texttt{ksize}=1, \texttt{blockSize}=37), \small(\texttt{ksize}=3, \texttt{blockSize}=21)  & 79.03\%   
\end{tabular}
\caption{Filter combinations and QR code detection accuracy}
\label{tab:filter_combinations}
\end{table}

\section{Applications}
We demonstrate how InfraredTags enable different use cases for interactions with objects and devices, storing data in them, and tracking them for sensing user input.

\subsection{Distant Augmented Reality (AR) Interactions with Physical Devices}
\label{Applications_Appliances}

InfraredTags can be embedded into physical devices and appliances to identify them individually through the embedded unique IDs and show the corresponding device controls that can be directly manipulated by the user.

\begin{figure*}[t]
  \centering
  \includegraphics[width=1\textwidth]{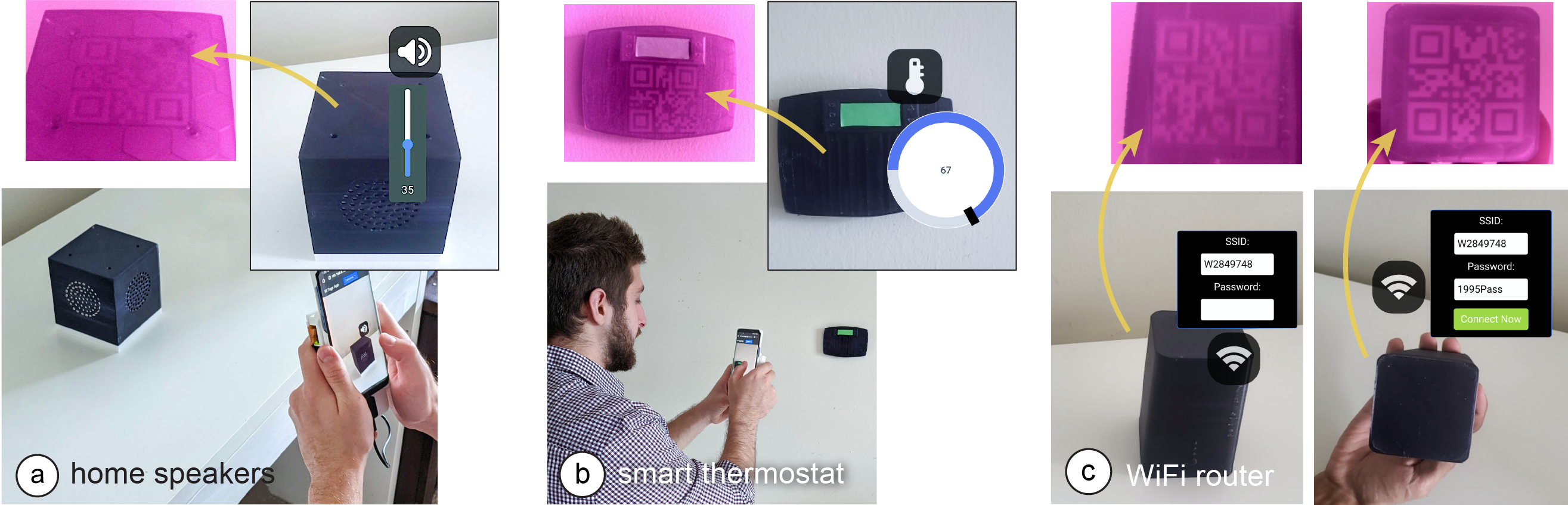}
  \caption{Controlling appliances using a mobile AR application. The user points at (a) the home  speakers to adjust its volume, and the (b) thermostat to adjust the temperature. The infrared camera in the phone's case identifies the appliances by reading the embedded QR codes. (c) Pairing a phone with a WiFi router, whose SSID is visible from all sides but the password is visible only from its bottom.}
  \label{fig:AppliancesAR}
\end{figure*}

In the application shown in Figure~\ref{fig:AppliancesAR}a and b, a user points their smartphone camera at the room and smart home appliances are identified through their InfraredTags, which are imperceptible to the human eye. A control menu is shown in the AR view, where the user can adjust the volume of the speaker or set a temperature for the thermostat.
InfraredTags could also allow multiple appliances of the same model (e.g., multiple smart speakers or lamps) in the room to be identified individually, which is not possible with standard computer vision-based object classification approaches.

\vspace{0.2cm}
\textit{Multiple tags on a single object for spatial awareness:} 
Furthermore, InfraredTags enable \textit{multiple tags} to be embedded in the same object. This enables different applications. For instance, when an object is partially occluded, multiple tags in the object can allow the capture of tags from different angles. Another application is to enable spatially aware AR controls where different settings appear at different locations within the same object. For example, Figure~\ref{fig:AppliancesAR}c illustrates how the front, side, and top faces of a WiFi router only have its network name (SSID) information, whereas its bottom also shows the password information, which can automatically pair the router to the phone. This enables quick pairing and authentication with devices without users having to type out complex character strings, while maintaining the physical use metaphors, such as the paper slip containing the password typically attached to the base of the router. 
While we demonstrate this application for mobile AR, InfraredTags could also enable lower friction, distant interactions with physical devices for head-mounted AR glasses.

\subsection{Embedding Metadata about Objects}
\label{Applications-Metadata}
\begin{figure}[h]
  \centering
  \includegraphics[width=0.6\columnwidth]{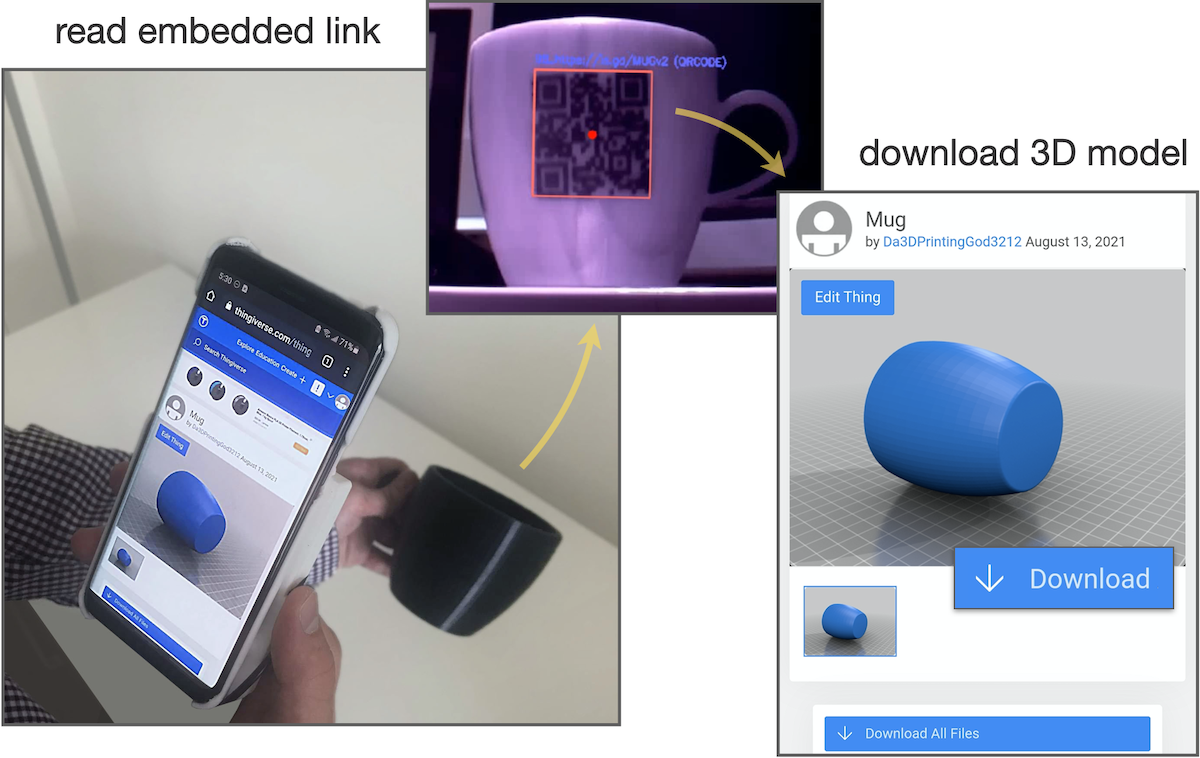}
  \caption{Embedded metadata about the object itself: The user is redirected to the \textit{Thingiverse} model that was used to fabricate the object.}
  \label{fig:FabricationURLMug}
\end{figure}

Spatially embedding metadata or documentation information within the object itself can provide richer contextualization and allow information sharing~\cite{ettehadi_documented_2021}. For example, we can embed the object's fabrication/origin link (e.g., a shortened \textit{Thingiverse} URL) as an InfraredTag  for users to look up in case they would like to get more information from its creator or 3D print it themselves as shown in
Figure~\ref{fig:FabricationURLMug}.
Other types of metadata that could be embedded include user manuals, expiry dates, date of fabrication, materials used to fabricate the object, weight, or size information.

\subsection{Tangible Interactions: Use Anything as a Game Controller}
\label{Applications-GameController}

Because fiducial markers can be embedded as InfraredTags, they can be used to track the object's movement. Thus, any passive object can be used as a controller that can be held by users when playing video games.

\begin{figure}[b]
  \centering
  \includegraphics[width=0.7\columnwidth]{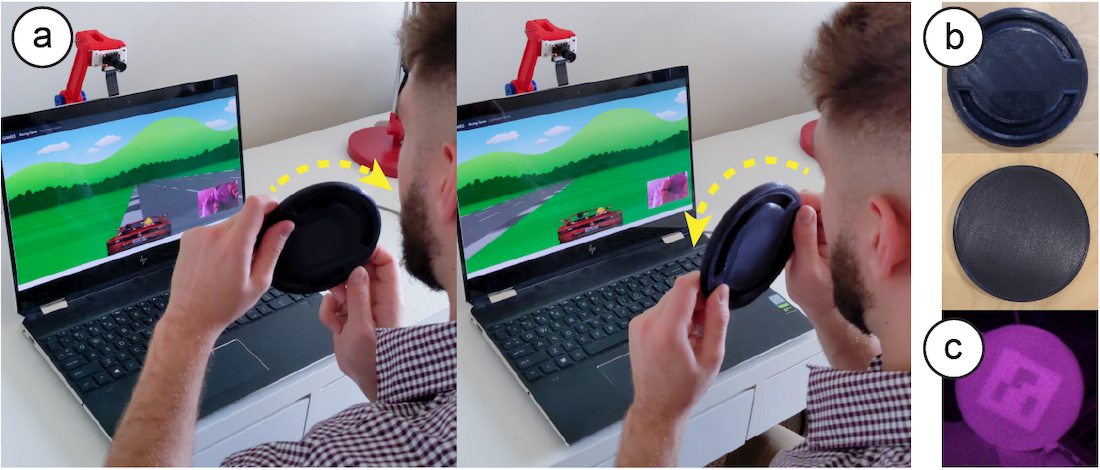}
  \caption{Using passive objects (a) as  a game controller. (b)~This wheel is black under visible light and has no electronic components. (c)~The fiducial marker embedded inside is only visible to an infrared camera.}
  \label{fig:CarGame}
\end{figure}

Figure~\ref{fig:CarGame} shows a 3D printed wheel with no electronics, being used as a game controller. The wheel contains an ArUco marker InfraredTag which is used to track the wheel's location and orientation. Even though the wheel is rotationally symmetric, the infrared camera can see the square marker inside and infer the wheel's position and orientation. Our method does not require any electronics as opposed to other approaches~\cite{yamaoka_foldtronics_2019}.

While we demonstrate an application where the user faces a screen with a camera behind it, this could be used to enable passive objects to serve as controllers for AR/VR headsets with egocentric cameras.
Such an application scenario could be particularly suitable for headsets like \textit{HoloLens 2}, which comes with an integrated infrared camera~\cite{ungureanu_hololens_2020} that could be utilized for InfraredTag detection in the future. Even though the tag would be facing the user, it would not be visible to the user but can still be identified by the headset.

\section{Evaluation of the Detection}
\label{TechnicalEvaluation}

In this section, we evaluate how InfraredTag detection is affected by fabrication- and environment-related factors.

\vspace{0.2cm}
\textbf{Marker size}: By following a test procedure similar to the one shown in Figure~\ref{fig:ShellThicknessValues}, we determined that the smallest detectable 4x4 ArUco marker printable is 9mm wide for single-material prints and 6mm wide for multi-material prints.
The resolution for multi-material prints is better than single-material ones because the large transmission difference between the two distinct materials makes it easier for the image sensor to resolve the border between the marker bits.
On the other hand, in single-material prints, the luminosity of an air gap resembles a 2D Gaussian distribution, i.e., the intensity gets higher towards the center. Thus, larger bits are needed to discern the borders between a single-material marker's bits.

\begin{figure}[h]
  \centering
  \includegraphics[width=0.67\columnwidth]{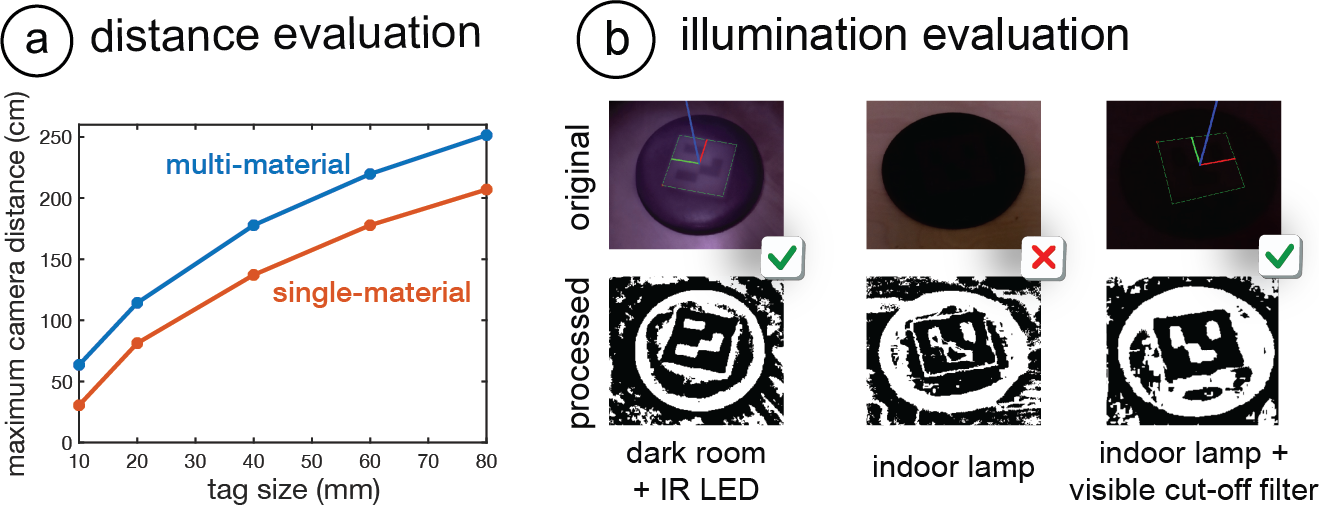}
  \caption{Detection evaluation. (a) Maximum detection distance for single- and multi-material ArUco markers. (b) Cases where the IR LED and the visible cut-off filter improve detection.}
  \label{fig:DetectionEvaluation}
\end{figure}

\vspace{0.2cm}
\textbf{Distance}: To test the limits of our detection method, we measured the maximum distance tags of different sizes can be detected.
This was done for both single-material (IR PLA) and multi-material (IR PLA + regular black PLA) prints.
The marker size range we evaluated was 10-80mm for 4x4 ArUco markers, which would translate to a range for 42-336mm for 21x21 QR codes (can store up to 25 numeric characters).
The results are given in Figure~\ref{fig:DetectionEvaluation}a, which shows that multi-material codes can be detected from further away than single single-material ones. The results are given for the filter parameters with the best detection outcome (Section~\ref{ImageProcessing}). 

\vspace{0.2cm}
\textbf{Lighting conditions:} 
For InfraredTags to be discernible in NIR camera images, there has to be enough NIR illumination in the scene. We measured the minimum NIR intensity needed to detect 4x4 ArUco markers using a lux meter which had a visible light cut-off filter (720nm) attached.
We found that just a tiny amount of NIR is sufficient for this, i.e., that at least 1.1 lux is needed for single-material prints, and 0.2 lux for multi-material prints.

Because sunlight also contains NIR wavelengths, the tags are detectable outdoors and also in an indoor areas that have windows during daytime.
We also noticed that many lamps used for indoor lighting emit enough NIR to detect the codes at nighttime (e.g., 1.5 lux in our office).
Furthermore, the IR LEDs on our imaging module (Section~\ref{DetectionInterfaceIRModule}) provide high enough intensity to sufficiently illuminate multi-material markers even in complete darkness (Figure~\ref{fig:DetectionEvaluation}b). In the future, brighter LEDs can be added to support single-material prints in such difficult detection scenarios. 

The visible light-cut off filter used on our IR imaging module also improves detection in spite of challenging lighting conditions. For instance, the last two columns in Figure~\ref{fig:DetectionEvaluation}b shows how certain print artifacts on the object's surface might create noise in the IR camera image, which is reduced when the cut-off filter is added. This is particularly helpful for single-material prints, which are more challenging to identify.

\section{Discussion}
In this section, we discuss the limitations of our approach and how it could be further developed in future research.

\vspace{5pt}
\noindent\textbf{Print Resolution:}
In this project, we used FDM printers, whose printing resolution is restricted by the size of its nozzle that extrudes the material, and a low-cost camera that has an 8MP resolution.
In the future, even smaller InfraredTags can be fabricated by applying our method to printing technologies with higher resolution, such as stereolithography (SLA).
Correspondingly, higher-resolution cameras with better aperture can be used to identify these smaller details (e.g., \textit{Samsung}’s latest 200MP smartphone camera sensor~\cite{samsung_isocell_2021}).
This would allow embedding more information in the same area.
 

\vspace{5pt}
\noindent\textbf{Discoverability vs. Unobtrusiveness:}
For InfraredTags to be detected, the user should orient the near-infrared camera such that the embedded marker is in the frame.
However, similar to related projects such as \textit{AirCode}~\cite{li_aircode_2017} and \textit{InfraStructs}~\cite{willis_infrastructs_2013}, this might be challenging since the marker is invisible to users and thus they might not know where exactly on the object to point the camera at.
For objects with flat surfaces, this can be compensated for by embedding a marker on each face (e.g., on the six faces of a cube). This way, at least one marker will always be visible to the camera. Similar to how a QR code printed on a sheet of paper is detectable from different angles, the flat InfraredTag will maintain its shape when viewed from different angles (e.g., the router Section~\ref{Applications_Appliances}c).

However, detection of codes on curved objects poses a bigger challenge. This is because a 2D code projected onto a curved surface (e.g., the mug in Section~\ref{Applications-Metadata}) has a warped outline when viewed from an angle far away from its center. 
As a solution, we plan to pad the whole object surface with the same code, similar to \textit{ChArUco} (a combination of ArUcos and chessboard patterns)~\cite{hu_deep_2019}, so that at least one of the codes appears undistorted in the captured image.
Also, for curved objects, other tag types that are more robust to deformations could be used~\cite{yaldiz_deepformabletag_2021} in the future.
Alternatively, a small visible or tactile marker in the form of a notch could be added to the surface of the object (corresponding to where the code is embedded) to help guide the user to the marker.

\vspace{5pt}
\noindent\textbf{Other Color and Materials:}
While we only used black IR PLA in this project, manufacturers could produce filaments of other colors that have similar transmission characteristics to create more customized or multi-material prints in rigid or flexible forms~\cite{forman_defextiles_2020}.
In the future, we also plan to combine the IR PLA filament with IR retro-reflective printing filaments to increase the marker contrast even more.

\section{Conclusion}
In this chapter, we presented InfraredTags, a low-cost method to integrate commonly used 2D tags into 3D objects by using infrared transmitting filaments. We explained how this filament can be used by adding air gaps inside the object or by combining it with regular, opaque filaments, which increases the tag contrast even more. We discussed what kind of camera is appropriate for detecting InfraredTags and what kind of code geometry should be used for best detection, while ensuring unobtrusiveness to the naked eye. After introducing our tag embedding user interface and mobile infrared imaging module, we presented a wide range of applications for identifying devices and interacting with them in AR, storing information in physical objects, and tracking them for interactive, tangible games. Finally, we evaluated our method in terms of marker size, detection distance, and lighting conditions.

\chapter{BrightMarkers: 3D Printed Fluorescent Markers for Object Tracking}
\label{thesis-BrightMarkers}




Existing methods for invisible object tagging and \textit{tracking} have several limitations that impede their widespread adoption. One significant limitation is their low signal-to-noise (SNR) ratio, i.e., the poor resolution and clarity of the imaged marker. 
Because invisible markers are embedded in the interior of objects, the markers are imaged from a weak signal, which needs to be amplified by optical and digital processes~\cite{li_aircode_2017, willis_infrastructs_2013}. These processes result in long capture or decoding times, typically ranging from seconds to minutes per frame.
This deteriorates further when objects are in motion, i.e., the captured markers appear blurrier and are thus unidentified in most frames.
In addition, 
existing invisible tags are often limited in terms of the variety of object colors they can be used with.
For example, \textit{InfraredTags}~\cite{dogan_infraredtags_2022} embedded 3D~printed codes in black objects, and \textit{AirCode}~\cite{li_aircode_2017} in white objects.

To address these limitations, we present BrightMarker, a fabrication method for passive invisible tags using fluorescent filaments that emit light in a specific near-infrared (NIR) wavelength, which NIR cameras can robustly detect. 
By isolating the markers from the rest of the scene using the matching filter, we are able to robustly \textit{track} markers even when objects are in motion.
Our work builds on \textit{InfraredTags} and addresses the limitations in regard to marker resolution 
and object colors by enhancing detection.
We were inspired by the motion capture system \textit{OptiTrack}, which uses passive retro-reflective markers that reflect the shined IR to the camera.


We demonstrate the potential of BrightMarker
by showcasing various applications, including product tracking on conveyor belts, flexible wearables for motion capture, 
tangible and haptic interfaces for AR/VR, and privacy-preserving night vision. 
Our technical evaluation shows that the markers embedded in a variety of surface colors can be detected robustly and in real time as they move. 

\begin{figure}[t]
  \centering
  \includegraphics[width=1\linewidth]{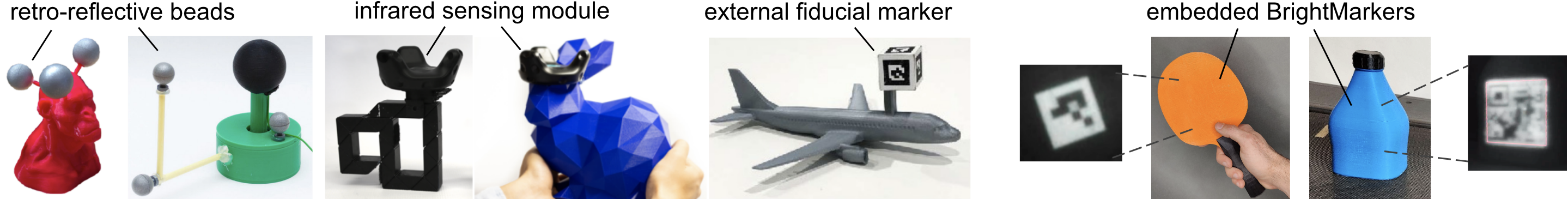}
  \caption{Ways to add tracking capabilities to objects as used in previous HCI projects. External tags rely on adding retro-reflective beads~\cite{schon_trackitpipe_2022, feick_designing_2022}, sensor modules~\cite{zhu_haptwist_2019, feick_tangi_2020}, or fiducial markers~\cite{shi_designing_2019}, but may result in bulky or visually obtrusive objects.
  To address this, BrightMarker (rightmost) embeds high-contrast markers using a fluorescent material.
  }
  \label{fig:BrightMarker-RelatedWork}
\end{figure}

We believe that BrightMarker is a promising approach that could significantly improve the performance and versatility of invisible object tagging and have a wide range of potential HCI applications.
The ability to track objects without the markers having a bulky form factor or being in the user's direct view allows for a more natural and immersive experience for users, and opens up new possibilities for tracking objects in real-life scenarios where visible markers would be impractical and obtrusive (Figure~\ref{fig:BrightMarker-RelatedWork}).
We note that this chapter was originally published at \textit{ACM UIST 2023}~\cite{dogan_brightmarker_2023}. 

\vspace{0.2cm}
In summary, we make the following contributions:

\begin{itemize}[leftmargin=0.5cm, noitemsep, topsep=2pt]
    \item  A novel method that uses fluorescent filaments and imaging to integrate invisible and passive tracking markers into 3D printed objects of different colors.
    \item A software interface that uniformly distributes markers onto 3D models based on their geometry.
    \item A detection hardware module for real-time fluorescence tracking that can be attached to existing AR/VR headsets. 
    \item An image processing pipeline that allows real-time tracking of the integrated markers.
    \item The evaluation of the design space, including object colors, under various distance and lighting conditions.
    \item A demonstration of the potential applications of our approach, including custom-fabricated motion capture wearables, AR/VR tangibles, and privacy-preserving night vision.
\end{itemize}

\section{Method: High-Contrast Markers Using Fluorescent Filaments}

BrightMarker has the benefit of using fluorescent filaments, which allow us to achieve high marker contrast in the filtered infrared images.
Fluorescence is the physical phenomenon in which a material \textit{absorbs} light at one wavelength and \textit{emits} it at a longer wavelength~\cite{guilbault_practical_1990}. This occurs when a molecule absorbs the energy of light and temporarily enters an excited state, before releasing the energy as a lower-energy photon. 
The difference between these two wavelengths is called
Stokes shift~\cite{rost_fluorescence_1992}.
By shifting the illumination from shorter wavelengths to longer ones, the fluorescent color can appear more saturated than it would by reflection alone, which enhances its detectability~\cite{schieber_modeling_2001}.
For this reason, fluorescent materials are particularly useful for imaging and sensing applications.

In this work, we utilize fluorescence as a material property to create objects with easily trackable markers via multi-material 3D printing. The key material we use is a fluorescent 3D printing filament which contains uniformly distributed fluorescent dye. The filament emits light in a specific wavelength when excited by an IR light source. By using an optical filter of the appropriate wavelength for the camera, we are able to capture exclusively the light emitted by the filament and thus the parts fabricated from it.

\subsection{Fluorescent Filament}
\label{fluorescence}
\begin{figure}[t]
  \centering
  \includegraphics[width=1\linewidth]{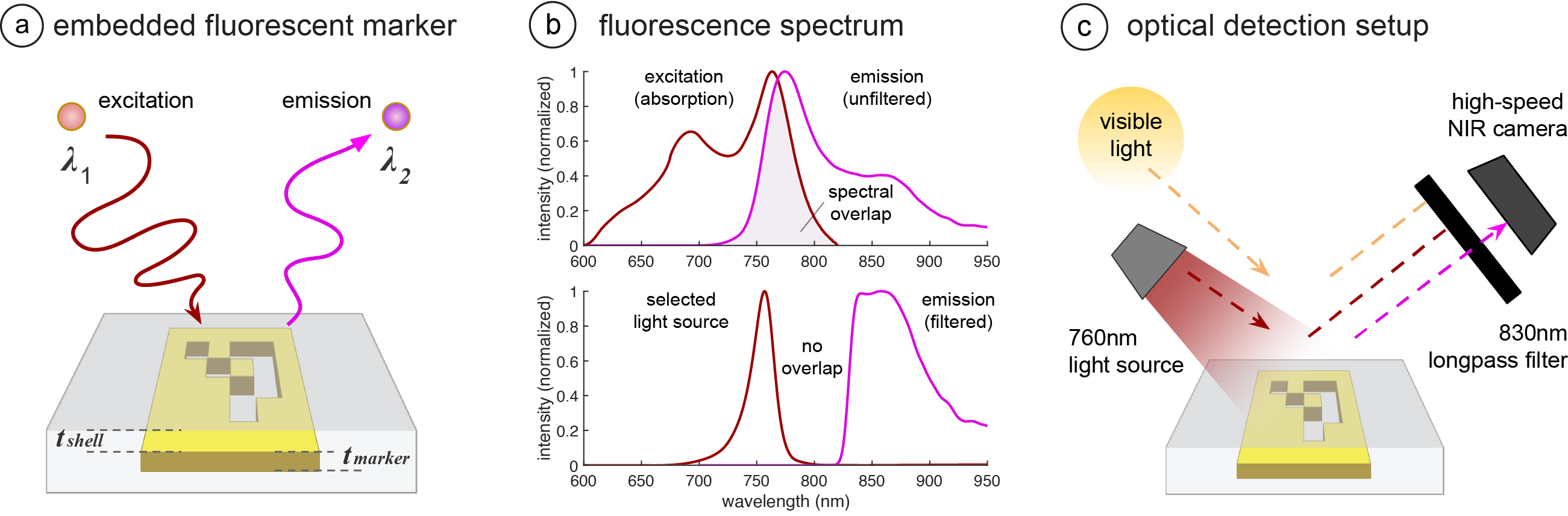}
  \caption{Fluorescence and our imaging approach. (a) BrightMarker embeds tracking markers with fluorescent filaments, which "shift" the wavelength of IR radiation. (b) Although the excitation and emission spectra of the filament overlap (top), they can be separated in practice using optical tools (bottom). (c) Our imaging setup filters for the marker's fluorescence.}
  \label{fig:fluorescence}
\end{figure}

We utilize an ABS (acrylonitrile butadiene styrene) filament developed by \textit{DIC Corporation}~\cite{dic_corporation_dic_2023} and Silapasuphakornwong et al.~\cite{silapasuphakornwong_technique_2019}. This filament contains a NIR-fluorescent dye, which reacts to the NIR light source by fluorescing as explained earlier. The fluoresced light can penetrate the exterior material of the object and thus be captured from the outside (Figure~\ref{fig:fluorescence}a).

\vspace{0.2cm} \noindent
\textbf{Fluorescence behavior}: 
To understand the fluorescence characteristics of this material and optimize our system for it, we measured its excitation and reaction using a fluorimeter 
(\textit{HORIBA Jobin Yvon Fluorolog-3}).
The top graph in Figure~\ref{fig:fluorescence}b shows both the \textit{excitation} (absorption, plotted in red) and \textit{emission} (fluorescence, plotted in purple) spectra of the material.
As can be seen in these unfiltered spectra, the emitted fluorescence has a longer wavelength than the absorbed light: While the material is most excited at wavelength 763nm, the peak of the emitted light is at 775nm.
This Stokes shift of 12nm between the two spectra allows us to separate the excitation and emission signals for the IR image capture.
However, due to the spectral overlap highlighted in the graph, the excitation and emission signals first need to be filtered using optical methods, which we explain in the imaging section.

\vspace{0.2cm} \noindent
\textbf{Multi-material 3D printing}: 
The NIR-fluorescent filament is used in one of the print heads of a multi-material FDM printer. We used \textit{Ultimaker 3} and \textit{S5} for our prints.
The printing method builds on \textit{InfraredTags}' multi-material printing approach in terms of the high-level CAD 
 modifications. When embedding a BrightMarker, however, the main object geometry is printed using the filament of the user's choice, while the fluorescent filament is used to print the marker placed in the interior of the object (Figure~\ref{fig:fluorescence}a).
Because the fluorescent filament's main polymer is ABS, it can be printed at standard ABS printing temperatures ($\sim$250$^{\circ}$C).

As shown in the cross-section in Figure~\ref{fig:fluorescence}a, we denote the thickness of the marker by $t_{marker}$, and the shell between the marker and the object surface by $t_{shell}$.
If $t_{marker}$ is too small, the fluorescence will be weak for robust marker capture. We found that one printed layer of the filament, i.e., $t_{marker}$= 0.15mm, is sufficient for our applications.
On the other hand, the value of $t_{shell}$ depends on the material used for the main object geometry, which we explain next.

\vspace{0.2cm} \noindent
\textbf{Main object material and color}:
Compared to \textit{InfraredTags}~\cite{dogan_infraredtags_2022} which only allows embedding into black objects, BrightMarker is compatible with multiple color options for the main object geometry (including object surface). Due to the high intensity of the fluoresced light, the emitted light can penetrate the shell above the fluorescent marker and reach the camera's image sensor.

We observed that the fluorescent marker can be combined with both PLA and ABS materials. This makes it possible to have a variety of colors for our fabricated objects. Since the fluorescent filament is ABS, we strive to use ABS for the main object geometry when possible, as it ensures similar printing parameters (e.g., temperature) among the two parts.
The only ABS color we determined that does not pass the fluoresced light is black. This is because the carbon black used in the conventional filament absorbs most wavelengths~\cite{bond_light_2006}.
Thus, for producing black objects, we use the IR-PLA filament~\cite{3dkberlin_pla_2021} used in \textit{InfraredTags}, which passes IR light.

\begin{figure*}[t]
  \centering
  \includegraphics[width=0.99\linewidth]{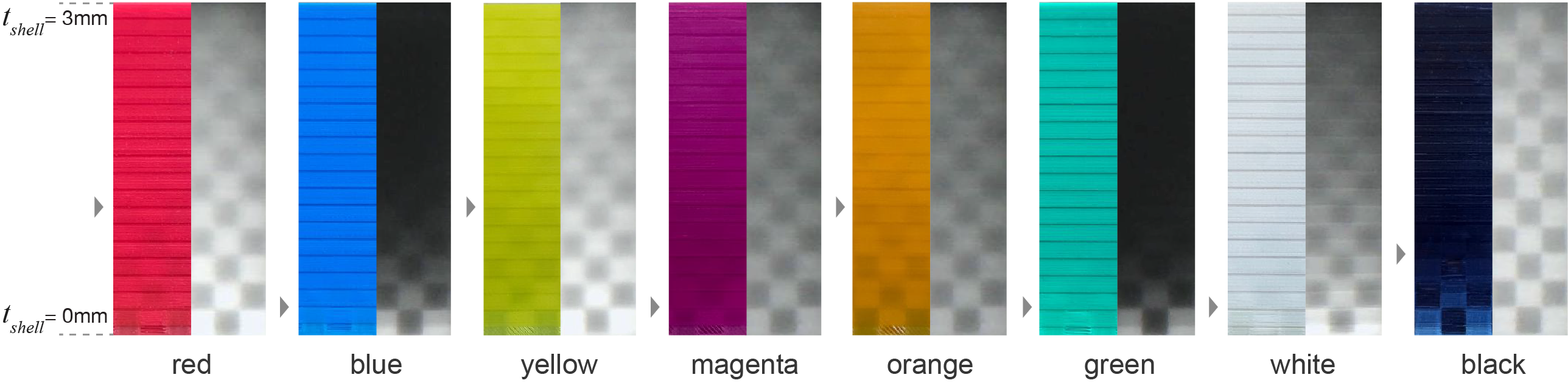}
  \caption{The design space of possible material colors. Each test slab was printed increasing shell thickness. The left image of each pair shows the visible camera capture, while the right image shows the NIR capture.}
  \label{fig:MaterialColor}
\end{figure*}

Figure~\ref{fig:MaterialColor} shows the design space of material colors. As shown in the NIR camera images, a fluorescent checkerboard pattern is embedded in each test slab to determine $t_{shell}$ for the different colors. The pattern is covered with increasing shell thickness, from $t_{shell}$= 0mm to $t_{shell}$= 3mm, thus the pattern contrast in the camera images gradually decreases.
We choose $t_{shell}$ such that at this value, the pattern is no longer in the visible camera image, i.e., its contrast is less than 5\%~\cite{bijl_visibility_1989}.
Further, if the shell is too thin, this can result in a poor surface finish, as observed in the lower end of the slabs.

We found that colors closer to infrared, i.e., red, yellow, and orange, tend to be less opaque than other colors due to their dye composition. Thus, we use $t_{shell}$ = 1.2mm for these colors.
For other colors printed with ABS, we use $t_{shell}$=0.3mm
For the IR-PLA (rightmost), we use $t_{shell}$= 0.8mm, which has a tinted appearance due to its translucent property.
These values are marked in the figure with a small triangle to the left of the slabs.
Larger thicknesses are also possible, but to ensure the detectability of the markers in IR, we recommend using values that maintain the bit binarization accuracy above 90\%~\cite{dogan_infraredtags_2022}.

We note that these values also depend on the specific filament vendor and that variations might exist even among different batches from the same vendor. Thus, users are recommended to determine the optimal values for their filaments by first printing test slabs as illustrated in this section.



\subsection{Infrared Imaging of Fluorescence}
\label{Imaging}
When coupled with wavelength-specific optical components, the light emitted by the fluorescent filament results in high contrast between the marker and the rest of the IR image. This allows us to robustly detect and track the objects in which the markers are embedded.

Our imaging system consists of three key components: (1) a light source for exciting the marker, (2) an optical filter for isolating marker fluorescence, and (3) a high-speed infrared camera.
We explain the role of these components in the next sections.

\vspace{0.2cm} \noindent
\textbf{Light source for excitation}:
Excitation of the NIR fluorescent material within the 3D printed object is necessary to read the embedded pattern.
There are two considerations that an appropriate light source should meet for this purpose.
First, the light emitted by the light source should be invisible to the user, i.e., there should be no emission below 700nm.
Second, the light should excite the fluorescent material as much as possible. To achieve this, the light source should have high power, and its peak wavelength should be as close to the material's peak excitation wavelength (i.e., 763nm as explained in Section~\ref{fluorescence}) as possible.

To satisfy these criteria in practice, we use LEDs that peak at 760nm and deliver high power.
The bottom graph in Figure~\ref{fig:fluorescence}b shows the spectrum of the LED used in our high-speed imaging module (Section~\ref{DetectionImagingModule}), marked red and labeled "selected light source."
We note that the intensity reaches zero at approximately 800nm.

\vspace{0.2cm} \noindent
\textbf{Optical filter for marker isolation}: Through a wavelength-specific filter, we can enhance the recognition of fluorescent markers by minimizing interference from other wavelengths.

Due to the natural characteristics of fluorescence, there is an overlap between the higher wavelength end of the excitation spectrum and the lower wavelength end of the emission spectrum.
This overlap, which is shown in the top graph of Figure~\ref{fig:fluorescence}b, must be eliminated to avoid overwhelming the weaker emitted fluorescence light with the brighter excitation light, which would significantly reduce marker contrast.

To separate these signals, 
we use a longpass filter with a cut-on wavelength of 830nm, which blocks any wavelength under this value from entering the camera.
As shown in Figure~\ref{fig:fluorescence}c, it blocks the excitation light emitted by the LED (marked red) and the visible environmental light (marked yellow) from reaching the camera.
The only wavelength range that can enter the camera corresponds to the fluorescence from the marker itself (marked purple).
Therefore, as shown in the top graph of Figure~\ref{fig:fluorescence}b, there is no longer an overlap between the excitation and emission spectra: Our selected light source cuts the emission at $\sim$810nm, while the filtered fluorescence that enters the camera starts at $\sim$820nm. 
This separation allows the high intensity of the markers in captured images.

\vspace{0.2cm} \noindent
\textbf{High-speed NIR camera}:
A high-speed camera allows the observation of the fluorescence emitted 
from the moving objects.

To capture images at a high frame rate and minimize motion blur, we use a 60-frames-per-second (fps) monochrome camera with a global shutter rather than a rolling shutter. Rolling shutters, which are commonly used in consumer-grade cameras, scan the image sensor line by line from top to bottom, resulting in a time delay between the capture of each row~\cite{grau_motion_2012}. This can lead to motion blur in fast-moving scenes, which makes it difficult to detect the BrightMarker accurately.

In contrast, a global shutter captures the entire image simultaneously, resulting in no time delay between rows. This is crucial for real-time tracking 
where every frame counts. Additionally, a monochrome camera achieves higher spatial resolution than color cameras, and eliminates the need for a color filter array, which reduces the amount of light reaching the image sensor and decreases the camera's sensitivity to NIR fluorescence~\cite{weber_color_2005}.

Together, these properties allow us to create a real-time video representation of the object with high temporal resolution.

\section{Fabricating, Capturing, and Tracking BrightMarkers}
In this section, we first describe the workflow in which BrightMarkers are added to objects for fabrication. We then explain the hardware and software we use for their detection once printed.

\subsection{Adding Markers for Fabrication}
\begin{figure}
  \centering
  \includegraphics[width=1\linewidth]{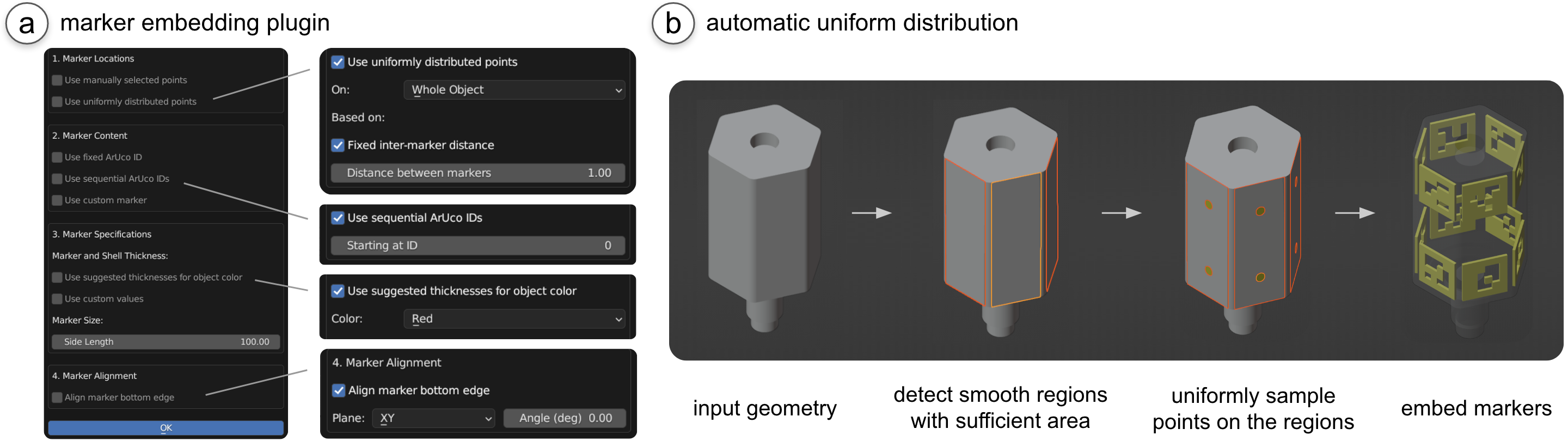}
  \vspace{-0.2cm}
  \caption{Marker embedding process. (a) Our tool allows users to (b) uniformly distribute markers based on the object geometry.}
  \label{fig:UserInterfacePlugin}
\end{figure}

Users start by importing a 3D model into the CAD editor (\textit{Blender}). Next, the user specifies how the fluorescent marker(s) should be implanted using our plugin, which offers a variety of options as shown in Figure~\ref{fig:UserInterfacePlugin}a.

\vspace{0.2cm} \noindent
\textit{Determining marker locations}: 
In tracking applications, it is important to pad the object surface with numerous markers, rather than a single one, to ensure the object can be tracked independently of the object orientation. 
The user can either manually pick exact target points on the object for the marker embedding, or use an automatic distribution mode. This mode uniformly distributes points on the object geometry based on a fixed inter-marker distance or a total number of desired markers.
The user can specify whether the target points should be distributed on the whole object, the areas other than the base, or a specific area selected by the user.

\vspace{0.2cm} \noindent
\textit{Determining marker content}: 
The plugin allows users to embed ArUco markers with the same ID ("Use fixed ArUco ID") or increasing IDs ("Use sequential ArUco IDs") based on the desired use case. Unique sequential IDs can help identify an object's location and orientation.
It is also possible to load a custom marker, such as a QR code or a Micro QR code, by selecting an image file.

\vspace{0.2cm} \noindent
\textit{Determining marker depth and dimensions}:
Next, the user specifies how deep the marker should be embedded under the object's surface. For this, the user can enter a custom value or simply select the desired object color from a dropdown, after which the plugin automatically retrieves the corresponding $t_{shell}$ value from Section~\ref{fluorescence}.
Next the user can specify the dimensions of the marker. The marker thickness is set to the recommended $t_{marker}$ value by default, and the user can specify a desired side length of the marker.

\vspace{0.2cm}
Finally, the user can specify whether the marker's bottom edge should be parallel to a certain plane. This helps users to align codes in the 3D printing direction, which can lower print time and avoid print failures that might occur due to frequent nozzle switching.
After setting up the parameters, users can start the embedding by clicking "OK". Once the markers are embedded, users can export the resulting STL files for 3D printing.

\vspace{0.2cm} \noindent
\textit{3D printing}:
The user imports the resulting STL files into the slicer software of the printer. For the popular slicer \textit{Cura}\footnote{\url{https://ultimaker.com/software/ultimaker-cura}}, we recommend disabling the "combing" parameter 
to ensure that the printer does not move over the already extruded pieces of fluorescent filament. This prevents the fluorescent filament from getting mixed into the outer surface layers and thus becoming visible.



\vspace{0.2cm} \noindent
\textbf{Implementation}:
The automatic uniform distribution of the markers is created by first determining the areas on the model that are large enough and close to being flat to accommodate a marker of the user's specified size. This is done by comparing the angles between the normals of adjacent faces with the method {\small \texttt{faces\_select\_ linked\_flat()}}, grouping the ones with an angle difference smaller than 0.1~rad, and ensuring that the contained surface area is greater than that of the marker. This threshold can be adjusted based on the user's preference and application if desired.
As shown in Figure~\ref{fig:UserInterfacePlugin}b for the lightsaber model from Section~\ref{applicationsARVR}, this gives the subregions on the object geometry that are not too curvy or sharp.

Next, to uniformly sample points, each subregion is temporarily duplicated, flattened, and its shape is approximated to a high degree of accuracy in an array of 1s and 0s created with repeated calls of the function 
{\small \texttt{ray\_cast()}}.
The array is analyzed row by row, choosing rows that are a distance of half the marker side length apart. Points in each row are picked if they are sufficiently far from each other and the edges of the subregion.
Finally, using the specifications set by the user, a knife projection is used to project the marker so it has the same curvature as the object at the sampled point. The projected marker is extruded to form a 3D model, which is exported separately.

\subsection{Detection Using Imaging Modules}
\label{DetectionImagingModule}

\begin{figure}[h]
  \centering
  \includegraphics[width=1\linewidth]{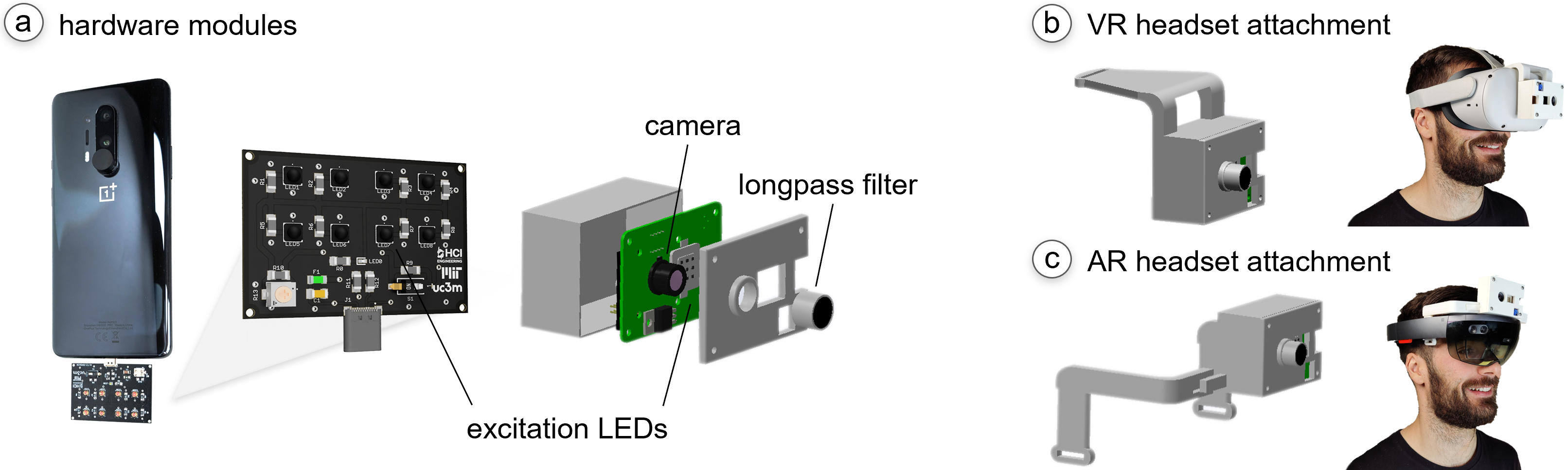}
  \caption{Hardware modules for tracking BrightMarkers.}
  \label{fig:HardwareModule}
\end{figure}

As shown in Figure~\ref{fig:HardwareModule}a, we built two hardware modules that fulfill the requirements of the fluorescence imaging principles explained in Section~\ref{Imaging}.

\vspace{0.2cm} \noindent
\textit{Smartphone attachment}: 
The module can be directly plugged into existing phones using a USB-C connection. It includes eight LEDs that correspond to the excitation wavelength of our filament. To ensure objects fit in the phone's field of view, half of the LEDs have a viewing angle of 60$^{\circ}$  and the other half 120$^{\circ}$  (\textit{Stanley 
FWR1107MS-TR} and \textit{FWR1108MS-TR}, \$2.5 each).
The attachment weighs 47 grams, and consumes 1.6W during operation.
We use it with a \textit{OnePlus 8 Pro} that comes with an embedded NIR camera (\textit{GalaxyCore GC5035}\footnote{\url{https://en.gcoreinc.com/products/index?cid=2&subcid=5}}), which we coupled with the appropriate longpass filter for fluorescence (\textit{MaxMax}, \$30).

\vspace{0.2cm} \noindent
\textit{Stand-alone module}: 
The module is built for high-speed fluorescence imaging and can be attached to existing AR/VR headsets.
It consists of a 60-fps NIR camera (\textit{ArduCam} with \textit{OV2311}\footnote{\url{https://www.ovt.com/products/ov2311/}}, \$98) coupled with the above-mentioned longpass filter, a 10W LED grid 
(\textit{Nagalugu}, \$18), and a small battery (generic 9V, \$2). 
A custom PCB utilizes a pulse-width modulated (PWM) signal to deliver battery power to the LEDs. It uses a timer IC (LM555) and an n-channel MOSFET to modulate the LED power via a constant current regulator (LM317), which can be adjusted using a potentiometer.
In total, the module costs \$148 and weighs 118g.
For prototyping purposes, we did not include a processing component on the module itself but connected its camera to an external computer using a cable. In the future, the module can be augmented with a small microprocessor to run the compute on-the-go, considering that our approach does not use compute-heavy ML models.

To be able to attach and detach the stand-alone module to existing AR/VR headsets, we designed and 3D printed an enclosure that encompasses all the parts. Figure~\ref{fig:HardwareModule}b,c shows how the module is mounted on the headsets.

\subsection{Image Processing for Marker Detection}
\label{ImageProcessing}
We developed an image processing pipeline to detect the markers in real time. 
The process of \textit{tracking} and \textit{decoding} the markers is shown in Figure~\ref{fig:ImageProcessing}, and is explained below.

\begin{figure}[h]
  \centering
  \includegraphics[width=0.6\linewidth]{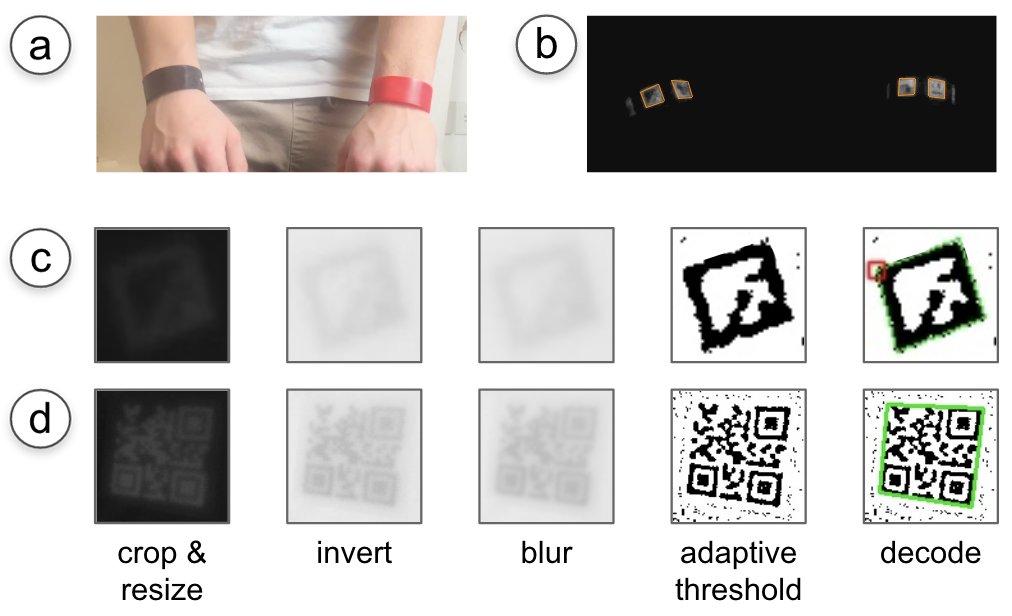}
  \caption{BrightMarker's image processing pipeline. 
  (a) The objects are tracked using (b) the outlines in IR capture. (c,~d)~The localized markers are decoded using a set of filters.}
  \label{fig:ImageProcessing}
\end{figure}

\vspace{0.2cm} \noindent
\textit{Tracking the marker}: 
To keep the detection script as lightweight as possible for fast tracking, we intend to apply a small number of processing steps on the main image frame when localizing the markers.
Thus, the grayscale input image is simply binarized using Otsu's thresholding method.
The binarization results in multiple identified contours.
We then approximate each contour as a polygon. The polygons with four sides are our target markers (Figure~\ref{fig:ImageProcessing}b).

This approach keeps track of all markers irrespective of the data encoded in them (i.e., ID/string). This keeps the detection lightweight if the encoded data is not relevant to the application, e.g., when all markers are known to have the same ID.

\vspace{0.2cm} \noindent
\textit{Decoding the marker}:
For data-relevant applications, the second part of the pipeline, marker decoding, is enabled.
In this part, we use each marker's bounding box from the previous step to crop out a smaller patch from the original frame (Figure~\ref{fig:ImageProcessing}c).
While cropping, we add a small padding around the marker, i.e., 1/8 of the detected bounding box length.
We then resize the patch to a certain size, i.e., 50px height.
We invert the image such that the markers are dark to follow the ArUco convention.
We apply a Gaussian blur to remove the noise, and apply an adaptive threshold with a block size of 15.
Next, we use the ArUco detection library to decode the marker ID from the binarized patch.

The patch resizing in this part ensures that the applied block size is appropriate regardless of marker location and distance. To determine this, we applied different filtering operations to a test recording consisting of 600 frames, where the marker is moved constantly (distance: 10-80cm).
When the adaptive threshold was applied to an uncropped frame, the markers could be decoded from only 85.2\% of the frames. On the other hand, applying the threshold after cropping and resizing allows 98.2\% of the frames to be decoded.

If no ArUco marker is found in the patch, the script checks if there is a QR code or a Data Matrix. For this, we follow the same steps, except for two differences: Because the codes have more bits than ArUcos, we use a rescaling height of 100px and an adaptive threshold block size of 9.
We then use a standard library to decode the codes.
An example is shown in Figure~\ref{fig:ImageProcessing}d.

If an ArUco or a QR code is found in the patch, the script records its data and corners in an array.


\vspace{0.2cm} \noindent
\textit{Tracking decoded markers across frames}: 
We also added a "caching" feature to our detection pipeline that helps track of marker data across frames, without having to run the decoding script in every new frame.
When caching is enabled, we attempt to match the newly localized markers to already decoded markers from the prior frame by comparing their corner coordinates.
If a new marker is found to be similar to a previous marker, we retrieve the marker data from the previous frame.
Otherwise, we run the remaining decoding script.
This is especially useful for QR codes, which traditionally take several frames to be \textit{decoded}, whereas the fluorescent outline can already be \textit{tracked} in all frames. However, to be conservative, we did not enable this feature during our evaluations (Section~\ref{evaluation}).

\vspace{0.2cm} \noindent
\textbf{Implementation}:
We use \textit{OpenCV}~\cite{bradski_opencv_2000}  for the implementation of our system and the \textit{Dynamsoft}\footnote{\url{https://www.dynamsoft.com/}} library for the detection of the 2D barcodes (i.e., QR, Micro QR, and Data Matrix). For rapid prototyping purposes, we ran the detection script on our laptop during the development of this project.
The processing of a single 640x360px frame takes 3.7 ms on average, corresponding to a 270fps on a 2020 \textit{MacBook Pro} with a 2GHz Quad-Core \textit{Intel Core i5} processor. Since our high-speed camera has a limit of 60fps, the detection in our applications was constrained to this rate, but our software pipeline can support higher rates as well.
In our AR/VR applications, after running \textit{OpenCV}'s ArUco pose estimation, we use \textit{Unity}'s coordinate transform features to convert the marker's location to a common local coordinate frame.

\section{Applications}

In this section, we show several applications of BrightMarker, in which the tracking of object locations is an integral part of a process.

\subsection{Rapid Product Tracking}
Because BrightMarkers can localize and track embedded objects, it can be used for industrial or commercial applications in which items need to be processed in a swift manner. This can be especially useful for product and packaging logging where, although the external labels may be intentionally or unintentionally removed, it is still crucial to keep track of the item origin and other supply chain or inventory-related data.

\vspace{0.2cm} \noindent
\textit{Tracing products on conveyor belts}:
In manufacturing and packaging industries, fluorescent markers could be integrated to enable product or part tracing as part of assembly lines (Figure~\ref{fig:teaser}a).

\vspace{0.2cm} \noindent
\textit{On-the-fly inventory logging}:
Users can use the smartphone attachment described in Section~\ref{DetectionImagingModule} to activate the fluorescence of BrightMarker, and scan a batch of products by moving the phone to quickly capture all codes.


\vspace{0.2cm} 
We note that currently, our approach is more suitable for products that are already being 3D printed (e.g. shoes~\cite{trapp_improving_2022}), and could be extended to those that will be printed in the future.

\subsection{Wearables for Tracking Human Motion}
\begin{figure}[t]
  \centering
  \includegraphics[width=1\linewidth]{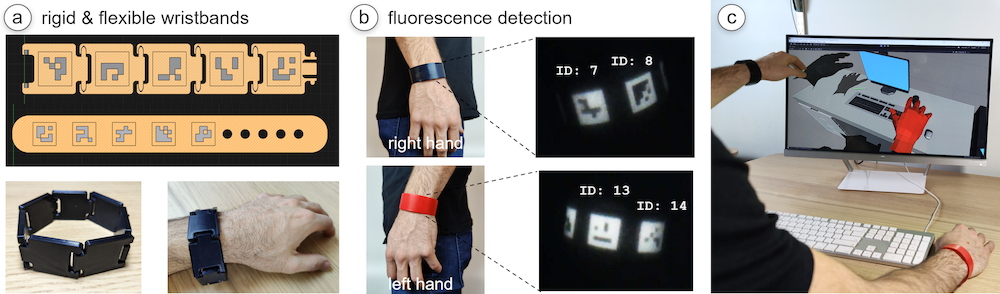}
  \caption{Wearable for hand tracking. (a) Rigid and flexible bracelet designs with embedded markers. (b) Fluorescence imaging is used to detect the unique tracking IDs. (c) The user's motion is digitized.}
  \label{fig:WearableApplication}
\end{figure}

One of the applications of BrightMarker is 3D printing custom wearables for tracking human motion.
Figure~\ref{fig:WearableApplication}a shows rigid and flexible wristbands printed with embedded markers.
Embedding unique markers allows us to digitize the user's motion and distinguish between right and left hands (Figure~\ref{fig:WearableApplication}b). 
In our 
setup, the detection module is mounted above the user's desk (Figure~\ref{fig:WearableApplication}c).

Such tracking wearables can allow various use cases unobtrusively, such as creating digital twins and animations, increasing safety in human-machine collaborations, posture correction warnings, or device control.
Compared to existing tracking methods, the use of BrightMarker preserves the user's privacy since the camera only captures the marker, not the user's face or environment. Furthermore, current methods, especially those based on machine learning, are usually tuned for able-bodied people's hands. Wearables with BrightMarker could support applications for people with limb differences or hand impairments. 
In our current implementation, we produced the black wristband using PLA and the red one using ABS.
Both wristbands have a small thickness, thus allowing them to be bent. In the future, more custom wearables could be printed using more flexible materials such as TPU or by utilizing FDM-based textile fabrication methods~\cite{forman_defextiles_2020, takahashi_3d_2019}.


\subsection{Tangible Interfaces for MR Experiences}
\label{applicationsARVR}

BrightMarkers  can be embedded into objects to turn them into tangibles with more precise tracking capabilities for mixed reality (MR)~\cite{dogan_fabricate_2022}. For instance, opportunistic tangible interfaces could enhance AR interactions~\cite{du_opportunistic_2022}, or serve as haptic placeholders in VR.



\vspace{0.2cm} \noindent
\textit{Appropriating physical parts as precise AR input tools}:
Figure~\ref{fig:teaser}c shows a loudspeaker that has been unused in an office. 
The user wants to make use of the unplugged speaker by transforming it into a passive tangible interface for controlling the volume and bass of his AR glasses. This allows him to have a more natural and tangible input method, while touching small buttons on the glasses can be cumbersome.

BrightMarker has the benefit that the marker objects also include the part's identifier, 
i.e., the top and bottom knobs can be distinguished and assigned unique functionalities.
Further, since the knobs have a uniform color and shape, it would be difficult to precisely track their rotation without the embedded BrightMarkers.

\begin{figure}[h]
  \centering
  \includegraphics[width=0.6\linewidth]{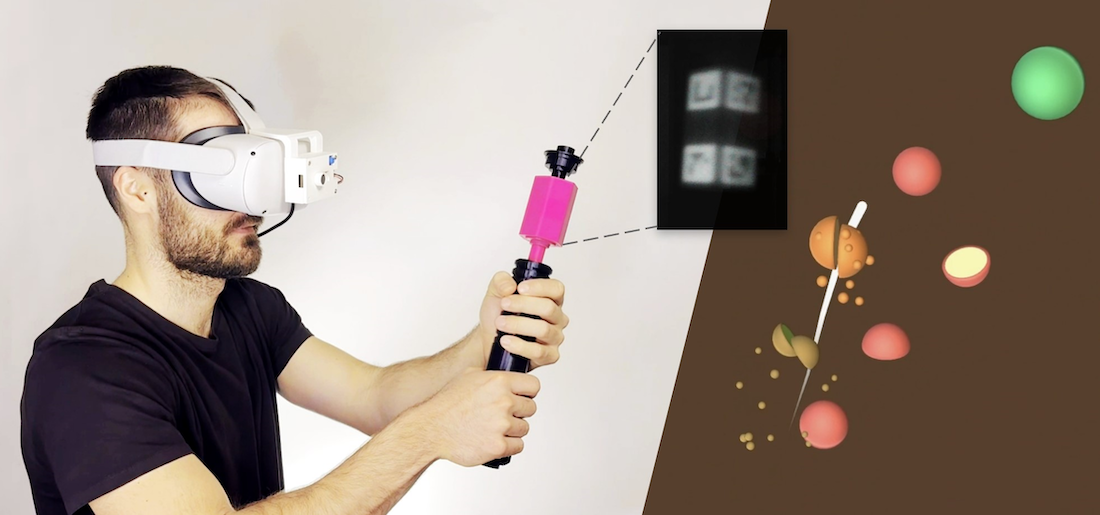}
  \caption{Using a lightsaber as a prop to slice fruits in a game.}
  \vspace{-0.2cm}
  \label{fig:VRHaptic}
\end{figure}

\vspace{0.2cm} \noindent
\textit{Real-life objects as VR haptic props}:
BrightMarker allows users to make use of the physical shape of existing real-world objects (e.g., toys, gadgets, sports gear) as haptic proxies in VR. For example, a "lightsaber" toy could come with integrated BrightMarkers, so it can be used as a different object in games.
As shown in Figure~\ref{fig:VRHaptic}, the lightsaber's hilt is used as a haptic placeholder for a sword to slice fruits in a game.
Another benefit is that the objects are fully passive, while typical VR game controllers contain infrared LEDs that need to be powered up.


\subsection{Privacy-Preserving Night Vision}
\begin{figure}[t]
  \centering
  \includegraphics[width=0.7\linewidth]{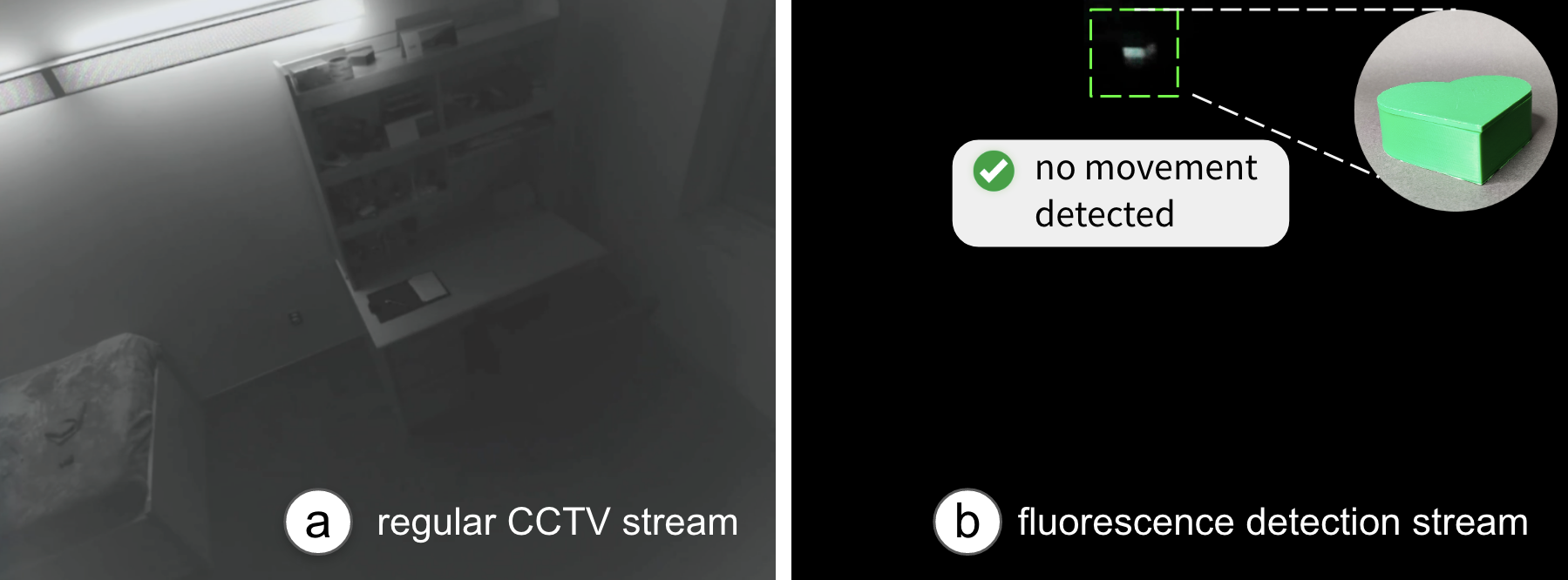}
  \caption{Privacy-preserving night vision. (a) Regular CCTVs help monitor important objects but may intrude on users' privacy. (b) Our detection setup allows tracking of solely the fluorescent objects.}
  \label{fig:NightVision}
\end{figure}

Traditional security cameras use infrared LEDs to monitor environments at night. However, these cameras might not be optimal for use in private environments, such as one's bedroom, although 
users may still want an alternative method to ensure the security of their valuable belongings.
BrightMarker's imaging system removes all details in the camera stream except for the marked objects. In Figure~\ref{fig:NightVision}, a box that stores valuables was tagged with a BrightMarker, and the rest of the object surface was uniformly inlaid with the fluorescent filament so it can be captured from afar. The camera detects the shiny outline of the object, it triggers an alarm. While doing this, it preserves the user's privacy. This fluorescence-based monitoring approach could be enabled by modifying existing home security cameras, i.e., by attaching the appropriate filter on the camera and an IR source next to it
as described in Section~\ref{Imaging}.

\section{Evaluation}
\label{evaluation}
In this section, we evaluate our system's performance under a variety of conditions, including detection distance, illumination, shell color, and speed.

\subsection{Detection Distance, Excitation Intensity, and Surface Color}
Because the embedded filament fluoresces more when exposed to a greater amount of IR excitation, the maximum distance at which a marker can be detected depends on the amount of excitation it receives. To evaluate how the IR intensity affects detection, we conducted an experiment determining the minimum amount of IR required to track vs. decode markers at varying distances from the camera. We consider tracking to be when the program finds the bounds of the BrightMarker; in contrast, decoding occurs when it is also able to extract the \textit{data} of the marker (Section~\ref{ImageProcessing}).

\vspace{0.2cm} \noindent
\textit{Procedure:}
We printed three square BrightMarkers with a 4x4 ArUco pattern of size 1"x1" (2.54cm x 2.54cm).
Each BrightMarker was printed using shell material of a different color (blue ABS, red ABS, and black IR-PLA).
We picked these to represent the variety in materials and fluorescence strength.
We placed a 220cm tape measure along a table in a dark room (3lux ambient light). The NIR camera was aligned at the 0cm mark.
As a baseline, we also recorded the maximum distance that a (visible) paper ArUco marker of the same size could be decoded using solely the off-the-shelf ArUco detection library.

For each trial, we started by placing a BrightMarker 10cm away from the camera. Then, we moved the excitation light as far away from the marker as possible until the marker is just able to be tracked. Keeping the light at this position, we measured the amount of illumination reaching the marker surface using a \textit{Tacklife LM01} digital luxmeter. We then repeated this, but for the amount of illumination required for decoding rather than tracking.
The minimum illumination needed to track and to decode the BrightMarker was recorded for each interval of 10cm until 220cm.

\begin{figure}[t]
  \centering
  \includegraphics[width=0.64\linewidth]{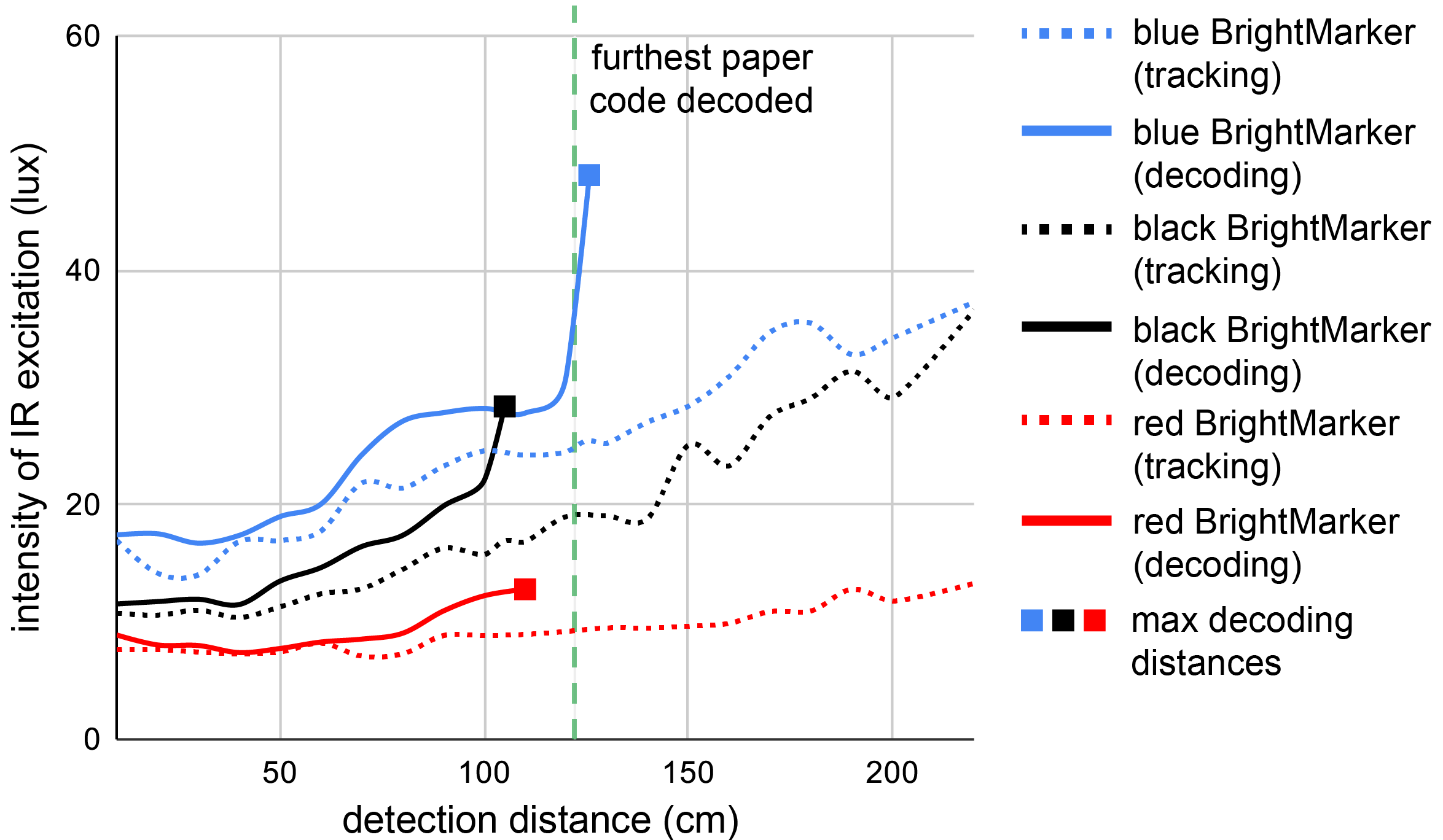}
  \caption{Excitation intensity required with increasing marker distance from the camera.}
  \label{fig:DistanceEvaluation}
\end{figure}

Since different shell colors allow for varying levels of fluorescence, we conducted trials for the different color prints.
The chosen colors represent a wider range of fluorescence, including those which only allow minimal fluorescence (e.g., blue) and those with substantial fluorescence (e.g., red).
We repeated the procedure three times for each color and plotted the average results in Figure~\ref{fig:DistanceEvaluation}.

\vspace{0.2cm} \noindent 
\textit{Results:}
The tracking and decoding results are plotted separately for each color in  Figure~\ref{fig:DistanceEvaluation}.
The dashed lines represent the intensity needed to \textit{track} the marker at each distance. The solid lines, however, represent the intensity to \textit{decode} the marker at each distance.

As a baseline, the dashed green line shows the maximum distance at which a standard paper marker can be tracked and decoded (i.e., 122cm).
We found that since the NIR capture filters out wavelengths lower than the  BrightMarker emission, markers are able to be tracked at distances far greater than a paper marker.
On the other hand, BrightMarker decoding distances were not expected to exceed this value, but were expected to be relatively close to it.
While the red and black samples exhibited this behavior, the blue sample exceeded the expectation and could be decoded just past the point where a paper code can be detected.
We believe this is because the filtered capture allows the decoding algorithm to ignore the noise that is in the unfiltered capture; thus, allowing it to decode the marker at a further distance.
However, this does not occur for other colors because unlike blue, which allows lower levels of fluorescence, the emitted light in the red and black markers slightly bleed into the shell of the markers.

The plots also indicate a correlation between material color and the intensity of light required for detection.
Overall, lower curves indicate easier detection than higher curves, since a low curve in this plot means that less light is needed given a marker distance.
The more fluorescence that the shell lets through, the less excitation is required to detect the marker, e.g., less light is required for red prints than blue prints. This is expected since the NIR camera is capturing the emitted light, so the materials that allow for more emission require less overall excitation. 
For this reason, we recommend using shell materials such as red ABS for tracking from far distances using weaker excitation sources.

\subsection{Detection Rate and Marker Speed}
\label{EvaluationDetectionRate}
Because the fluorescence of BrightMarkers is imaged using a longpass filter, our detection setup eliminates non-marker elements from the scene, thus reducing the chance of undetected frames. We conducted an experiment to compare how the fluorescence-based detection approach improves the detection rate compared to regular \textit{InfraredTags}~\cite{dogan_infraredtags_2022} as the tags are being moved.

\vspace{0.2cm} \noindent
\textit{Procedure:}
We printed a red and blue BrightMarker with a 4x4 ArUco pattern  
(size: 2.54cm x 2.54cm)
to represent high and low fluorescence.
We then used the printer as a CNC tool: We attached each marker separately on the printhead of an \textit{Ultimaker 3}, and placed our detection module on the printbed, with its camera looking upward at the marker.
We moved the printhead in an upward conical helix trajectory to cover the remaining printer volume (21.6x21.6x130cm), while also staying in the camera's field of view. We repeated this for 12 different head speeds (range: 20mm/s - 240mm/s).
We computed the detection rate throughout the trajectory from the resulting 60-fps videos.

\vspace{0.2cm} \noindent
\textit{Results:}
Across the 12 speed values, the \textit{InfraredTag} could be tracked and decoded in only 60.73\% of the frames on average (std=2.26).
For BrightMarkers, the red and blue sample could be \textit{tracked} in 100\% of the frames. 
The red sample could be \textit{decoded} in 99.41\% of the frames (std=0.48) and the blue sample 99.83\% (std=0.19). The small standard deviation values across the different speeds show that the high-speed camera employed in our imaging module successfully avoids motion blur due to its global shutter (Section~\ref{Imaging}).

\subsection{Marker Size}
To test what the smallest detectable BrightMarker size is, we printed samples in 1mm dimension increments (range: 5-10mm) separately
 for red and blue ABS surface materials.
For both colors,  the smallest decodable marker was 6mm wide.
However, the camera had to be 2cm close to the blue marker to decode it, while the red marker could be decoded from as far as 7cm.
This is likely due to the higher IR absorbance of the blue shell material.

 \section{Discussion}
In this section, we discuss the limitations of BrightMarker and potential directions for future research.

\subsection{Concentration of Fluorescent Dye}

The concentration of the fluorescent dye used in the filament of BrightMarkers is an important factor in its detectability.


To test how the dye concentration affects the marker intensity, we obtained two small test plates from the filament manufacturer, since a high-concentration filament is not readily available yet.
Each plate was doped with fluorochrome, one with the same concentration as our filament, and another one with eight times the standard concentration.
Given the same level of excitation, we measured the intensity of the emitted light when the plates are uncovered. 
We repeated this after covering both samples with a 1.2mm sheet made from regular red ABS filament, mimicking BrightMarker's $t_{shell}$.

In the uncovered comparison, we observed that the highly doped sample resulted in 33\% more IR intensity than the regular sample.
In the covered comparison, which represents the conditions for BrightMarker,  the intensity of the highly doped sample was 20\% higher than the regular sample. 
Thus, to further increase detection performance,
further research and optimization could lead to the production of filaments with higher fluorochrome concentrations.

\subsection{Mass Production}
In the future, the BrightMarker approach may be adapted for mass-market manufacturing. For instance, instead of 3D printing, plastic overmolding could be utilized to include fluorescent materials in mass-manufactured products without any post-processing steps.

\subsection{NIR Power}
While the LED component we used in the AR/VR attachment module (Section~\ref{DetectionImagingModule}) is rated 10W, we did not use it at full power for the headset. The potentiometer in the PCB allows us to adjust the power from 1.8W up to 3.5W. At the lowest power, a black IR PLA-covered sample can be tracked from up to 50cm away. At max power, it can be tracked at 90cm away.
While, as far as we know, there are no longitudinal studies yet on NIR and eye safety,
commonly used examples of NIR light include \textit{iPhone FaceID}, \textit{Microsoft Kinect}, and \textit{Intel RealSense} depth cameras. For instance, RealSense D400's IR projector~\cite{intel_realsense_2020} can consume up to 4.25W.

\subsection{Embedding Circular Markers}
In this project, we embedded \textit{ArUco} markers for tracking purposes. Another way to add trackers would be to use the fluorescent filament insert small circular markers, similar to \textit{OptiTrack}'s retro-reflective beads, underneath the object surface. Such triangulation-based tracking systems are typically deemed more robust than square marker-based methods. However, since triangulation-based methods require multiple cameras to be set up, a costly and cumbersome process, we chose to use ArUco markers for their simplicity which suits our everyday applications. Interested researchers can use our embedded fabrication approach for other tracking methods based on their needs and constraints. In the future, this could also be used to improve detection for more intricate objects, i.e., those with high-frequency surface details that would lead to local occlusion of the markers based on the viewing angle.

\subsection{Occlusion}
Object occlusion is a typical problem for most tracking methods.
Similarly, BrightMarker could be occluded by the user's hands and other objects.
One way to overcome this is to couple the optical tracking with magnetic tracking methods (i.e., hybrid tracking)~\cite{state_superior_1996}. A magnetic material could be embedded in the printed object using magnetic filaments, which would enhance the tracking when the object is occluded by utilizing magnetic sensors.

\section{Conclusion}

We presented BrightMarker, a novel method for embedding and tracking hidden high-contrast markers using fluorescent 3D~printing filaments.
Our approach offers an easy-to-use solution for marker-based tracking without affecting the object's look or shape.
We showed that  BrightMarkers can be embedded in various object colors, and can be easily localized using a light source and camera filter that match the fluorescence characteristics of the material.
Our CAD tool allows users to add markers to their 3D models before printing, and our optical detection hardware can be attached to existing AR/VR headsets for marker tracking.
BrightMarker's image processing pipeline uses the captured images to robustly localize the markers.


Our applications demonstrate rapid product tracking, custom-fabricated wearables, tangible interfaces in AR/VR, and privacy-preserving night vision.
Our evaluation shows that markers of different colors can be detected from afar and at various object speeds. 
We believe that our work demonstrates the potential of using fluorescence as an effective and versatile method for embedding invisible markers, and we hope it fosters further exploration and innovation in this area by the HCI and fabrication community.


\chapter{Discussion}
\label{thesis-Discussion}

Together, the presented tagging approaches (\textit{natural}, \textit{structural}, and \textit{internal}) provide effective ways of how to include identifiers and markers in objects without disrupting their look, feel, and functionality. This is especially true for natural markers as they, by definition, do not alter any feature of the object. However, for structural and internal markers, the design needs a more meticulous thought process so as to not impact these aspects.


While our presented approaches align with the design criteria outlined in Section~\ref{DesignCriteria},
they also represent a targeted exploration of the broader design space for embedded markers. The precise dimensions of this design space have not been the primary focus of this thesis.
Nonetheless, future investigations could extend their scope to provide a more thorough exploration of the specific dimension axes.

These embedded markers contribute to the bigger vision of enabling users to more seamlessly identify products, devices, and other items in their surroundings and access more information related to them rapidly. Further, these techniques can be used in various industries that inherently rely on tags, such as retail, packaging, shipping, robotics, and manufacturing.

The following sections explore various aspects and implications of the embedded markers and the vision of ubiquitous metadata. We explore key topics related to scaling up for mass production, privacy considerations, the markers' use in consumer products, and interaction and personalization in AR. Through these discussions, we aim to gain deeper insights into the challenges and potential solutions that arise in the context of digital fabrication, mixed reality, and ubiquitous metadata.

\section{Scaling Up and Mass Production}
The successful implementation of embedded machine-readable markers extends beyond 3D printing and requires consideration for generalization to more common and scalable manufacturing methods. In this section, we discuss the challenges and opportunities associated with extending the use of embedded tags to diverse manufacturing processes to allow for broader adoption and mass-scale production.

\textit{Scalability} is a key aspect to consider when generalizing embedded markers to more commonly used manufacturing methods. While 3D printing offers flexibility and customization, its relatively slow production speed makes it less suitable for large-scale manufacturing. Therefore, it is necessary to explore and adopt faster and more scalable methods, such as injection molding, casting, and CNC machining, to meet an increasing demand for tagged objects.
To ensure compatibility with these scalable fabrication methods, the design of embedded tags should be adaptable and compatible with the specific constraints, material properties, and production workflows associated with each technique. This requires careful consideration of factors such as the choice of materials and geometric design optimizations. 


\textit{Standardization} also plays a vital role in the generalization of embedded markers to more common manufacturing methods. Developing industry-wide standards for tag designs and integration processes fosters interoperability and facilitates seamless adoption across different manufacturers and industries. Standardization efforts should consider aspects such as tag specifications and encoding formats. By establishing common practices, manufacturers could more easily incorporate embedded markers into their existing production workflows.

\textit{Cost considerations} are another important factor for scalable manufacturing.  Exploring cost-efficient materials, optimizing fabrication processes, and identifying opportunities for economies of scale are essential to make embedded markers economically viable. Finding a balance between cost and functionality ensures that manufacturers can adopt embedded markers without significant financial barriers while still reaping the benefits of enhanced object identification, tracking, and interaction.

\section{Privacy}
\label{DiscussionPrivacy}
The widespread adoption of embedded machine-readable markers entails important privacy implications for end users~\cite{cheng_seeing_2022}. In this section, we discuss the privacy challenges and potential solutions related to embedded markers.

One of the primary privacy concerns is the unauthorized access or misuse of data encoded in the markers. As embedded markers become more prevalent in various domains, including retail, robotics, and smart homes, the potential risks associated with data breaches and unauthorized tracking of individuals become more significant. It is crucial to implement robust encryption and authentication mechanisms to protect the integrity and confidentiality of the information stored in the markers. This includes techniques such as data anonymization, secure key management, and access control protocols~\cite{govinda_identity_2012, zhong_privacy-enhancing_2005} to prevent unauthorized access and ensure data privacy.

Another privacy consideration is the potential for tracking and profiling of individuals based on the information collected through embedded markers. In certain contexts, such as retail or marketing, embedded markers may be used to track user behavior, preferences, and interactions. To address this concern, privacy-enhancing technologies and techniques should be utilized, such as data minimization, privacy-preserving algorithms, and user consent mechanisms~\cite{nouwens_dark_2020}. By allowing individuals to have control over their data and providing transparent information about data collection and usage, privacy risks can be mitigated.

To address the privacy implications of embedded tags, interdisciplinary collaborations between researchers, policymakers, and industry experts are essential. Establishing privacy guidelines, standards, and best practices specific to embedded tags can help ensure responsible and privacy-preserving deployment.

\section{Embedded Markers in Consumer Products}
One of the main applications of embedded machine-readable markers is the concept of Digital Product Passports (DPP). DPPs serve as comprehensive digital records that accompany physical consumer products throughout their lifecycle, providing valuable information and insights to various stakeholders, including manufacturers, retailers, and consumers~\cite{king_proposed_2023, walden_digital_2021}.

By embedding machine-readable markers into products during the manufacturing process, a wealth of information can be captured and associated with each item. This information can include details about the product's origin, materials used, manufacturing processes, quality certifications, and sustainability attributes. Thus, DPPs could allow the metadata to be securely stored and accessed throughout the product's journey.

For \textit{manufacturers}, DPPs could enable enhanced traceability, supply chain management, and quality control. Manufacturers could monitor the entire lifecycle of their products, ensuring compliance with regulations, identifying potential issues, and tracking product performance. This comprehensive data could enable manufacturers to make informed decisions, optimize processes, and improve sustainability practices.

\textit{Retailers and distributors} could also benefit from DPPs. By accessing the embedded metadata, retailers can verify product authenticity, ensure compliance with standards, and provide consumers with detailed information about the products they offer. This transparency builds trust and enables consumers to make more informed purchasing decisions based on their preferences and values.

\textit{Consumers} may make sustainable and ethical choices, supporting brands that align with their values, using the information from the embedded markers. Additionally, DPPs can facilitate after-sales services, such as warranty registration, maintenance guides, and product recalls, enhancing the overall user experience.

However, the implementation of DPPs faces certain challenges. Data privacy and security are key concerns, as sensitive information is associated with each product. Robust measures must be in place to ensure the secure and controlled access to this information, protecting both the integrity of the data and the privacy of individuals.
Additionally, the standardization and interoperability of metadata formats and protocols are essential to enable seamless exchange of information across different stakeholders and systems.

\section{Interaction and Personalization in AR}
Enabling the detection of physical objects' metadata on AR headsets, such as smart glasses, holds great potential for advancing personalized AR experiences. By incorporating sensing capabilities into these headsets, users can effortlessly interact with their surroundings and receive additional context and information about tagged objects. The continuous scanning of the environment using \textit{always-on} sensors  would eliminate the need for users to point their mobile devices in specific directions to read the tags.

However, it is essential to display the information from the marker only when it is relevant and beneficial to the user. Avoiding information overload is crucial to prevent distractions and maintain focus. Inspired by previous research~\cite{lindlbauer_context-aware_2019}, an optimization-based approach could be pursued to determine which tags should be visible in the AR view. Factors such as the user's gaze, attention level, and current activity will be taken into consideration. Leveraging tools and sensors already present in AR headsets, such as eye tracking, will aid in the computational determination of information visibility in the AR environment. This personalized approach ensures that users receive the most relevant and contextually appropriate information.

\begin{figure}[t]
  \centering
  \includegraphics[width=0.87\linewidth]{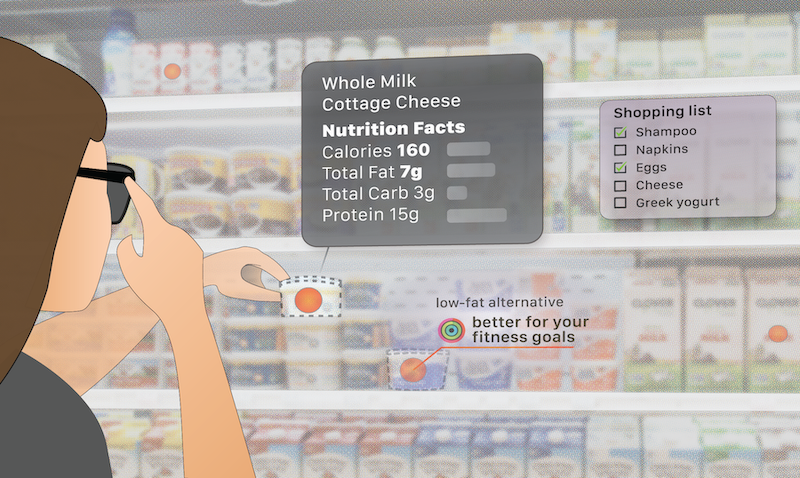}
  \caption{Adaptive display of related information acquired from unobtrusive tags could be utilized to create a more personalized and assistive 
  mixed reality environment.}
  \label{fig:futureWork}
\end{figure}

Moreover, the AR content displayed can be tailored to the user's daily tasks and goals, further enhancing \textit{personalization}~\cite{dogan_standarone_2023}. For instance, a tagged artwork in a museum can trigger personalized information about the artist, the painting's history, or related artworks, providing a more engaging and informative experience for each visitor. When shopping, the embedded tags on product packaging can be spatially displayed and filtered based on the user's predetermined shopping list and fitness goals (Figure~\ref{fig:futureWork}). This tailored approach would provide users with targeted recommendations and relevant information that aligns with their specific needs and preferences. Throughout these interactions, ensuring privacy and obtaining user consent should be prioritized (Section~\ref{DiscussionPrivacy}).

\chapter{Conclusion}


In this thesis, we have explored the design, fabrication, and detection of embedded machine-readable markers that serve as a bridge between the physical and virtual worlds. These tags offer a means to identify, track, and interact with real-world objects in mixed reality and spatial computing. Our research explores three distinct approaches: natural markers, structural markers, and internal markers.

Natural markers utilize the inherent fingerprints of objects and materials.
Structural markers leverage the structural artifacts that emerge during the fabrication process for unique identification.
Internal markers are intentionally engineered within the interior of objects and are made from specialized materials. 
These approaches have been optimized for minimal impact on the object's appearance, cost-effective fabrication using commonly available tools like 3D printers and laser cutters, and detection using readily accessible sensors such as RGB and near-infrared cameras.

Our research envisions a future where every real-world object carries metadata that can be accessed effortlessly using mobile devices or augmented reality glasses, a concept we refer to as "ubiquitous metadata." By combining techniques from computer vision, machine learning, computational imaging, and material science, our approaches provide robust and versatile solutions toward this vision.

In conclusion, our work brings us closer to a future where real-world objects are seamlessly integrated into virtual worlds. By embedding embedded machine-readable markers into real-world objects, we empower users with a wealth of multimedia content, interactive experiences, and contextual information. The implications and applications of our research span a wide range of domains, including product design, manufacturing, entertainment, marketing, logistics, security, and sustainability. With responsible implementation, we can revolutionize object identification and interaction, paving the way for an interconnected and intelligent future.

\appendix
\begin{singlespace}
\bibliography{main}
\bibliographystyle{plainurl}
\end{singlespace}

\end{document}